%% file: Hossein_TCOM_Final_arXiv.tex
\documentclass[lettersize,journal]{IEEEtran}
\usepackage{float}
\usepackage{color}
\usepackage{graphicx}
\usepackage{mathtools}
\usepackage{amsmath,amssymb,amsthm}

\usepackage[caption=false,subrefformat=parens,labelformat=parens]{subfig}
\usepackage{balance}
\usepackage[noadjust]{cite}
\usepackage{epstopdf}
\usepackage{psfrag,pst-pdf}
\usepackage[nolist]{acronym}

\usepackage{algcompatible}
\usepackage{algorithm}
\usepackage{dsfont}

\usepackage{algpseudocode}

\makeatletter
\newcounter{phase}[algorithm]
\newlength{\phaserulewidth}

\makeatother

\usepackage{hyperref}
\input{acronyms.tex}

\begin{document}

\title{A Novel Terabit Grid-of-Beam Optical Wireless Multi-User Access Network with Beam Clustering}

% \author{\IEEEauthorblockN{Hossein~Kazemi~\IEEEmembership{Member,~IEEE}, Elham~Sarbazi, Michael~Crisp~\IEEEmembership{Member,~IEEE}, Taisir~E.~H.~El-Gorashi, \\Jaafar~M.~H.~Elmirghani~\IEEEmembership{Fellow,~IEEE}, Richard~V.~Penty~\IEEEmembership{Senior Member,~IEEE}, Ian~H.~White~\IEEEmembership{Fellow,~IEEE}, Majid~Safari~\IEEEmembership{Senior Member,~IEEE} and~Harald~Haas~\IEEEmembership{Fellow,~IEEE}\vspace{-27pt}}}

\author{\IEEEauthorblockN{Hossein~Kazemi, Elham~Sarbazi, Michael~Crisp, Taisir~E.~H.~El-Gorashi, \\Jaafar~M.~H.~Elmirghani, Richard~V.~Penty, Ian~H.~White, Majid~Safari and~Harald~Haas\vspace{-15pt}}}

\maketitle

%%%%%%%%%%%%%%%%%%%%%%%%%%%%%%%%%%%%%%%%%%%%%%%%%%%%%%%%%%%%%%%%%%%%%%%%%%%%%%%%%%%%%%%%%%%%%%%%%%%%
%%%%%%%%%%%%%%%%%%%%%%%%%%%%%%%%%%%%%%%%%%%%%%%%%%%%%%%%%%%%%%%%%%%%%%%%%%%%%%%%%%%%%%%%%%%%%%%%%%%%
\begin{abstract}
In this paper, we put forward a proof of concept for sixth generation (6G) Terabit infrared (IR) laser-based indoor optical wireless networks. We propose a novel double-tier access point (AP) architecture based on an \textit{array of arrays} of vertical cavity surface emitting lasers (VCSELs) to provide a seamless grid-of-beam (GoB) coverage with multi-Gb/s per beam. We present systematic design and thorough analytical modeling of the AP architecture, which are then applied to downlink system modeling using non-imaging angle diversity receivers (ADRs). We propose static beam clustering with coordinated multi-beam joint transmission (CoMB-JT) for network interference management and devise various clustering strategies to address inter-beam interference (IBI) and inter-cluster interference (ICI). Non-orthogonal multiple access (NOMA) and orthogonal frequency division multiple access (OFDMA) schemes are also adopted to handle intra-cluster interference, and the resulting signal-to-interference-plus-noise ratio (SINR) and achievable data rate are derived. The network performance is studied in terms of spatial distributions and statistics of the downlink SINR and data rate through extensive computer simulations. The results demonstrate that data rates up to $15$~Gb/s are achieved within the coverage area and a properly devised clustering strikes a balance between the sum rate and fairness depending on the number of users.
\end{abstract}


\vspace{-5pt}

%%%%%%%%%%%%%%%%%%%%%%%%%%%%%%%%%%%%%%%%%%%%%%%%%%%%%%%%%%%%%%%%%%%%%%%%%%%%%%%%%%%%%%%%%%%%%%%%%%%%
%%%%%%%%%%%%%%%%%%%%%%%%%%%%%%%%%%%%%%%%%%%%%%%%%%%%%%%%%%%%%%%%%%%%%%%%%%%%%%%%%%%%%%%%%%%%%%%%%%%%
\section{Introduction} \label{Introduction}
The rapid technological advancement in today's digital world has expedited the availability of diverse Internet-driven premium services such as live 4K and 8K \ac{UHD} video streaming, immersive experience and \ac{3D} stereoscopic vision using \ac{VR} and \ac{AR}, holographic telepresence and multi-access edge computing, which will exceedingly push wireless connectivity limits over the coming years \cite{Giordani2020toward6G}. The real time operation of these technologies will require an unprecedented system capacity beyond $1$ \ac{Tb{/}s}. Achieving \ac{Tb{/}s} aggregate data rates indeed constitutes one of the key performance indicators for developing the future \ac{6G} wireless systems \cite{Strinati20196G}.

Optical wireless communication (OWC)\acused{OWC} technology based on narrow-beam \ac{IR} lasers has demonstrated the potential to provide ultra-high-speed transmissions for indoor multi-user wireless networks \cite{Koonen2018OWC,TKoonen2020ultra,TKoonen2018high,KWang2018,KWang2011high}. Beam-steered \ac{IR} light communication is tailored to provide multi-Gb/s data rates per user by directing narrow \ac{IR} laser beams at mobile devices \cite{Koonen2018OWC,YHong2021, RSingh2020}. One of the main challenges for the realization of indoor narrow-beam \ac{IR} systems is beam steering as addressed in published research \cite{TKoonen2016ultra,TKoonen2018high,KWang2018}. In \cite{TKoonen2016ultra}, Koonen \textit{et al.} have proposed and investigated a technique for \ac{2D} steering of directional \ac{IR} pencil beams (i.e., beams with very low divergence) by means of a passive diffractive optical module composed of crossed grating elements whose output beam angle is controlled by the wavelength of the laser beam. The authors have demonstrated downstream delivery with data rates of $32$ Gb/s and $42.8$ Gb/s per beam using $4$-\ac{PAM} and adaptive \ac{DMT} modulation, respectively, for a wavelength of $1550$ nm, with an angular coverage of $6^\circ\times12^\circ$ at a $3$ m distance, equivalent to an area of $32\times64$~cm$^2$. In \cite{TKoonen2018high}, Koonen \textit{et al.} have proposed an alternative approach to implement wavelength-controlled \ac{2D} steering of \ac{IR} pencil beams by using a high port-count \ac{AWGR} encompassing a bundle of \acp{SMF} with distinct wavelengths. To this end, they have rearranged the output fibers of the \ac{AWGR} as a \ac{2D} square array and placed a lens next to the array, and the beam direction for each fiber is thereby determined by its position with respect to the optical axis of the lens. The authors have shown $112$~Gb/s $4$-\ac{PAM} transmission based on an $80$-port C-band \ac{AWGR} and a $9\times9$ fiber array, aiming to cover an area of $0.75\times0.75$~m$^2$ with beam spots of diameter $8.5$~cm at a distance of $3.4$~m. They have then anticipated that this system can potentially achieve a total throughput of $8.9$~Tb/s (i.e., $80\times112$~Gb/s). In \cite{KWang2018}, Wang \textit{et al.} have designed a silicon photonic integrated phased array as an optical beam steering device, whereby they have achieved $12.5$ Gb/s error-free transmission over a distance of $1.4$ m using \ac{OOK} modulation.

The aforementioned works apply advanced optical designs for beam steering to provide \ac{IR} laser-class indoor coverage. However, the high complexity and expensive components associated with beam steering systems do not suit large-scale indoor network deployments. Taking a different approach, in \cite{CSun2019beam,CSun2020networked}, Sun \textit{et al.} have proposed a beam-domain massive \ac{MIMO} \ac{OWC} system in which a large transmit lens is employed in front of a \ac{LED} array. The lens refracts lights emitted from the \ac{LED} array elements toward different directions, enabling the optical \ac{AP} to simultaneously communicate with a large number of user terminals in a wide coverage area. The authors have presented performance analysis for the optical massive multi-user \ac{MIMO} transmission system with linear precoding and derived the optimal transmit covariance matrix under the total and per \ac{LED} power constraints. Based on their findings, as the number of \acp{LED} approaches infinity, the asymptotically optimal multi-user transmission policy under both power constraints is to use non-overlapping light beams, which is referred to as \ac{BDMA}. According to \cite{CSun2020networked}, considering a scenario with four \acp{AP} each made of a $12\times12$ \ac{LED} array and a transmit lens that cover $12$ users within a $5\times5$~m$^2$ cell in a room of height $3$~m, the data rate per user of $15$ bits per channel use is achieved based on \ac{BDMA} for a $30$~dB transmitted \ac{SNR}.

Aiming for ultra-high transmission capacities and a full coverage for indoor environments, we have considered laser-based \ac{IR} \ac{OWC} systems by proposing an access network design in \cite{ESarbazi2020tb}, and a backhaul system design in \cite{HKazemi2022tb}. In both designs, we have used \ac{VCSEL} arrays to attain aggregate data rates beyond $1$ Tb/s. The choice of \acp{VCSEL} over \acp{LED} or other types of laser diodes is because of their prominent features including a high power efficiency, a high modulation bandwidth and well-controlled output beam properties. \acp{VCSEL} are easier to fabricate and to be precisely arranged as \ac{2D} arrays in contrast to edge emitting lasers. They also have a low manufacturing cost, enhanced reliability, and circularly symmetric output beam profile \cite{RMichalzik2012}. In \cite{YLiu2023}, Liu \textit{et al.} have proposed an optical wireless transmitter system design based on a \ac{VCSEL} array to provide a multi-beam coverage with uniform beam spots for indoor applications. The authors have adopted an optical design with multiple cascaded components for beam collimation, homogenization and expansion, by using micro-lens arrays, and plano-concave and plano-convex lenses. Based on a $5\times5$ \ac{VCSEL} array, they have demonstrated a total coverage of $1\times1$~m$^2$ with each beam spot of size $20\times20$~cm$^2$ in a $3$ m distance using Zemax OpticStudio simulations. As a proof of concept, the authors have also presented experimental results for a linear $1\times4$ \ac{VCSEL} array, confirming a data rate of $8$~Gb/s within each beam spot at a pre-\ac{FEC} \ac{BER} of $3\times10^{-3}$. They have anticipated that this transmitter design could achieve an aggregate data rate of $200$~Gb/s. However, the proposed structure is complex and requires meticulous design procedures to ensure a precise alignment between the optical components.

In this paper, we propose a novel double-tier \ac{AP} architecture incorporating an array of \ac{VCSEL} arrays, which we also refer to as \textit{array of arrays} for brevity. The \ac{AP} architecture comprises multiple transmitter elements with each one having a predefined orientation to cover a specific area of the network. Each transmitter element is composed of a \ac{VCSEL} array and a plano-convex lens. This design offers a rather simple and scalable solution suitable for establishing an indoor network of densely deployed laser beams to provide a seamless \ac{GoB} coverage through a single \ac{AP} without the need for beam steering. To unlock the ultra-high capacity of the proposed \ac{AP} and to eliminate strict alignment requirements, we employ a non-imaging \ac{ADR} design based on \acp{CPC} coupled with high-speed \ac{PD} arrays \cite{ESarbazi2022Design2}. Besides, the contributions of this paper are summarized as follows:

\begin{itemize}
    \item The in-depth analytical modeling of the proposed \ac{AP} is carried out, taking into account various design parameters including those of the \ac{VCSEL} array and lens as well as the geometric variables of the \ac{AP}.
    \item Static beam clustering strategies with \ac{CoMB} \ac{JT} are devised and investigated for downlink interference management.
    \item The multi-user performance of the \ac{GoB} optical wireless network is thoroughly analyzed. To this end, the downlink system is elaborated in terms of the received \ac{SINR} and achievable rate for both \ac{NOMA} and \ac{OFDMA} schemes, involving beam clustering and \ac{CoMB}-\ac{JT} in conjunction with non-imaging \ac{ADR} and \ac{MRC}.
    \item The system performance is evaluated through extensive computer simulations, providing key insights into the spatial \ac{GoB} coverage and the impact of the distribution of \ac{UE} devices in various beam clustering scenarios. The results evince how a judicious choice of the clustering layout besides the number of clusters can realize the best performance by maintaining the balance between the sum rate and fairness.
\end{itemize}

The rest of the paper is organized as follows. In Section~\ref{BeamModel}, the propagation model of a Gaussian laser beam is concisely described. In Section~\ref{AccessPointDesign}, the design and analysis of the double-tier \ac{AP} architecture are presented. In Section~\ref{BeamClustering}, the downlink transmission modeling of beam clustering with \ac{CoMB}-\ac{JT} and the performance analysis are carried out. In Section~\ref{NumericalResults}, the proposed beam clustering scenarios along with numerical results and discussions are provided. In Section~\ref{Conclusions}, concluding remarks and future research directions are discussed.

\vspace{-3pt}

%%%%%%%%%%%%%%%%%%%%%%%%%%%%%%%%%%%%%%%%%%%%%%%%%%%%%%%%%%%%%%%%%%%%%%%%%%%%%%%%%%%%%%%%%%%%%%%%%%%%
%%%%%%%%%%%%%%%%%%%%%%%%%%%%%%%%%%%%%%%%%%%%%%%%%%%%%%%%%%%%%%%%%%%%%%%%%%%%%%%%%%%%%%%%%%%%%%%%%%%%

\section{Gaussian Beam Propagation} \label{BeamModel}
A single-mode \ac{VCSEL} generates the fundamental \ac{TEM} mode, designated as TEM$_{00}$ mode, which is a spherical Gaussian beam with a circularly symmetric intensity profile. A Gaussian beam is characterized by the radius of the beam waist $w_0$ and the wavelength $\lambda$. The beam waist is where the wavefront is planar and the beam diameter is minimum. Assuming that the laser beam is propagating along the $z$ axis and the beam waist is located at the origin of cylindrical coordinates, the intensity distribution is expressed as \cite[p.~84]{Saleh2019}:
\begin{equation}
	I(r,z) = \frac{2P_\mathrm{t}}{\pi w^2(z)}{\exp{\left(-\dfrac{2r^2}{w^2(z)} \right)}} ,
	\label{Eq:I_G}
\end{equation}
where $P_\mathrm{t}$ is the transmit optical power, and $r$ and $z$ are the radial and axial positions, respectively. The radius of the beam spot at distance $z$ from the beam waist is given by \cite[p.~82]{Saleh2019}:
\begin{equation}
    w(z) = w_0 \sqrt{1+ {\left(\dfrac{z}{z_\mathrm{R}}\right)}^2}, \quad z_\mathrm{R}=\dfrac{\pi {w^2_0}}{\lambda},
    \label{Eq:w_z}
\end{equation}
where $z_\mathrm{R}$ represents the Rayleigh range.
Although the tail of the Gaussian beam intensity never actually reaches zero, the \textit{edge} of the beam is determined by the radial distance of $r = w(z)$. That is where the intensity drops to a fraction $e^{-2}$ (i.e., about $\%14$) of its value on the propagation axis. The wavefront radius of curvature is given by \cite[p.~82]{Saleh2019}:
\begin{equation}
    R(z) = z \left[ 1+ {\left( \dfrac{z_\mathrm{R}}{z}\right)}^2 \right].
    \label{Eq:R_z}
\end{equation}
By using \eqref{Eq:w_z}, the divergence angle of the beam in the far field is obtained as \cite[p.~85]{Saleh2019}:
\begin{equation}
	\theta = \lim_{z \to \infty} \tan^{-1}\left(\frac{w(z)}{z}\right) = \frac{\lambda}{\pi w_0} .
	\label{Eq:2_5}
\end{equation}
A Gaussian beam can also be fully characterized based on its complex $q$-parameter, which is defined as \cite[p.~81]{Saleh2019}:
\begin{equation}
    q(z) = z + jz_\mathrm{R}.
    \label{Eq:q}
\end{equation}
In fact, the real and imaginary parts of the reciprocal of $q(z)$ encompass $R(z)$ and $w(z)$ as follows \cite[p.~81]{Saleh2019}:
\begin{equation}
\frac{1}{q(z)}=\frac{1}{R(z)}-j\frac{\lambda}{\pi w^2(z)} .
\label{Eq:q_inverse}
\end{equation}

\vspace{-5pt}

%%%%%%%%%%%%%%%%%%%%%%%%%%%%%%%%%%%%%%%%%%%%%%%%%%%%%%%%%%%%%%%%%%%%%%%%%%%%%%%%%%%%%%%%%%%%%%%%%%%%
%%%%%%%%%%%%%%%%%%%%%%%%%%%%%%%%%%%%%%%%%%%%%%%%%%%%%%%%%%%%%%%%%%%%%%%%%%%%%%%%%%%%%%%%%%%%%%%%%%%%
\section{Double-Tier Access Point Design} \label{AccessPointDesign}
\subsection{Array of Arrays Architecture}
\begin{figure}[!t]
	\centering
	\includegraphics[width=0.45\textwidth]{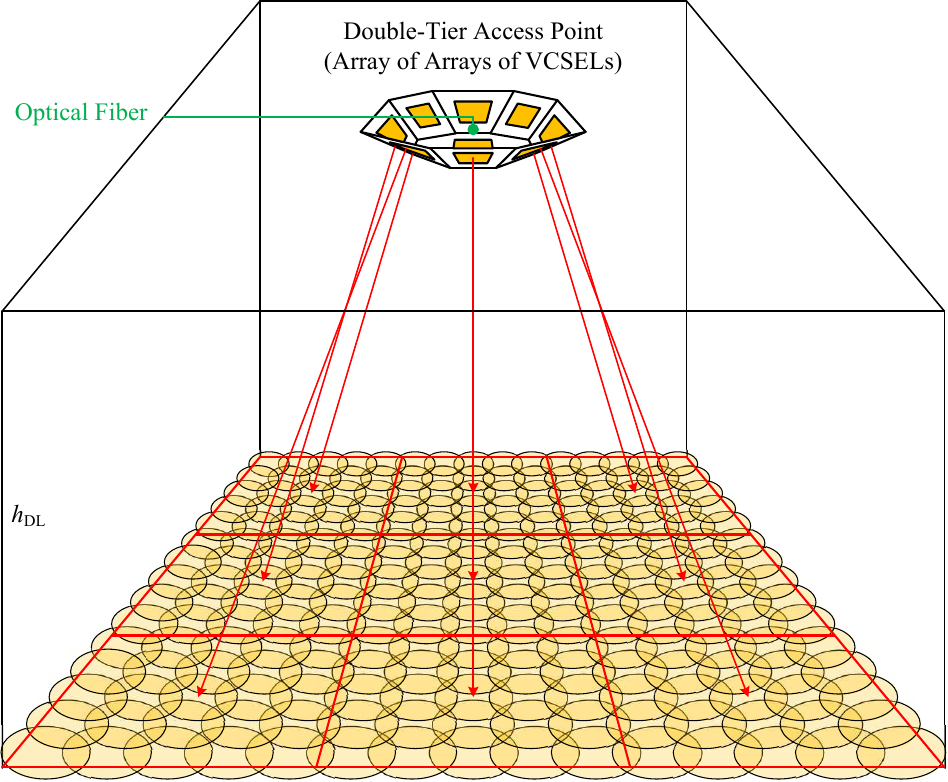}
	\caption{Indoor grid-of-beam optical wireless multi-user access network using the proposed double-tier AP design based on an array of arrays of VCSELs.}
	\label{Fig:NetworkArchitecture}
	\vspace{-15pt}
\end{figure}

Fig.~\ref{Fig:NetworkArchitecture} depicts an indoor optical wireless network with full beam coverage by deploying a massive number of laser beams based on an array of arrays of \acp{VCSEL}. The proposed double-tier \ac{AP} design is composed of $9$ identical transmitter elements arranged as a $3\times3$ array with a $5\times 5$ \ac{VCSEL} array incorporated in each element\footnote{The term `double-tier' refers to the two-tier addressing process required to reach each \ac{VCSEL} in the array of arrays architecture.}. The total number of \acp{VCSEL} used in this array of arrays architecture is $N_\mathrm{VCSEL}=225$. By choosing appropriate orientation angles for transmitter elements, each \ac{VCSEL} array covers almost a ninth of the network area, as shown in Fig.~\ref{Fig:NetworkArchitecture}. Note that the middle element of the \ac{AP} that covers the central region of the network is not titled.

\begin{figure}[!t]
    \centering
    \subfloat[\label{Fig:AP_a}]{\includegraphics[height=0.15\textheight, keepaspectratio=true]{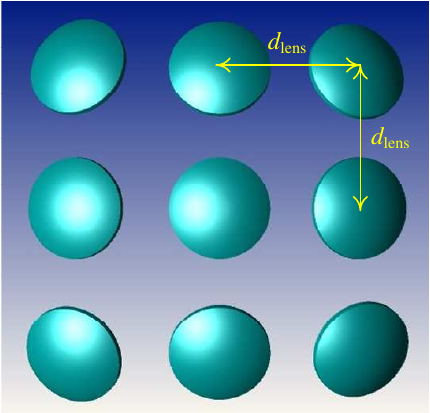}} \quad
    \subfloat[\label{Fig:AP_b}]{\includegraphics[height=0.15\textheight, keepaspectratio=true]{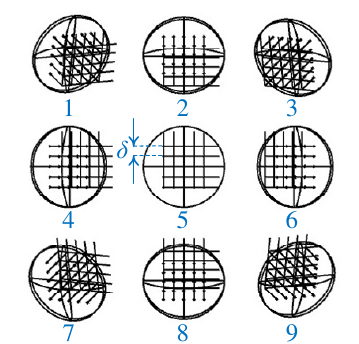}}
    \caption{Double-tier access point architecture based on a $3\times3$ array of $5\times5$ VCSEL arrays: (a) Shaded top view, (b) Detailed top view.}
    \label{Fig:OpticalDesign}
    \vspace{-10pt}
\end{figure}

Figs.~\subref*{Fig:AP_a} and~\subref*{Fig:AP_b} show the optical design of the double-tier \ac{AP}. The adjacent transmitter elements are placed at a center-to-center distance of $d_\mathrm{lens}$. In each array, the \acp{VCSEL} are placed close to each other with a pitch distance of $\delta$. As a result, the beam spots may significantly overlap on the receiver plane. The beams can be separated from one another by using a plano-convex lens in front of each \ac{VCSEL} array. This allows the light beams to be to refracted towards different directions so as to provide a high spatial resolution on the receiver plane. The refraction angle experienced by each beam depends on the lens thickness and the path it travels through within the lens. In order to achieve the desired spatial resolution, the lens parameters need to be properly adjusted based on the design requirements. The adjustable parameters of the plano-convex lens include the diameter $L$, the radius of curvature $R_\mathrm{lens}$, and the center thickness $\tau_\mathrm{c}$ \cite{Lens_EdmundOptics}.

%---------------------------------------------------------------------------------------------------
\subsection{Transmitter Element Modeling}
When a Gaussian beam travels through a lens, its Gaussian characteristics are preserved \cite{Saleh2019}. If the propagation axis of the beam is aligned with the optical axis of the lens, it continues to propagate on the same optical axis after the lens. Otherwise, its propagation direction changes. In the following, we derive the parameters of the transformed Gaussian beam and the direction vector for each \ac{VCSEL} in a transmitter element. Subsequently, we extend the analysis for the array of arrays architecture.

% \vspace{-10pt}

\subsubsection{Light Refraction} When light traverses the boundary of two homogeneous media with refractive indices of $n_1$ and $n_2$, it is refracted according to Snell's law so that \cite[p.~7]{Saleh2019}:
\begin{equation}
n_1 \sin{\theta_1} = n_2\sin{\theta_2},
\label{Eq:Snell_Scalar}
\end{equation}
where $\theta_1$ is the angle of incidence, and $\theta_2$ is the angle of refraction, with respect to the normal line to the boundary. Snell's law can be extended into a vector formulation to model the behavior of light rays as they traverse a curved surface separating two homogeneous media in \ac{3D} space. Sommerfeld and Runge demonstrated that this behavior can be analyzed using principles similar to those employed in electrodynamics for deriving the conditions that govern the changes of vector fields across a surface discontinuity \cite{Sommerfeld1911Optics}. Consequently, the law of refraction is derived analogously to the derivation of the continuity conditions for the tangential components of vector fields, leading to the vector form of Snell's law \cite[p.~133]{Born2019Optics}:
\begin{equation}
n_1\left(\mathbf{n}\times\mathbf{v}_1\right) = n_2\left(\mathbf{n}\times\mathbf{v}_2\right),
\label{Eq:Snell_Vector}
\end{equation}
or equivalently:
\begin{equation}
\mathbf{n}\times\mathbf{v}_2 = \mu\left(\mathbf{n}\times\mathbf{v}_1\right), \quad \mu=\frac{n_1}{n_2},
\label{Eq:Law_of_Refraction}
\end{equation}
where $\mathbf{n}$ is the unit vector normal to the boundary surface, $\mathbf{v}_1$ and $\mathbf{v}_2$ are the normalized incidence and refraction vectors, respectively, and $\times$ denotes the cross product.
Note that the scaler form of Snell's law is obtained by taking the Euclidean norm of both sides of \eqref{Eq:Snell_Vector}. A key result is deduced from \eqref{Eq:Snell_Vector}: the refracted ray lies in the same plane as the incident ray and the normal to the surface, which is referred to as the plane of incidence. The refraction vector $\mathbf{v}_2$ can be derived in terms of the incidence vector $\mathbf{v}_1$ and the normal vector $\mathbf{n}$ as follows:
\begin{equation}
   \mathbf{v}_2 = \sqrt{1-\mu^2\left(1-\left(\mathbf{n}{\cdot}\mathbf{v}_1 \right)^2\right)}\mathbf{n}+\mu\left(\mathbf{v}_1-\left(\mathbf{n}{\cdot}\mathbf{v}_1\right)\mathbf{n}\right),
   \label{Eq:Snell_v2_general}
\end{equation}
where ${\cdot}$ denotes the inner product. The derivation of \eqref{Eq:Snell_v2_general} is provided in Appendix~\ref{Appendix:A}.

\begin{figure}[!t]
    \centering
    \includegraphics[width=0.45\textwidth]{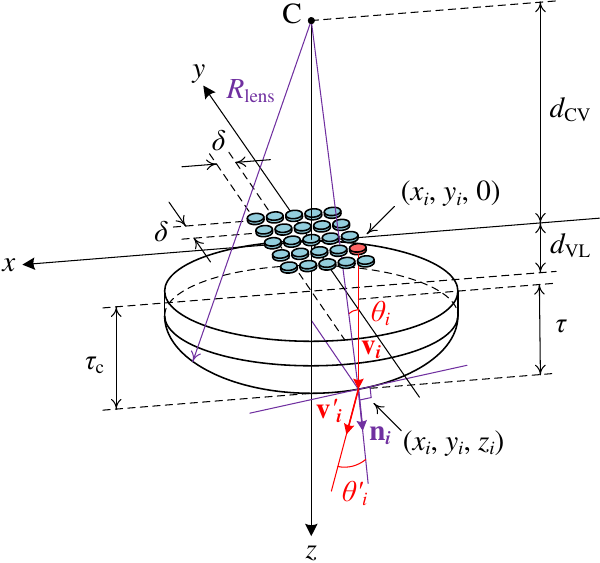}
    \caption{The 3D modeling of a transmitter element. Light refraction through the plano-convex lens for the $i$th \ac{VCSEL} is illustrated using the law of refraction.}
    \label{Fig:BeamRefraction}
    \vspace{-15pt}
\end{figure}

Fig.~\ref{Fig:BeamRefraction} illustrates the \ac{3D} geometry of the light refraction for the \ac{VCSEL} array through the plano-convex lens based on the law of refraction. To elaborate, the \acp{VCSEL} are labeled by a single index indicating their position in the array such that the $i$th \ac{VCSEL} signifies $\text{VCSEL}_{mn}$ where $i=5(m-1)+n$ for $m,n=1,2,\dots,5$. Assuming that the center of the array sits at the origin of the $xyz$ system, the coordinates of the $i$th \ac{VCSEL}, for $i=1,2\dots,25$, are calculated as:
\begin{subequations}
    \begin{align}
        x_i &= (-3+n)\delta, \\
        y_i &= (3-m)\delta,
    \end{align}
    \label{Eq:VCSEL_i_middle}%
\end{subequations}
where $m=\Big\lceil\dfrac{i}{5}\Big\rceil$ and $n = i-5\Big(\Big\lceil\dfrac{i}{5}\Big\rceil-1\Big)$, and $\lceil s\rceil$ denotes the smallest integer value satisfying $\lceil s\rceil\geq s$. The distance between the \ac{VCSEL} array and the lens is represented by $d_\mathrm{VL}$. For the $i$th \ac{VCSEL}, the beam waist is assumed to be located at $(x_i, y_i, 0)$. The beam is incident on the lens at $(x_i, y_i, d_\mathrm{VL})$ with the propagation vector $\mathbf{v}_i = \mathbf{n}_z$, where $\mathbf{n}_z$ is the unit vector of the $z$ axis, exiting the lens at $(x_i, y_i, z_i)$. Since the incident angle on the planar surface of the lens equals zero, the beam propagation axis remains unchanged upon entering the lens. When the beam reaches the convex surface of the lens, its propagation vector is changed due to the curvature of the surface as well as the change in the refractive index. By applying the law of refraction at the boundary for $n_1=n_\mathrm{lens}$, $n_2=1$, $\mathbf{v}_1=\mathbf{v}_i$, $\mathbf{v}_2=\mathbf{v}'_i$ and $\mathbf{n}=\mathbf{n}_i$, , as shown in Fig.~\ref{Fig:BeamRefraction}, the propagation vector of the refracted beam is derived as:
\begin{equation}
    \mathbf{v}'_i = {n_\mathrm{lens}}\mathbf{v}_i + \left(\cos{\theta'_i} - n_\mathrm{lens}\cos{\theta_i}\right)\mathbf{n}_i,
   \label{Eq:Snell_v_prime_i}
\end{equation}
through the use of $\mathbf{v}_i\cdot\mathbf{n}_i=\cos{\theta_i}$ and $\mathbf{v}'_i\cdot\mathbf{n}_i=\cos{\theta'_i}$. We need to determine $\theta_i$, $\theta'_i$ and $\mathbf{n}_i$. With the aid of Fig.~\ref{Fig:BeamRefraction}, the angle of incidence and the normal vector on the convex surface of the lens are calculated as:
\begin{equation}
\theta_i = \sin^{-1} \left(\dfrac{\sqrt{x_i^2 + y_i^2}}{R_\mathrm{lens}}\right),
\label{Eq:theta_i}
\end{equation}
\begin{equation}
\mathbf{n}_i = \frac{x_i\mathbf{n}_{x} + y_i\mathbf{n}_{y} + (z_i+d_\mathrm{CV})\mathbf{n}_{z}}{R_\mathrm{lens}},
\end{equation}
with $\mathbf{n}_{x}$ and $\mathbf{n}_{y}$ representing the unit vectors of the $x$ and $y$ axes, respectively. Note that: 
\begin{equation}
z_i+d_\mathrm{CV} = \sqrt{R_\mathrm{lens}^2-x_i^2 - y_i^2}.
\label{Eq:z_i}
\end{equation}
Based on \eqref{Eq:Snell_Scalar} and \eqref{Eq:theta_i}, the refraction angle is obtained as:
\begin{equation}
\theta'_i = \sin^{-1}\left(n_\mathrm{lens}\sin\theta_i\right) = \sin^{-1}\left(\dfrac{{n_\mathrm{lens}}\sqrt{x_i^2 + y_i^2}}{R_\mathrm{lens}}\right).
\label{Eq:theta_prime_i}
\end{equation}
Combining \eqref{Eq:theta_i}--\eqref{Eq:theta_prime_i} with \eqref{Eq:Snell_v_prime_i}, the expression of $\mathbf{v}'_i$ becomes:
\begin{equation}
    \mathbf{v}'_i = Q x_i\mathbf{n}_x + Q y_i\mathbf{n}_y + \left(n_\mathrm{lens}+Q\sqrt{R_\mathrm{lens}^2-x_i^2 - y_i^2}\right)\mathbf{n}_z,
   \label{Eq:v_prime_i}
\end{equation}
through defining:
\begin{equation}
    Q = \frac{\sqrt{R_\mathrm{lens}^2-n_\mathrm{lens}^2\left(x_i^2 + y_i^2\right)}-n_\mathrm{lens}\sqrt{R_\mathrm{lens}^2-x_i^2 - y_i^2}}{R_\mathrm{lens}^2}.
   \label{Eq:v_prime_i_Q}
\end{equation}
%

%---------------------------------------------------------------------------------------------------
\subsubsection{Beam Transformation}
\begin{figure}[!t]
    \centering
    \includegraphics[width=0.5\textwidth]{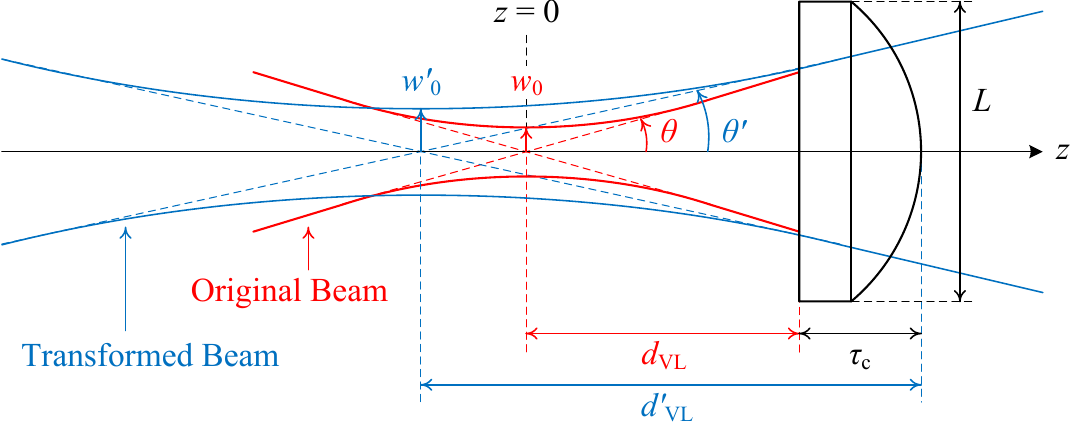}
    \caption{The transformation of a Gaussian beam by the plano-convex lens with a virtual image of the beam waist behind the lens.}
    \label{Fig:BeamTransformation}
    \vspace{-15pt}
\end{figure}

The propagation of light beams in optical systems is often described by ray transfer or ABCD matrices \cite{Saleh2019}. In this method, a $2 \times 2$ matrix relates the position and angle of paraxial rays at the input and output planes of an optical system via linear algebraic equations. In the case of a Gaussian beam, the ABCD law applies to the $q$-parameters of the original and transformed beams. Fig.~\ref{Fig:BeamTransformation} shows the Gaussian beam propagation through a plano-convex lens, assuming that the beam axis is parallel to the optical axis of the lens. The parameters of the transformed beam are denoted by $w'_0$, $w'(z')$, $\theta'$, $z'_\mathrm{R}$ and $q'(z')$, as defined in \eqref{Eq:w_z}--\eqref{Eq:q}, and $d'_\mathrm{VL}$ represents the distance of the new beam waist to the lens surface. Depending on the location of the input beam waist with respect to the lens, the output beam waist (i.e., the image) may be either real or virtual. We consider the case where a virtual beam waist is created behind the lens, since a real beam waist that is formed in front of the lens imposes a strict limitation on the maximum allowable transmit power due to eye safety \cite{HKazemi2023MultiBeam}.

\newcounter{mycounter1}
% Store the current equation number.
\setcounter{mycounter1}{\value{equation}}
\setcounter{equation}{29}
% \addtocounter{equation}{1}
% Set the equation number to one less than the one
% desired for the first equation here.
\begin{figure*}[t!]
% ensure that we have normalsize text
\normalsize
% IEEE uses as a separator
% \hrulefill
%
\begin{equation}
\mathbf{q}_v = \left[\begin{matrix}
x_v \\ y_v \\ z_v
\end{matrix}\right] = 
\mathbf{R}_{y}(\pi+\beta_v)\mathbf{R}_{x}(\alpha_v)\left[\begin{matrix}
0 \\ 0 \\ d_\mathrm{c}
\end{matrix}\right] + 
\mathbf{R}_{y}(\pi)\left[\begin{matrix}
x'_v \\ y'_v \\ 0
\end{matrix}\right] + \left[\begin{matrix}
0 \\ 0 \\ h_\mathrm{DL}+d_\mathrm{c}
\end{matrix}\right] = 
\left[\begin{matrix}
-x'_v-d_\mathrm{c}\sin\alpha_v\cos\beta_v \\
y'_v-d_\mathrm{c}\sin\alpha_v \\ 
h_\mathrm{DL}+d_\mathrm{c}\left(1-\cos\alpha_v\cos\beta_v\right)
\end{matrix}\right],
\label{Eq:TxEl_Coordinates}
\end{equation}
\hrulefill
% The spacer can be tweaked to stop underfull vboxes.
% \vspace*{4pt}
\vspace{-15pt}
\end{figure*}
% Restore the current equation number.
\setcounter{equation}{\value{mycounter1}}

In order to derive the parameters of the transformed beam, we use the $q$-parameter transformation \cite[p.~97]{Saleh2019}, \cite{Nemoto1990}:
\begin{equation}
    q' = \frac{Aq+B}{Cq+D} ,
    \label{Eq:q_trasnform}
\end{equation}
where:
\begin{subequations}
\begin{align}
A &= 1, \\ 
B &= \frac{\tau}{n_\mathrm{lens}}, \\
C &= \frac{1-n_\mathrm{lens}}{R_\mathrm{lens}}, \\
D &= 1+\frac{\tau}{R_\mathrm{lens}}\left(\frac{1}{n_\mathrm{lens}}-1\right),
\end{align}
\label{Eq:ABCD}%
\end{subequations}
where $\tau$ is the thickness of the lens through which the beam travels. From the geometry shown in Fig.~\ref{Fig:BeamRefraction}, $\tau=z_i-d_\mathrm{VL}$ and $d_\mathrm{CV}=R_\mathrm{lens}-d_\mathrm{VL}-\tau_\mathrm{c}$. Combining these with \eqref{Eq:z_i} yields:
\begin{equation}
\tau = \sqrt{R_\mathrm{lens}^2-x_i^2 - y_i^2} + \tau_\mathrm{c} - R_\mathrm{lens}.
\end{equation}
Now, we apply the transformation in \eqref{Eq:q_trasnform} between the original beam at $z = d_\mathrm{VL}$ and the transformed beam at $z = d_\mathrm{VL} +\tau$ (i.e., at $z' = d'_\mathrm{VL}$), as shown in Fig.~\ref{Fig:BeamTransformation}, so that:
\begin{equation}
    q'(d'_\mathrm{VL}) = d'_\mathrm{VL} + jz'_\mathrm{R} = \frac{A(d_\mathrm{VL}+jz_\mathrm{R})+B}{C(d_\mathrm{VL}+jz_\mathrm{R})+D}.
    \label{Eq:q_trasnform2}
\end{equation}
By substituting the ABCD parameters from \eqref{Eq:ABCD} into \eqref{Eq:q_trasnform2}, the distance of the transformed beam waist from the tangent plane on the lens surface and the Rayleigh range of the transformed beam are derived as:
\begin{equation} \label{Eq:d2_transformed}
    \begin{aligned}
    d'_\mathrm{VL} &= \operatorname{Re}\left\{q'(d'_\mathrm{VL})\right\} \\
    &= \frac{\left(d_\mathrm{VL}+\dfrac{\tau}{n_\mathrm{lens}}\right)\left[1-\dfrac{1}{f}\left(d_\mathrm{VL}+\dfrac{\tau}{n_\mathrm{lens}}\right)\right]-\dfrac{z_\mathrm{R}^2}{f}}{\left[1-\dfrac{1}{f}\left(d_\mathrm{VL}+\dfrac{\tau}{n_\mathrm{lens}}\right)\right]^2+\dfrac{z_\mathrm{R}^2}{f^2}},
    \end{aligned}
\end{equation}
\begin{equation} \label{Eq:zR_transformed}
    z'_\mathrm{R} = \operatorname{Im}\left\{q'(d'_\mathrm{VL})\right\} = \frac{z_\mathrm{R}}{\left[1-\dfrac{1}{f}\left(d_\mathrm{VL}+\dfrac{\tau}{n_\mathrm{lens}}\right)\right]^2+\dfrac{z_\mathrm{R}^2}{f^2}},
\end{equation}
where the operators $\mathrm{Re}$ and $\mathrm{Im}$ take the real and imaginary parts of a complex variable, respectively, and $f$ is the effective focal distance of the lens which, according to the lensmaker's equation \cite[p.~249]{Hecht2017Optics}, is defined as: 
\begin{equation}
    f = \frac{R_\mathrm{lens}}{n_\mathrm{lens}-1}.
\end{equation} 
Based on \eqref{Eq:zR_transformed}, and by using \eqref{Eq:w_z}, the radius of the transformed beam waist is obtained as:
\begin{equation}
w'_0 = \frac{w_0}{\sqrt{\left[1-\dfrac{1}{f}\left(d_\mathrm{VL}+\dfrac{\tau}{n_\mathrm{lens}}\right)\right]^2+\left(\dfrac{\pi w_0^2}{\lambda f}\right)^{\!2}}}.
\label{Eq:w0_transformed}
\end{equation}

\vspace{-10pt}

%---------------------------------------------------------------------------------------------------
\subsection{Total Spatial Intensity Distribution}
In order to derive the total spatial intensity of the \ac{AP}, the transmitter elements are indexed according to Fig.~\subref*{Fig:AP_b}. Based on Euler angles with clockwise rotations, the tilting of the $v$th transmitter element about the $x$ and $y$ axes can be modeled using the following rotation matrices:
\begin{subequations}\label{Eq:RotationMatrices}
\begin{align}
\mathbf{R}_{x}(\alpha_v) &= 
\left[\begin{matrix}
1 & 0 & 0 \\
0 & \cos\alpha_v & -\sin\alpha_v \\
0 & \sin\alpha_v & \cos\alpha_v
\end{matrix}\right], \label{Eq:RotationMatrix_x}\\
\mathbf{R}_{y}(\beta_v) &= 
\left[\begin{matrix}
\cos\beta_v & 0 & \sin\beta_v \\
0 & 1 & 0 \\ 
-\sin\beta_v & 0 & \cos\beta_v
\end{matrix}\right], \label{Eq:RotationMatrix_y}
\end{align}
\end{subequations}
where the rotation angles $\left[\alpha_v~\beta_v\right]^T$ are given as the $v$th column of $\mathbf{\Theta}_\mathrm{t}\in\mathbb{R}^{2\times9}$ by:
\begin{equation}
    \mathbf{\Theta}_\mathrm{t} = \theta_\mathrm{tilt}\left[\begin{matrix}
                            -1 & -1 & -1 & 0 & 0  & 0  & 1  & 1 & 1 \\ 
                            -1 & 0  & 1  & 1 & 0  & -1 & -1 & 0 & 1
                            \end{matrix}\right].
\end{equation}
Let the \acp{VCSEL} in the array of arrays architecture be labeled by a global index of $j\in\{1,2,\dots,225\}$, and the $j$th \ac{VCSEL} lies in the $v$th array for $v=1,2,\dots,9$. The propagation vector of the refracted beam for the $j$th \ac{VCSEL} projected onto the coordinate system of the receiver plane can be expressed as:
\begin{equation}
    \begin{aligned}
    \mathbf{v}'_{j} &= \mathbf{R}_{y}(\pi)\mathbf{R}_{y}(\beta_v)\mathbf{R}_{x}(\alpha_v)\mathbf{v}'_i, \\
                   &= \mathbf{R}_{y}(\pi+\beta_v)\mathbf{R}_{x}(\alpha_v)\mathbf{v}'_i,
    \end{aligned}
\end{equation}
where $\mathbf{v}'_i$ is evaluated for $i=j-25(v-1)$ based on \eqref{Eq:v_prime_i}. The coordinates of the $v$th transmitter element are obtained as \eqref{Eq:TxEl_Coordinates}, given at the top of the page, where $h_\mathrm{DL}$ represents the vertical separation between the \ac{AP} and the receiver plane, as shown in Fig.~\ref{Fig:NetworkArchitecture}, $d_\mathrm{c}=d_\mathrm{VL}+\tau_\mathrm{c}$, and $\left[x'_v~y'_v\right]^T$ is equal to the $v$th column of $\mathbf{Q}_\mathrm{t}\in\mathbb{R}^{2\times9}$ given by:
\addtocounter{equation}{1}
\begin{equation}
    \mathbf{Q}_\mathrm{t} = d_\mathrm{lens}\left[\begin{matrix}
                            1 & 0 & -1 & 1 & 0  & -1 & 1  & 0  & -1 \\ 
                            1 & 1 & 1  & 0 & 0  & 0  & -1 & -1 & -1
                            \end{matrix}\right].
\end{equation}
For a given point of $\mathbf{p}=[x~y~0]^T$ on the receiver plane, the distance vector from $\mathbf{q}_v$ to $\mathbf{p}$ is given by $\mathbf{d}_v=\mathbf{p}-\mathbf{q}_v$, and the corresponding Euclidean distance is:
\begin{equation}
    d_v = \lVert\mathbf{d}_v\rVert=\sqrt{(x-x_v)^2+(y-y_v)^2+h_\mathrm{DL}^2}.
    \label{Eq:d_v}
\end{equation}
Based on \eqref{Eq:I_G}, the intensity distribution of the $i$th beam on the receiver plane at $\mathbf{p}$ can be expressed as:
\begin{equation}
	I_i(x,y) = \frac{2P_\mathrm{t}}{\pi w'^2(z_i)}{\exp{\left(-\frac{2r_i^2}{w'^2(z_i)}\right)}},
    \label{Eq:I_i}
\end{equation}
where $r_i=d_v\sin\phi_i$ and $z_i=d_v\cos\phi_i$ for $i=1,2,\dots,225$ and $v=\Big\lceil\dfrac{i}{25}\Big\rceil$, and $\phi_i$ is the radiance angle of the $i$th beam relative to $\mathbf{p}$, which is calculated as:
\begin{equation}
    \phi_i = \cos^{-1}\left(\frac{\mathbf{d}_v{\cdot}\mathbf{v}'_i}{d_v}\right).
    \label{Eq:phi_i}
\end{equation}
The total spatial intensity of the \ac{AP} is obtained as:
\begin{equation}
    \mkern-18mu I_\mathrm{AP}(x,y) = \sum_{i=1}^{225}\frac{2P_\mathrm{t}}{\pi w'^2(d_v\cos\phi_i)}{\exp{\left(-\frac{2d_v^2\sin^2\phi_i}{w'^2(d_v\cos\phi_i)}\right)}}.
    \label{Eq:I_AP}
\end{equation}
Note that $d_v$ and $\phi_i$ are functions of $x$ and $y$ through \eqref{Eq:d_v} and \eqref{Eq:phi_i}, respectively.

\begin{figure*}[!t]
	\centering
	\begin{minipage}[b]{\textwidth}
		\centering
		\subfloat[\label{Fig:LensParameters1_a} $R_\mathrm{lens}=25$~mm]{\includegraphics[width=0.3\linewidth, keepaspectratio=true]{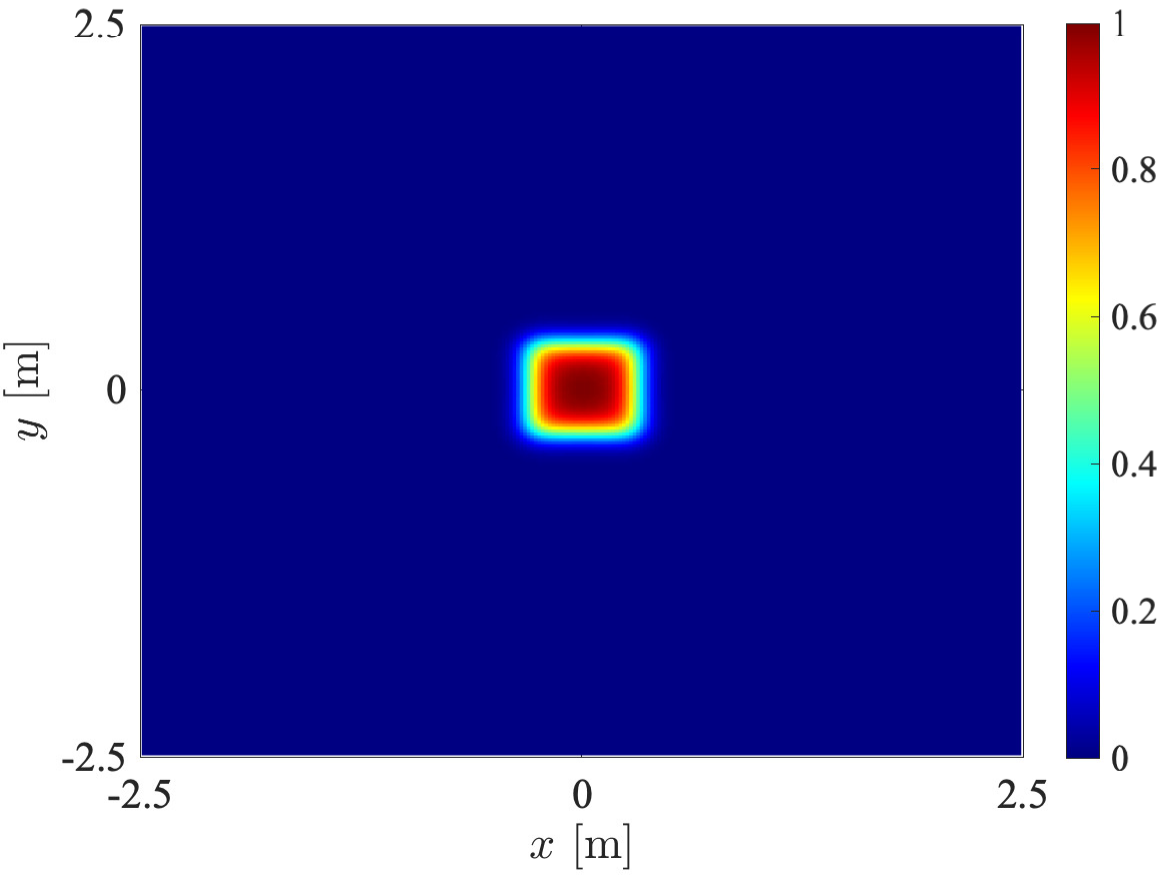}}            
		\subfloat[\label{Fig:LensParameters1_b} $R_\mathrm{lens}=17$~mm]{\includegraphics[width=0.3\linewidth, keepaspectratio=true]{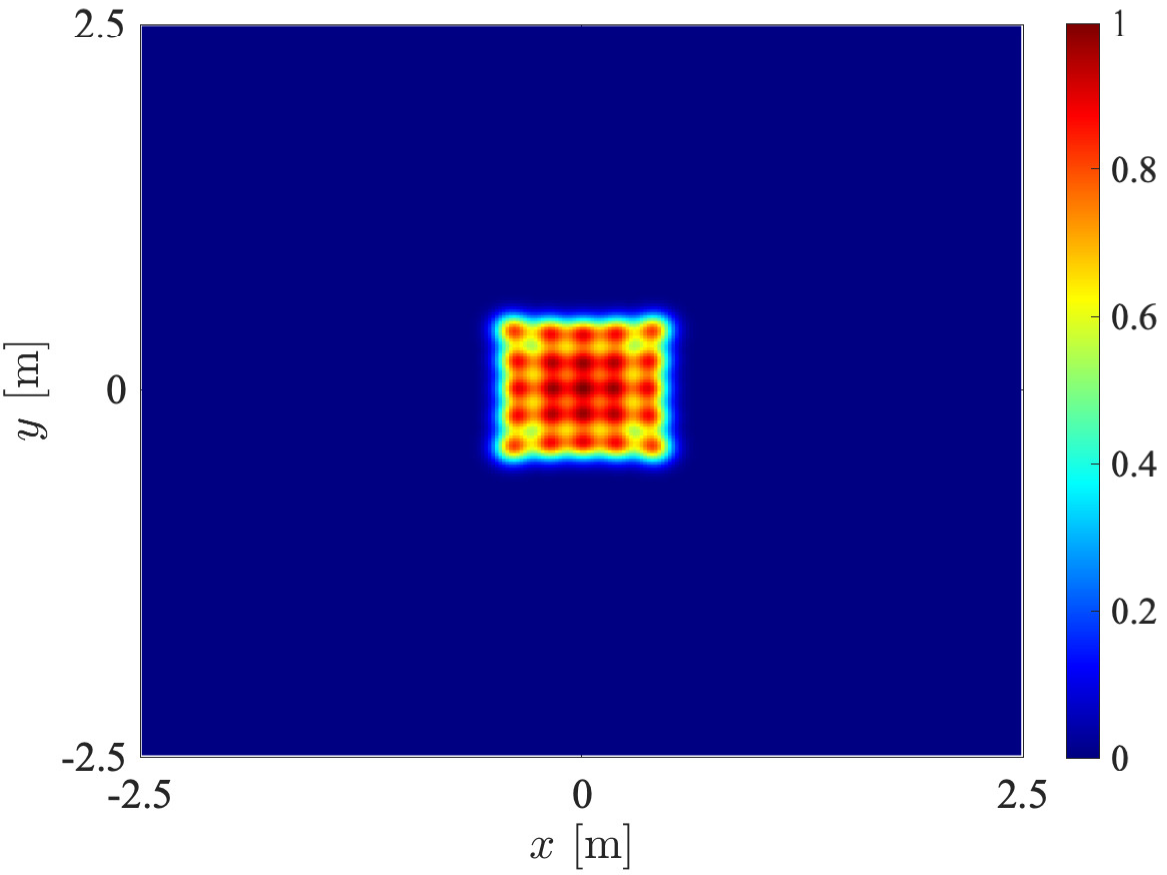}}          
		\subfloat[\label{Fig:LensParameters1_c} $R_\mathrm{lens}=15$~mm]{\includegraphics[width=0.3\linewidth, keepaspectratio=true]{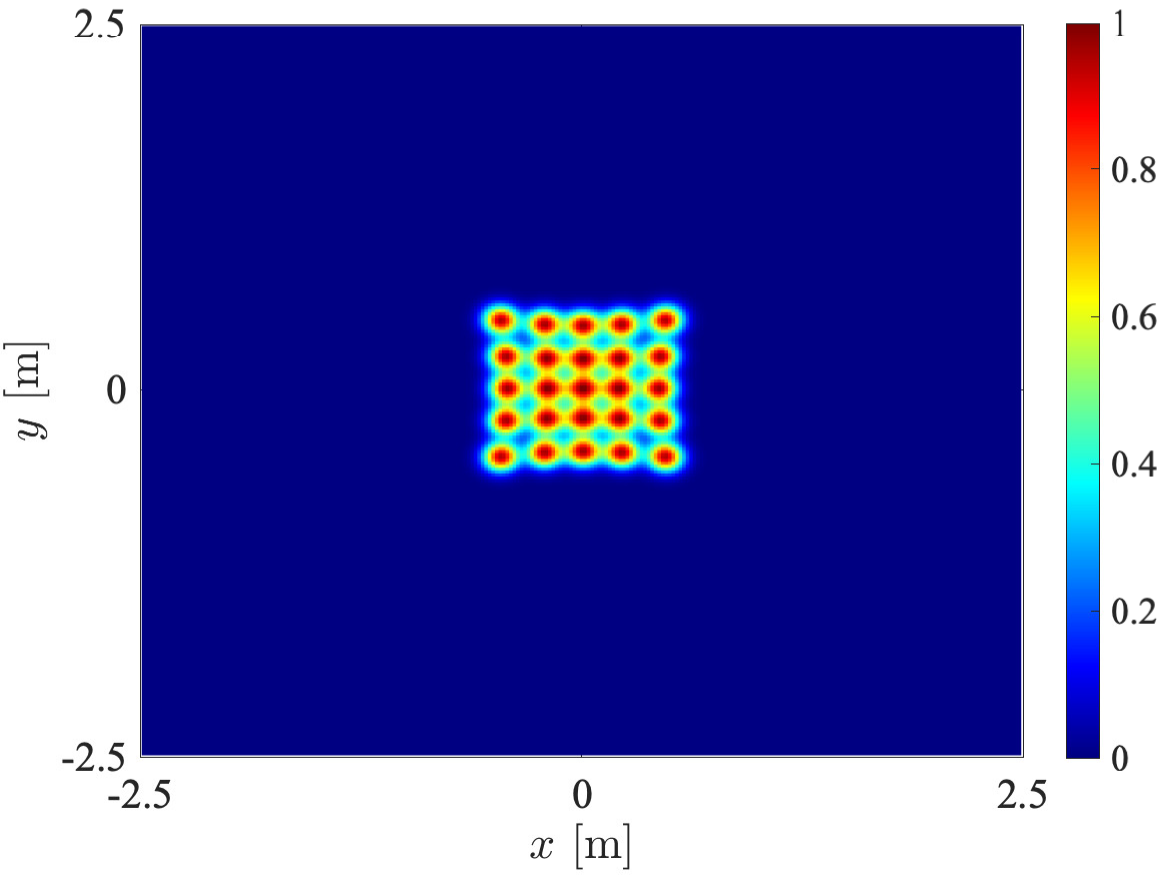}}
		\caption{Normalized intensity distribution on the receiver plane at $h_\mathrm{DL}=3$~m for three lenses with $L=24$~mm and different values of $R_\mathrm{lens}$.}
		\label{Fig:LensParameters1}
	\end{minipage}\hfill
	\begin{minipage}[b]{\textwidth}
		\centering
		\subfloat[\label{Fig:LensParameters2_a} $R_\mathrm{lens}=15$~mm]{\includegraphics[width=0.3\linewidth, keepaspectratio=true]{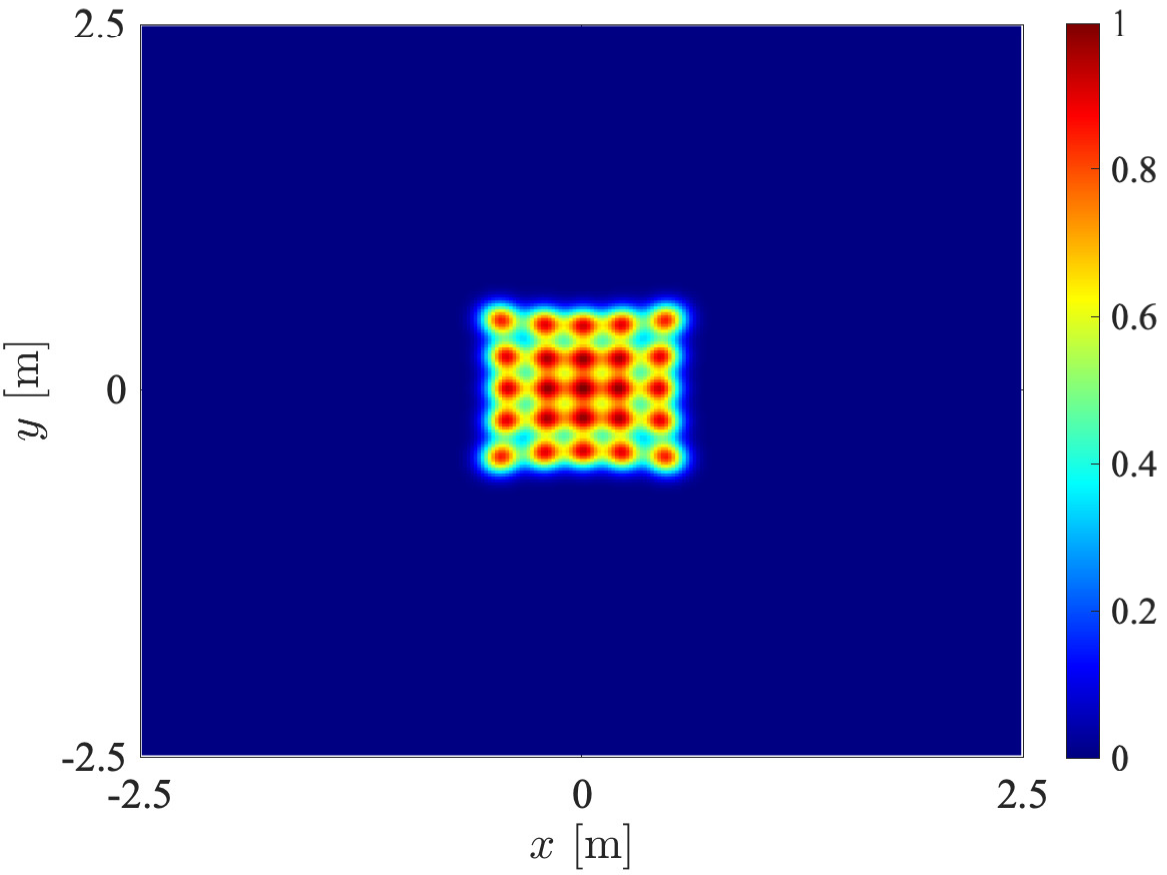}}            
		\subfloat[\label{Fig:LensParameters2_b} $R_\mathrm{lens}=13$~mm]{\includegraphics[width=0.3\linewidth, keepaspectratio=true]{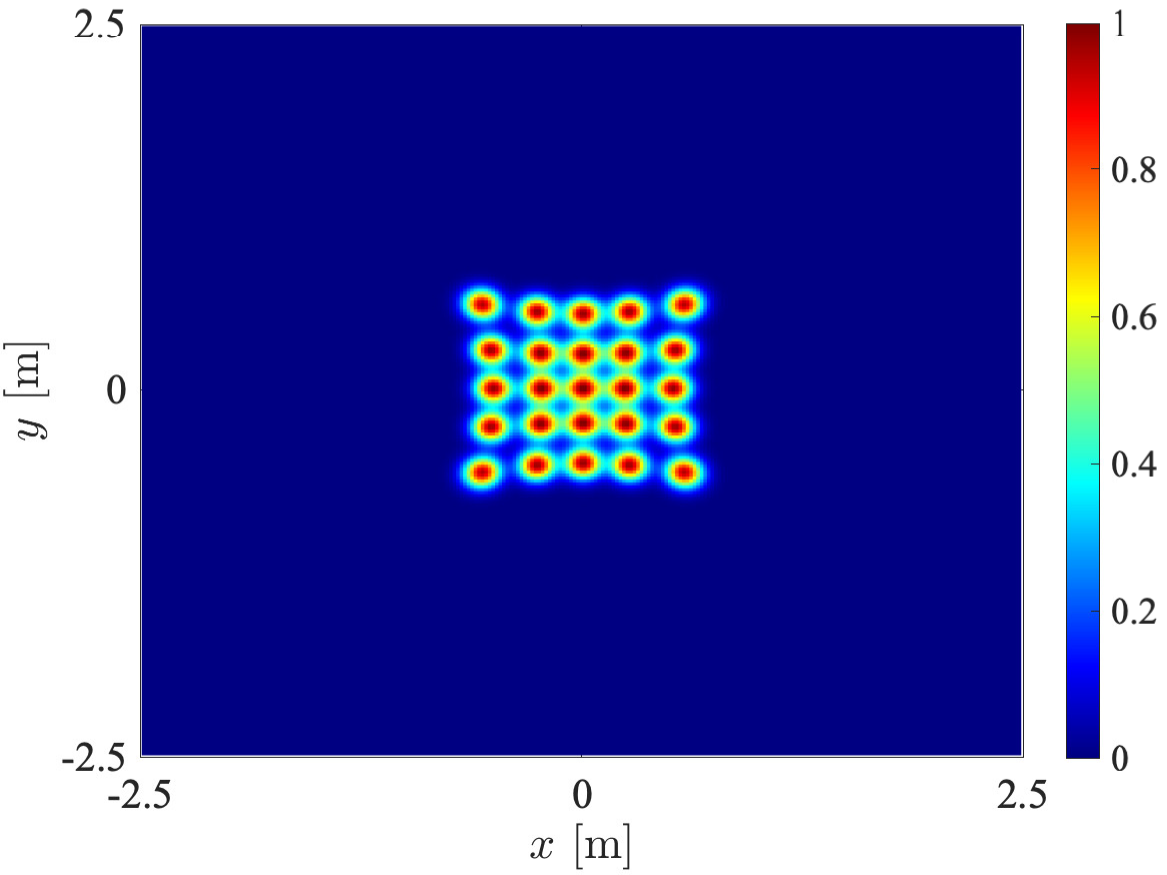}}            
		\subfloat[\label{Fig:LensParameters2_c} $R_\mathrm{lens}=10$~mm]{\includegraphics[width=0.3\linewidth, keepaspectratio=true]{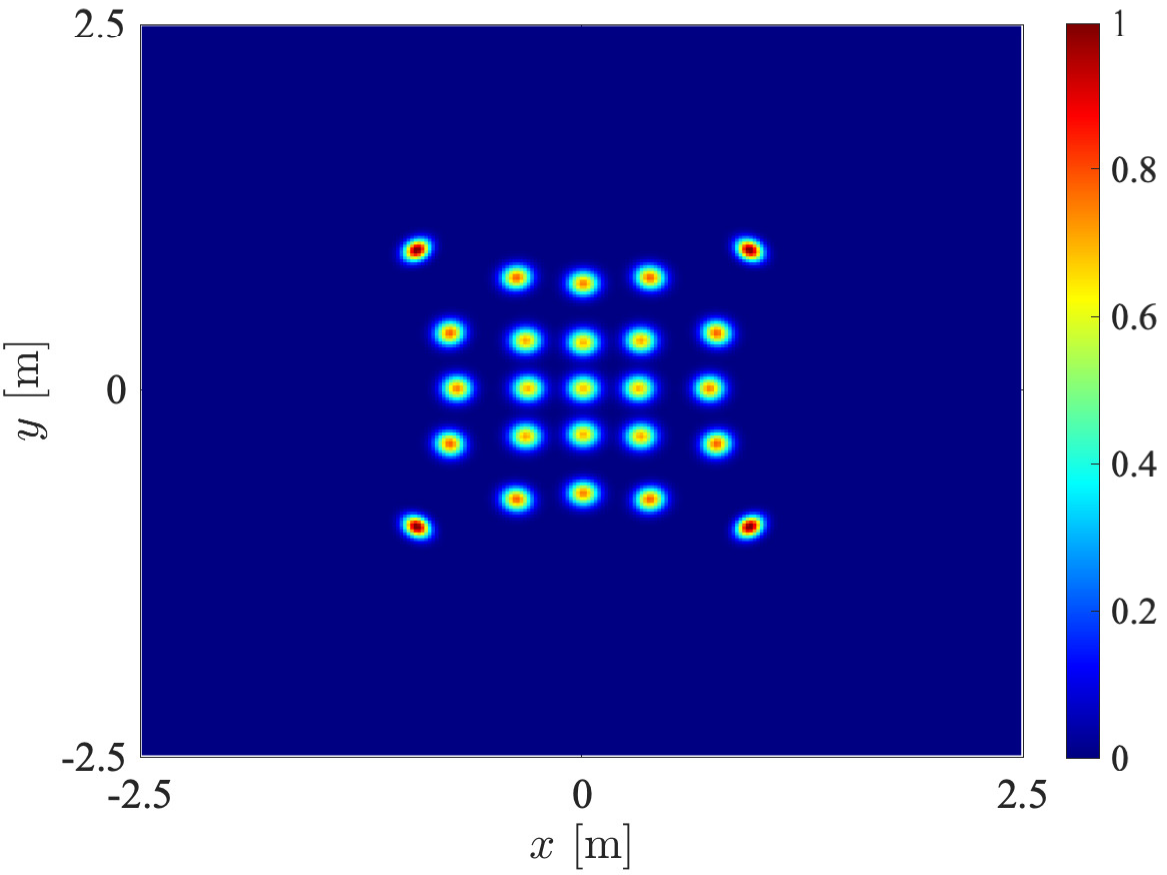}}
		\caption{Normalized intensity distribution on the receiver plane at $h_\mathrm{DL}=3$~m for three lenses with $L=16$~mm and different values of $R_\mathrm{lens}$.}
		\label{Fig:LensParameters2}
	\end{minipage}
	\vspace{-16pt}
\end{figure*}

\vspace{-10pt}
%---------------------------------------------------------------------------------------------------
\subsection{Design Examples}
We provide numerical examples to demonstrate the \ac{AP} design. The results are presented by assuming a wavelength\footnote{Most commercially available \ac{VCSEL} devices operate in an \ac{IR} wavelength range of $800-1000$~nm and they are generally cheaper than those operating at higher \ac{IR} wavelengths such as $1550$~nm. Nevertheless, the choice of higher wavelengths is beneficial in terms of eye safety considerations.} of $\lambda=950$ nm and a beam waist radius of $w_0=5$~{\textmu}m. For each \ac{VCSEL} array, the total size is $1$~cm$^2$ with a pitch distance of $\delta = 2$~mm, and the array-to-lens distance is $d_\mathrm{VL}=5$~mm. To understand the effect of the design parameters on the spatial separation between the beam spots, Figs.~\ref{Fig:LensParameters1} and \ref{Fig:LensParameters2} illustrate the normalized intensity distribution of the middle transmitter element on the receiver plane at $h_\mathrm{DL}=3$ m underneath the \ac{AP}. In Fig.~\ref{Fig:LensParameters1}, a relatively large lens of diameter $L=24$~mm is used. In this case, for $R_\mathrm{lens}=25$~mm, the beams constitute a quasi-square coverage area, and for $R_\mathrm{lens}=17$~mm, they are still largely overlapping. When using $R_\mathrm{lens}=15$~mm, despite a slight overlap between the adjacent beam spots, the beams are sufficiently separated. In Fig.~\ref{Fig:LensParameters2}, a smaller lens with $L=16$~mm is used. It can be observed that by reducing the radius of curvature of the lens, refraction angles increase and the beams move considerably further apart.

\begin{figure}[!t]
    \vspace{-15pt}
    \centering
    \subfloat[\label{Fig:AP_Coverage_Analytical} Analytical Modeling]{\includegraphics[width=0.48\linewidth, keepaspectratio=true]{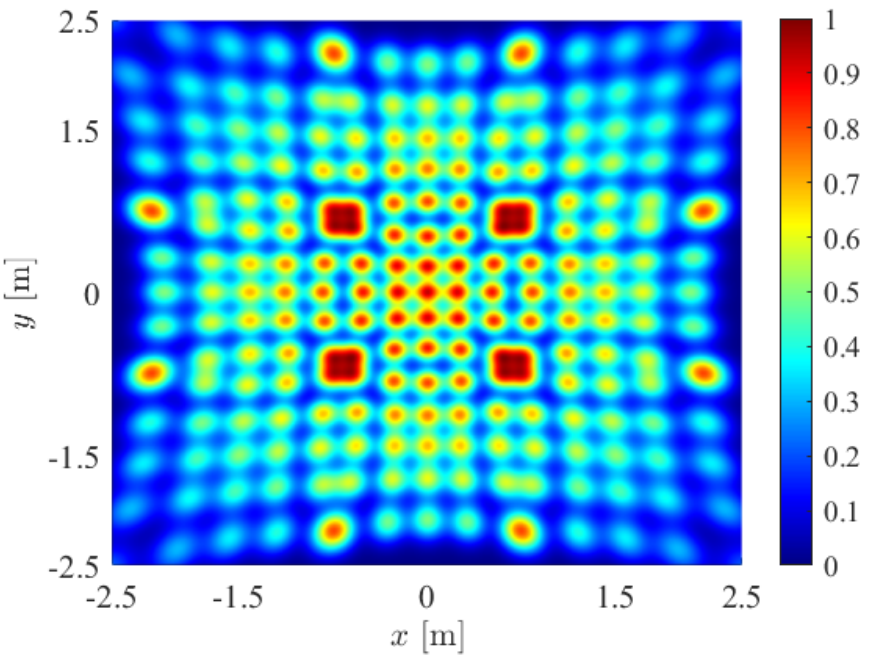}}
    \quad
    \subfloat[\label{Fig:AP_Coverage_Zemax} Zemax OpticStudio]{\includegraphics[width=0.48\linewidth, keepaspectratio=true]{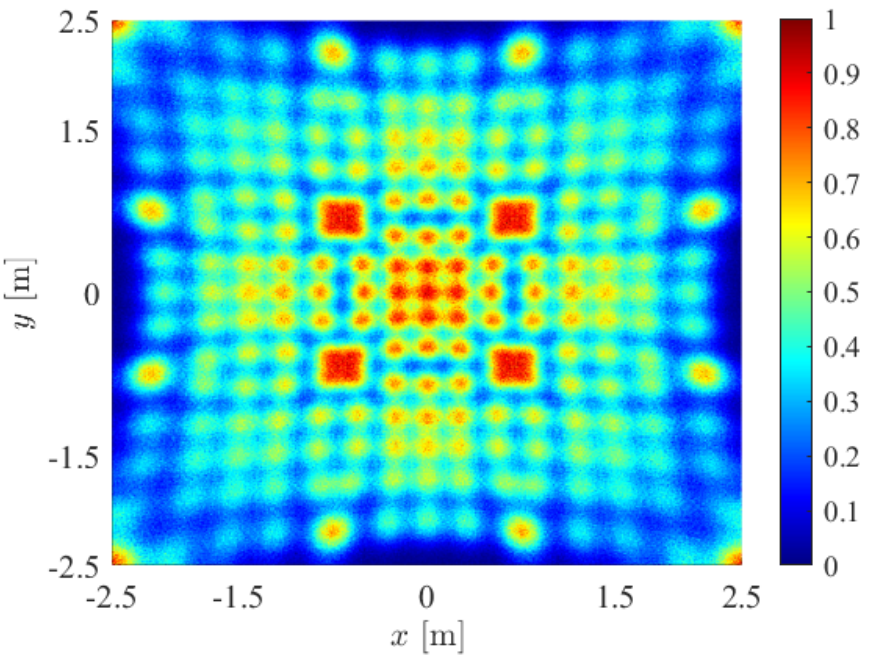}}
    \caption{Normalized intensity distribution of the double-tier AP on the receiver plane at $h_\mathrm{DL}=3$~m for $\lambda=950$~nm and $w_0=5~\mu$m.}
    \label{Fig:AP_Coverage}
    \vspace{-15pt}
\end{figure}

Aiming to achieve a full beam coverage while maintaining minimum overlaps between the beam spots, we opt for $L=16$~mm and $R_\mathrm{lens}=15$~mm. With this configuration, the beam spots cover the receiver plane in a complementary manner while the area covered by each of them is almost a circle of radius $w'(z)$ (i.e., the transformed beam spot radius containing $86\%$ of the transmit power carried by the beam), as shown in Fig.~\subref*{Fig:LensParameters2_a}. We use this configuration for a room of size $5 \times 5 \times 3$~m$^3$. To extend the \ac{AP} coverage over the entire room area of $25$~m$^2$, the transmitter elements are arranged next to each other at a distance of $d_\mathrm{lens}=2$~cm, as shown in Fig.~\subref*{Fig:AP_a}. To obtain the required tilt angle for each transmitter element, the beam spot center of the central \ac{VCSEL} in each array is assumed to be nearly landing on the intersection of the diagonals of the corresponding quadrilateral formed at the corners of the room, and this leads to $\theta_\mathrm{tilt}=21^\circ$. Fig.~\ref{Fig:AP_Coverage} shows the resulting spatial distribution of the normalized intensity on the receiver plane. To examine the accuracy of the proposed analytical modeling, the results are compared with Zemax OpticStudio simulations based on non-sequential ray tracing \cite{Zemax}. For each \ac{VCSEL}, $10^5$ rays are randomly generated and each ray is independently traced until it reaches the receiver plane. Comparing Figs.~\subref*{Fig:AP_Coverage_Analytical} and \subref*{Fig:AP_Coverage_Zemax}, it can be seen that the analytical results and those obtained by using Zemax OpticStudio are closely matching.

\vspace{-6pt}

%%%%%%%%%%%%%%%%%%%%%%%%%%%%%%%%%%%%%%%%%%%%%%%%%%%%%%%%%%%%%%%%%%%%%%%%%%%%%%%%%%%%%%%%%%%%%%%%%%%%
%%%%%%%%%%%%%%%%%%%%%%%%%%%%%%%%%%%%%%%%%%%%%%%%%%%%%%%%%%%%%%%%%%%%%%%%%%%%%%%%%%%%%%%%%%%%%%%%%%%%
\section{Beam Clustering for Interference Management} \label{BeamClustering}

\subsection{Coordinated Multi-Beam Transmission}
In order to achieve higher network throughput, it is desirable for every beam to fully utilize the available spectrum. However, the dense deployment of \acp{VCSEL} in the double-tier \ac{AP} and small sizes of beam spots unavoidably give rise to \ac{IBI} when the frequency spectrum is shared among multiple beams. Therefore, advanced interference mitigation techniques are required for improving the spectral efficiency and hence unlocking the capacity of the indoor access network. In this work, we use \ac{CoMB}-\ac{JT}, aiming to reduce \ac{IBI} levels by coordinating transmissions from a cluster of beams. This is similar to \ac{CoMP}-\ac{JT} \cite{Ma2015coordinated}, though \ac{CoMB}-\ac{JT} proposed in this work does not necessarily originate from \textit{multiple points}, but it is always formed with \textit{multiple beams} which may come from the same transmitter element of the double-tier \ac{AP}. This way, \ac{IBI} can be effectively mitigated and converted into useful signals to enhance the downlink \ac{SINR} especially for spot-edge users.

For beam clustering based on \ac{CoMP}, three main categories are identified \cite{Bassoy2017coordinated}: static clustering, semi-dynamic clustering, and dynamic clustering. Static clustering provides a means to cancel intra-cluster interference by using a predefined set of clusters which once formed do not change with any variations in the network. Static clusters are designed based on the network topology such as the geometry of the environment, the number of \acp{AP} and their positions \cite{Bassoy2017coordinated}. 
Semi-dynamic clustering is an enhanced version of static clustering where several layers of static clusters are formed to avoid \ac{ICI} and users are associated with the best serving cluster. While semi-static clustering can improve potential performance gains compared to static clustering, the overlapping nature of clusters increases the system complexity \cite{Bassoy2017coordinated}. 
In dynamic clustering, the system continuously adapts its clustering strategy to any changes in the network such as user mobility, inactive beams or load variations. Although with this scheme \ac{ICI} can be minimized by dynamically adjusting cluster sizes for individual users, it comes with a higher complexity due to optimal scheduling.

The implementation complexity of \ac{CoMB} increases with the number of coordinated beams and it can only be realized for a small cluster of beams in practice. The key for maximizing the benefits of \ac{CoMB} lies in the optimum formation of clusters. By comparison, static clustering incurs significantly less complexity and is a suitable candidate for initial \ac{CoMB} deployment in conjunction with the proposed \ac{AP} design. In fact, the formation of predefined clusters of beams in static clustering is compatible with the centralized architecture of the double-tier \ac{AP} and it greatly facilitates the implementation of \ac{CoMB}-\ac{JT} as all the \acp{VCSEL} are reachable at the same time.

\vspace{-10pt}

%---------------------------------------------------------------------------------------------------
\subsection{Downlink System Modeling} \label{Sec:5_1}
\begin{figure}[!t]
	\centering
	\includegraphics[width=0.4\textwidth]{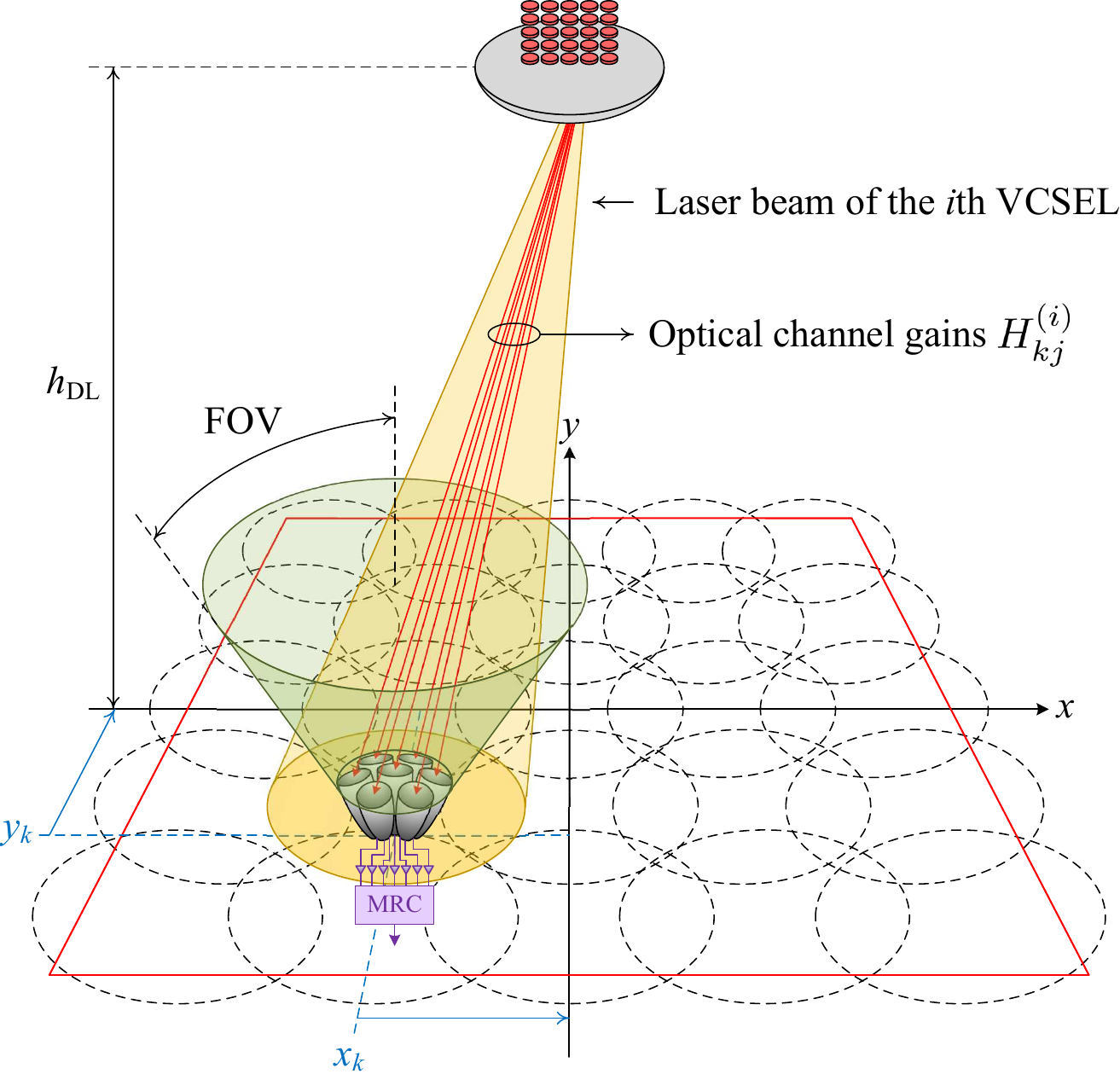}
	\caption{Downlink system model representing the optical channel gains $H_{kj}^{(i)}$ from the $i$th VCSEL to the $j$th element of the ADR, $j=1,2,\dots,7$, for the $k$th \ac{UE}. Note that only the central element of the AP is shown.}
	\label{Fig:Downlink}
	\vspace{-15pt}
\end{figure}

Fig.~\ref{Fig:Downlink} depicts the downlink system configuration where only the middle transmitter element of the double-tier \ac{AP} is shown for convenience. There are a total of $K$ \acp{UE} distributed uniformly over a horizontal communication plane. The coverage area of the network is partitioned into $N_\mathrm{c}$ non-overlapping regions, and each region is covered by an exclusive cluster of beams. According to \ac{CoMB}-\ac{JT}, each \ac{UE} is assumed to be served by \textit{one} cluster within a predefined region that provides the highest \ac{SINR}. Coordinated \acp{VCSEL} jointly send multiple copies of the same signal for each \ac{UE}. By using \ac{IM{/}DD}, these signals are then detected at the same time by the receiver, and therefore their received powers are combined. Each \ac{UE} is equipped with an \ac{ADR}, the details of which are provided as follows.

\subsubsection{Angle Diversity Receiver}
A multi-element non-imaging \ac{ADR} based on \acp{CPC} is adopted from \cite{ESarbazi2022Design2}, where we put forward the optimized multi-tier \ac{ADR} design under the constraints on the overall \ac{FOV} and dimensions. The rationale behind choosing this receiver design is that the use of \acp{CPC} allows the maximum concentration gain to be achieved due to the law of conservation of Etendue \cite{RWinston2005}, and it offers a robust performance by relieving strict alignment requirements and enabling connectivity to multiple light beams \cite{ZChen2018angle,ESarbazi2023ARobust}. We consider a single-tier \ac{ADR} composed of seven identical elements oriented in the desired spatial directions, as shown in Fig.~\ref{Fig:Downlink}. Each receiver element comprises a dielectric \ac{CPC} coupled with a \ac{2D} array of PIN \acp{PD}. A two-stage array processing is performed by applying \ac{EGC} to the output signals of each \ac{PD} array and \ac{MRC} to the output signals of the receiver elements. The \ac{ADR} parameters include $\theta_\mathrm{CPC}$, $A_\mathrm{PD}$ and $N_\mathrm{PD}$, which denote the \ac{CPC} acceptance angle, the \ac{PD} area, and the number of \acp{PD} on the array, respectively. Also, $A_1$ and $A_2$ represent the areas of the \ac{CPC} entrance and exit apertures, respectively. For each receiver element \cite{ESarbazi2022Design2}:
\begin{equation}
A_2 = \dfrac{N_\mathrm{PD}A_\mathrm{PD}}{\mathrm{FF}},
\label{Eq:CPC_A2}
\end{equation} 
where $\mathrm{FF}$ is the \ac{FF} of the \ac{PD} array. In addition, the geometrical concentration gain of a \ac{CPC} is given by \cite{RWinston2005}:
\begin{equation}
    G_\mathrm{CPC} = \frac{A_1}{A_2} = \frac{n_\mathrm{CPC}^2}{\sin^2{\theta_\mathrm{CPC}}} ,
    \label{Eq:CPC_gain}
\end{equation}
where $n_\mathrm{CPC}$ is the refraction index of the dielectric material. The half-angle \ac{FOV} of the \ac{ADR} is $\mathrm{FOV} = 3\theta_\mathrm{CPC}$ \cite{ESarbazi2022Design2}. The normal vector of each receiver element can be described by using the azimuth and elevation angles, donated by $\theta_j$ and $\omega_j$, respectively, for the $j$th receiver element. According to the \ac{ADR} structure, the elevation angle of the six elements around the middle one is the same as their tilting angle which is equal to $2\theta_\mathrm{CPC}$ \cite{ESarbazi2022Design2}, assuming that the \ac{ADR} is oriented vertically upward. Consequently, $\theta_j=2\theta_\mathrm{CPC}$ for $j=1,2,\dots,6$, and $\theta_7=0$. The azimuth angle of the $j$th receiver element is given by $\omega_j=\dfrac{(j-1)\pi}{3}$ for $j=1,2,\dots,6$, and $\omega_7=0$. Hence, the normal vector of the $j$th receiver element can be expressed in Cartesian coordinates as follows:
\begin{equation}
    \mathbf{e}_j = [\sin\theta_j\cos\omega_j~\sin\theta_j\sin\omega_j~\cos\theta_j]^T.
    \label{Eq:e_j}
\end{equation}

\subsubsection{Optical Channel Gain}
We use the coordinate system shown in Fig.~\ref{Fig:Downlink} for analysis. The $k$th \ac{UE} is located at $\mathbf{p}_k=[x_k~y_k~0]^T$. The distance vector from the $v$th transmitter element of the \ac{AP} to $\mathbf{p}_k$ is given by $\mathbf{d}_{kv}=\mathbf{p}_k-\mathbf{q}_v$, and:
\begin{equation}
    d_{kv}=\lVert\mathbf{d}_{kv}\rVert=\sqrt{(x_k-x_v)^2+(y_k-y_v)^2+h_\mathrm{DL}^2}.
    \label{Eq:d_ku}
\end{equation}
The incident beam intensity of the $i$th \ac{VCSEL} with a radiance angle of $\phi_{ki}$ with respect to the \ac{UE} position is:
\begin{equation}
	I_i(x_k,y_k) = \frac{2P_\mathrm{t}}{\pi w'^2(d_{kv}\cos\phi_{ki})}{\exp{\left(-\frac{2d_{kv}^2\sin^2\phi_{ki}}{w'^2(d_{kv}\cos\phi_{ki})}\right)}},
    \label{Eq:I_v}
\end{equation}
where $\phi_{ki} = \cos^{-1}\left(\dfrac{\mathbf{d}_{kv}{\cdot}\mathbf{v}'_i}{d_{kv}}\right)$, for $v=\Big\lceil\dfrac{i}{25}\Big\rceil$.
Therefore, the optical power received by the $j$th \ac{ADR} element of the $k$th \ac{UE} can be expressed as:
\begin{equation}
    P_{kj}^{(i)} = I_i(x_k,y_k)A_1\mathrm{FF}\cos\psi_{ji}\mathds{1}_{\theta_\mathrm{CPC}}(\psi_{ji}),
    \label{Eq:P_kj}
\end{equation}
where $\psi_{ji}=\cos^{-1}(\mathbf{e}_j{\cdot}\mathbf{v}'_i)$ is the incidence angle of the beam with respect to the normal vector of the $j$th receiver element. The last factor in \eqref{Eq:P_kj} is an indicator function which is defined as $\mathds{1}_{\theta_\mathrm{CPC}}(\psi_{ji})=1$ if $0\leq\psi_{ji}\leq\theta_\mathrm{CPC}$, and $0$ otherwise. By using \eqref{Eq:CPC_A2}--\eqref{Eq:P_kj}, the \ac{DC} gain of the downlink channel between the $i$th \ac{VCSEL} and the $j$th receiver element of the $k$th \ac{UE} can be derived as: 
\begin{equation}
\begin{aligned}
    H_{kj}^{(i)} = \frac{2N_\mathrm{PD}A_\mathrm{PD}G_\mathrm{CPC}}{\pi w'^2(d_{kv}\cos\phi_{ki})}&{\exp{\left(-\frac{2d_{kv}^2\sin^2\phi_{ki}}{w'^2(d_{kv}\cos\phi_{ki})}\right)}} \\ 
    & \times \cos\psi_{ji}\mathds{1}_{\theta_\mathrm{CPC}}(\psi_{ji}).
\end{aligned}
\label{Eq:Hkj_i}
\end{equation}

Let $\mathcal{C}_u$ represent the index set of the \acp{VCSEL} participating in the $u$th cluster for $u=1,2,\dots,N_\mathrm{c}$. Once the \acp{UE} joined their clusters, the $u$th cluster serves a subset of them with $K_u$ members such that $\sum\limits_{u=1}^{N_\mathrm{c}}K_u=K$. The \acp{UE} in the network are globally indexed from $1$ to $K$ and those served by the $u$th cluster are indicated by the index set $\mathcal{K}_u$ with $\lvert\mathcal{K}_u\rvert=K_u$. The optical power received by the $j$th receiver element of the $k$th \ac{UE} from the $u$th cluster and the corresponding total received power are given by:
\begin{subequations}
    \begin{align}
        P_{kj}^{\mathcal{C}_u} &= \sum_{i\in\mathcal{C}_u}H_{kj}^{(i)}P_\mathrm{t}, \\
        P_k^{\mathcal{C}_u} &= \sum_{j=1}^{7}P_{kj}^{\mathcal{C}_u} = \sum_{j=1}^{7}H_{kj}^{\mathcal{C}_u}P_\mathrm{t} = H_{k}^{\mathcal{C}_u}P_\mathrm{t},
    \end{align}
\end{subequations}
where $H_{k}^{\mathcal{C}_u}$ is the equivalent \ac{DC} gain of the \ac{CoMB}-\ac{JT} channel between the $u$th cluster and the $k$th \ac{UE} such that:
\begin{equation}
    H_{k}^{\mathcal{C}_u} = \sum_{j=1}^{7}H_{kj}^{\mathcal{C}_u} = \sum_{j=1}^{7}\sum_{i\in\mathcal{C}_u}H_{kj}^{(i)}.
    \label{Eq:Hk_Cu}
\end{equation}
\subsubsection{Noise Model}
Noise sources at the receiver consist of the thermal noise and shot noise, and the laser noise arising from the \acp{VCSEL}. For the $j$th receiver element of the $k$th \ac{UE}, the total noise is modeled as a white Gaussian noise with a single-sided \ac{PSD} of \cite{ESarbazi2020tb}:
\begin{equation}
\begin{split}
	S_{kj} =\frac{4\kappa T}{R_\mathrm{L}}F_\mathrm{n}N_\mathrm{PD} &+2q\sum_{u=1}^{N_\mathrm{c}}R_\mathrm{PD} P_{kj}^{\mathcal{C}_u} \\
	&+\mathrm{RIN}\sum_{u=1}^{N_\mathrm{c}}\left(R_\mathrm{PD} P_{kj}^{\mathcal{C}_u}\right)^{\!2},
 \end{split}
	\label{Eq:S_kj}
\end{equation}
where $\kappa$ is the Boltzmann constant, $T$ is temperature in Kelvin, $R_\mathrm{L}$ is the load resistance, $F_\mathrm{n}$ is the preamplifier noise figure, $q$ is the elementary charge, and $\mathrm{RIN}$ is the \ac{PSD} of the \ac{RIN}, defined as the mean square of intensity fluctuations normalized to the squared average intensity \cite{LColdren2012}. The total noise variance is given by $\sigma_{kj}^2=S_{kj}B$, where $B$ is the single-sided bandwidth of the system which is determined by the receiver bandwidth, as \acp{VCSEL} typically have a very large modulation bandwidth (e.g., in excess of $30$~GHz \cite{NLedentsov2022VCSEL}).

\vspace{-10pt}

%---------------------------------------------------------------------------------------------------
\subsection{Multi-User Performance Analysis}
To ensure a high spectral efficiency, we assume the use of \ac{DCO{-}OFDM} based on an $N$-point \ac{FFT} with adaptive \ac{QAM}. Thus, $\frac{N}{2}-1$ data-carrying subcarriers are available for each \ac{VCSEL}. The average electrical power, $P_\mathrm{elec}$, is related to the average transmit optical power, $P_\mathrm{t}$, by $P_\mathrm{elec}=\frac{1}{9}P_\mathrm{t}^2$, so that the clipping distortion remains negligible \cite{Dimitrov20122}. We apply \ac{NOMA} and \ac{OFDMA} techniques for the downlink multi-user access. The received \ac{SINR} and achievable rate expressions for these techniques are provided in the following.

\subsubsection{Multi-User Access}
In \ac{NOMA}, spectral resources (i.e., \acs{OFDM} subcarriers) are fully shared by every user in a cluster and downlink transmissions for different users are multiplexed in the power domain based on superposition coding. In principle, for each cluster, the superposed signal is broadcast to all the involved users and \ac{SIC} is applied at the receiver side for multi-user detection. For each user, \ac{SIC} is performed by detecting and canceling the interference caused by other users with weaker channel conditions while treating the interference originating from those with stronger channel conditions as noise.
\newcounter{mytempeqncnt1}
\setcounter{mytempeqncnt1}{\value{equation}}
\setcounter{equation}{46}
\begin{figure*}[t!]
\normalsize
% IEEE uses as a separator
% \hrulefill
\begin{equation}\label{NOMA:eq2}
    Y_{kj}^{\mathcal{C}_u} = R_\mathrm{PD}\zeta\sqrt{P_\mathrm{elec}}\Bigg(\underbrace{\sum_{\ell=1}^{k-1}a_\ell^{\mathcal{C}_u} X_\ell^{\mathcal{C}_u}}_{\text{SIC}}+a_k^{\mathcal{C}_u} X_k^{\mathcal{C}_u}+\underbrace{\sum_{\ell=k+1}^{K_u}a_\ell^{\mathcal{C}_u} X_\ell^{\mathcal{C}_u}}_{\text{residual MUI}}\Bigg)H^{\mathcal{C}_u}_{kj}
    +R_\mathrm{PD}\zeta\sqrt{P_\mathrm{elec}}\underbrace{\sum_{v\neq u}\sum_{\ell\in\mathcal{K}_v} a_\ell^{\mathcal{C}_v} X_\ell^{\mathcal{C}_v} H^{\mathcal{C}_v}_{kj}}_{\text{ICI}}+\sqrt{\xi}V_{kj},
\end{equation}
\hrulefill
% The spacer can be tweaked to stop underfull vboxes.
% \vspace*{4pt}
\vspace{-10pt}
\end{figure*}
\setcounter{equation}{\value{mytempeqncnt1}}

\newcounter{mytempeqncnt2}
\setcounter{mytempeqncnt2}{\value{equation}}
\setcounter{equation}{47}
\begin{figure*}[t!]
\normalsize
% IEEE uses as a separator
% \hrulefill
\begin{equation}
    \mathsf{SINR}_k^{\mathcal{C}_u} =
    \begin{cases}
    \dfrac{\bigg(\sum\limits_{j=1}^{7}w_{kj} a_k^{\mathcal{C}_u} H^{\mathcal{C}_u}_{kj}\bigg)^{\!2}}
    {\sum\limits_{\ell=k+1}^{K_u} {\bigg(\sum\limits_{j=1}^{7}w_{kj} a_\ell^{\mathcal{C}_u} H^{\mathcal{C}_u}_{kj}\bigg)^{\!2}} 
    + \sum\limits_{v \neq u}\sum\limits_{\ell\in\mathcal{K}_v} {\bigg(\sum\limits_{j=1}^{7}w_{kj} a_\ell^{\mathcal{C}_v} H^{\mathcal{C}_v}_{kj}\bigg)^{\!2}}
    +\gamma\sum\limits_{j=1}^{7} w_{kj}^2 \sigma_{kj}^2}, 
    & k = 1,2,\dots,K_u-1\\
    \dfrac{\bigg(\sum\limits_{j=1}^{7}w_{kj} a_k^{\mathcal{C}_u} H^{\mathcal{C}_u}_{kj}\bigg)^{\!2}}
    {\sum\limits_{v \neq u}\sum\limits_{\ell\in\mathcal{K}_v} {\bigg(\sum\limits_{j=1}^{7}w_{kj} a_\ell^{\mathcal{C}_v} H^{\mathcal{C}_v}_{kj}\bigg)^{\!2}} 
    +\gamma\sum\limits_{j=1}^{7} w_{kj}^2 \sigma_{kj}^2},  & k = K_u 
    \end{cases}
    \label{Eq:SINR_NOMA}
\end{equation}
\hrulefill
\begin{equation}
    w_{kj} =
    \begin{cases}
    \dfrac{\zeta\sqrt{\gamma} a_k^{\mathcal{C}_u} H^{\mathcal{C}_u}_{kj}}
    {\sum\limits_{\ell=k+1}^{K_u} {\left(a_\ell^{\mathcal{C}_u} H^{\mathcal{C}_u}_{kj}\right)^{\!2}} 
    + \sum\limits_{v \neq u}\sum\limits_{\ell\in\mathcal{K}_v} {\left(a_\ell^{\mathcal{C}_v} H^{\mathcal{C}_v}_{kj}\right)^{\!2}}
    +\gamma \sigma_{kj}^2}, 
    & k = 1,2,\dots,K_u-1\\
    \dfrac{\zeta\sqrt{\gamma} a_k^{\mathcal{C}_u} H^{\mathcal{C}_u}_{kj}}
    {\sum\limits_{v \neq u}\sum\limits_{\ell\in\mathcal{K}_v} {\left(a_\ell^{\mathcal{C}_v} H^{\mathcal{C}_v}_{kj}\right)^{\!2}} 
    +\gamma \sigma_{kj}^2},  & k = K_u 
    \end{cases}
    \label{Eq:w_jk_MRC_NOMA}
\end{equation}
\hrulefill
% The spacer can be tweaked to stop underfull vboxes.
% \vspace*{4pt}
\vspace{-10pt}
\end{figure*}
\setcounter{equation}{\value{mytempeqncnt2}}

Without loss of generality, suppose that the channel gains of the \acp{UE} associated with the $u$th cluster are sorted in an ascending order such that:
\begin{equation}\label{NOMA:eq1}
    H_1^{\mathcal{C}_u} \leq \cdots \leq H_k^{\mathcal{C}_u} \leq \cdots \leq H_{K_u}^{\mathcal{C}_u},
\end{equation}
for $u=1,2,\dots,N_\mathrm{c}$. Let $a_k^{\mathcal{C}_u}$ be the optical power allocation coefficient corresponding to the $k$th \ac{UE} in the $u$th cluster. The total power constraint implies that $\sum\limits_{k=1}^{K_u}\left(a_k^{\mathcal{C}_u}\right)^{\!2}=1$. To maintain fairness among \acp{UE}, a suitable power allocation policy is one that prioritizes the \acp{UE} that have weaker channel conditions. In this case, it follows from \eqref{NOMA:eq1} that $a_1^{\mathcal{C}_u} \geq \cdots \geq a_k^{\mathcal{C}_u} \geq \cdots \geq a_{K_u}^{\mathcal{C}_u}$. The power allocation coefficients are chosen as $a_k^{\mathcal{C}_u}=\sqrt{\dfrac{K_u-k+1}{\sigma}}$, where $\sigma=\dfrac{K_u(K_u+1)}{2}$ \cite{ZhiguoNOMA2014}. Upon removing the \ac{DC} term, the received signal at the $j$th receiver element of the $k$th \ac{UE} in the $u$th cluster is expressed as \eqref{NOMA:eq2}, given at the top of the page, where the message symbols $X_\ell^{\mathcal{C}_v}$ are assumed to be drawn from a normalized \ac{QAM} constellation, and the factor $\zeta=\sqrt{\dfrac{N}{N-2}}$ ensures that the average power of the time domain signal is normalized to unity. Also, the noise terms $V_{kj}$ are independent Gaussian random variables with a zero mean and variance $\sigma_{kj}^2$, and the factor $\xi=\dfrac{N-2}{N}$ is the subcarrier utilization ratio. The received signals, for $j=1,2,\dots,7$, are combined together in the form of $Y_k^{\mathcal{C}_u} = \sum\limits_{j=1}^{7}w_{kj} Y_{kj}^{\mathcal{C}_u}$, where $w_{kj}$ is the corresponding combining weight. The receiver fully decodes the \ac{SIC} term, and then subtracts its remodulated version from the received signal before proceeding to decode the $k$th \ac{UE}'s message. During the whole decoding process, the receiver treats both the residual \ac{MUI} in the same cluster and the \ac{ICI} from the rest of the clusters as noise. Assuming perfectly error-free \ac{SIC} operations, the \ac{SINR} of the $k$th \ac{UE} in the $u$th cluster is given by \eqref{Eq:SINR_NOMA} at the top of the page, where $\gamma=\dfrac{\xi^2}{R_\mathrm{PD}^2 P_\mathrm{elec}}$.
%---------------------------------------------------------------------------------------------------
Based on \ac{MRC}, the combining weight for each receiver element is proportional to its received photocurrent to noise ratio \cite{JCarruther2000angle}, resulting in \eqref{Eq:w_jk_MRC_NOMA} at the top of the page.

When using \ac{OFDMA}, a fraction $b_k^{\mathcal{C}_u}$ of the total bandwidth and a fraction $p_k^{\mathcal{C}_u}$ of the total power are allocated to the $k$th \ac{UE} in the $u$th cluster subject to $\sum\limits_{k=1}^{K_u} b_k^{\mathcal{C}_u} \!=\! 1$ and $\sum\limits_{k=1}^{K_u} p_k^{\mathcal{C}_u} \!=\! 1$. Hence, under \ac{OFDMA}, the intra-cluster \ac{MUI} is eliminated and the \ac{ICI} remains as the residual interference. The \ac{SINR} of the $k$th \ac{UE} in the $u$th cluster can be written as:
\addtocounter{equation}{3}
\begin{equation}
    \mathsf{SINR}_k^{\mathcal{C}_u} =
    \frac{p_k^{\mathcal{C}_u} \bigg(\sum\limits_{j=1}^{7}w_{kj} H^{\mathcal{C}_u}_{kj}\bigg)^{\!2}}
    {\sum\limits_{v \neq u} p_\ell^{\mathcal{C}_v} \bigg(\sum\limits_{j=1}^{7}w_{kj} H^{\mathcal{C}_v}_{kj}\bigg)^{\!2}
    +\gamma b_k^{\mathcal{C}_u} \sum\limits_{j=1}^{7} w_{kj}^2 \sigma_{kj}^2}.
    \label{Eq:SINR_OFDMA}
\end{equation}
The \ac{MRC} weights for \ac{OFDMA} are given by:
\begin{equation}
    w_{kj} =
    \frac{\zeta\sqrt{\gamma p_k^{\mathcal{C}_u}}H^{\mathcal{C}_u}_{kj}}
    {\sum\limits_{v \neq u} p_\ell^{\mathcal{C}_v} \left(H^{\mathcal{C}_v}_{kj}\right)^{\!2}
    +\gamma b_k^{\mathcal{C}_u} \sigma_{kj}^2}.
    \label{Eq:w_kj_MRC_OFDMA}
\end{equation}

%---------------------------------------------------------------------------------------------------
\subsubsection{Sum Rate Analysis}
For an \ac{AWGN} channel, using adaptive \ac{QAM}, a tight upper bound of the achievable rate for $0\leq\mathsf{SNR}\leq30$ dB is given by \cite{Goldsmith1997}:
\begin{equation}
R = \xi B\log_2\left(1+\frac{\mathsf{SNR}}{\Gamma}\right),
\label{Eq:Rate}
\end{equation}
and $\Gamma = -\dfrac{\ln\left(5\ \!\mathrm{BER}\right)}{1.5}$ is the \ac{SNR} gap accounting for the required \ac{BER} performance.

The achievable rate of the $k$th \ac{UE} in the $u$th cluster under \ac{NOMA} and \ac{OFDMA} is deduced from \eqref{Eq:Rate} as:
\begin{equation}\label{Eq:RatePerUE}
R_k^{\mathcal{C}_u} = 
\begin{cases}
\xi B\log_2\left(1+\dfrac{\mathsf{SINR}_k^{\mathcal{C}_u}}{\Gamma}\right), & \mathrm{NOMA}\\
b_k^{\mathcal{C}_u}\xi B\log_2\left(1+\dfrac{\mathsf{SINR}_k^{\mathcal{C}_u}}{\Gamma}\right), & \mathrm{OFDMA}
\end{cases}
\end{equation}
where $\mathsf{SINR}_k^{\mathcal{C}_u}$ for \ac{NOMA} and \ac{OFDMA} is given by \eqref{Eq:SINR_NOMA} and \eqref{Eq:SINR_OFDMA}, respectively. The sum rate for all clusters of the network is computed by:
\begin{equation}\label{Eq:SumRate}
R = \sum_{u=1}^{N_\mathrm{c}}\sum_{k=1}^{K_u}R_k^{\mathcal{C}_u}.
\end{equation}
The multi-user fairness is measured by means of Jain's fairness index \cite{RJain1984}, which is defined as follows:
\begin{equation}\label{Eq:FairnessIndex}
    J \triangleq \frac{\left(\sum\limits_{u=1}^{N_\mathrm{c}}\sum\limits_{k=1}^{K_u}R_k^{\mathcal{C}_u}\right)^{\!\!2}}{K\sum\limits_{u=1}^{N_\mathrm{c}}\sum\limits_{k=1}^{K_u}\left(R_k^{\mathcal{C}_u}\right)^{\!2}}.
\end{equation}

\vspace{-10pt}

%%%%%%%%%%%%%%%%%%%%%%%%%%%%%%%%%%%%%%%%%%%%%%%%%%%%%%%%%%%%%%%%%%%%%%%%%%%%%%%%%%%%%%%%%%%%%%%%%%%%
%%%%%%%%%%%%%%%%%%%%%%%%%%%%%%%%%%%%%%%%%%%%%%%%%%%%%%%%%%%%%%%%%%%%%%%%%%%%%%%%%%%%%%%%%%%%%%%%%%%%

\begin{figure*}[!t]
    \vspace{-10pt}
    \centering
    \subfloat[Borders of SDMA clusters ($N_\mathrm{c}=225$) \label{Fig:6_NoClustering_a}]{\includegraphics[width=0.3\linewidth, keepaspectratio=true]{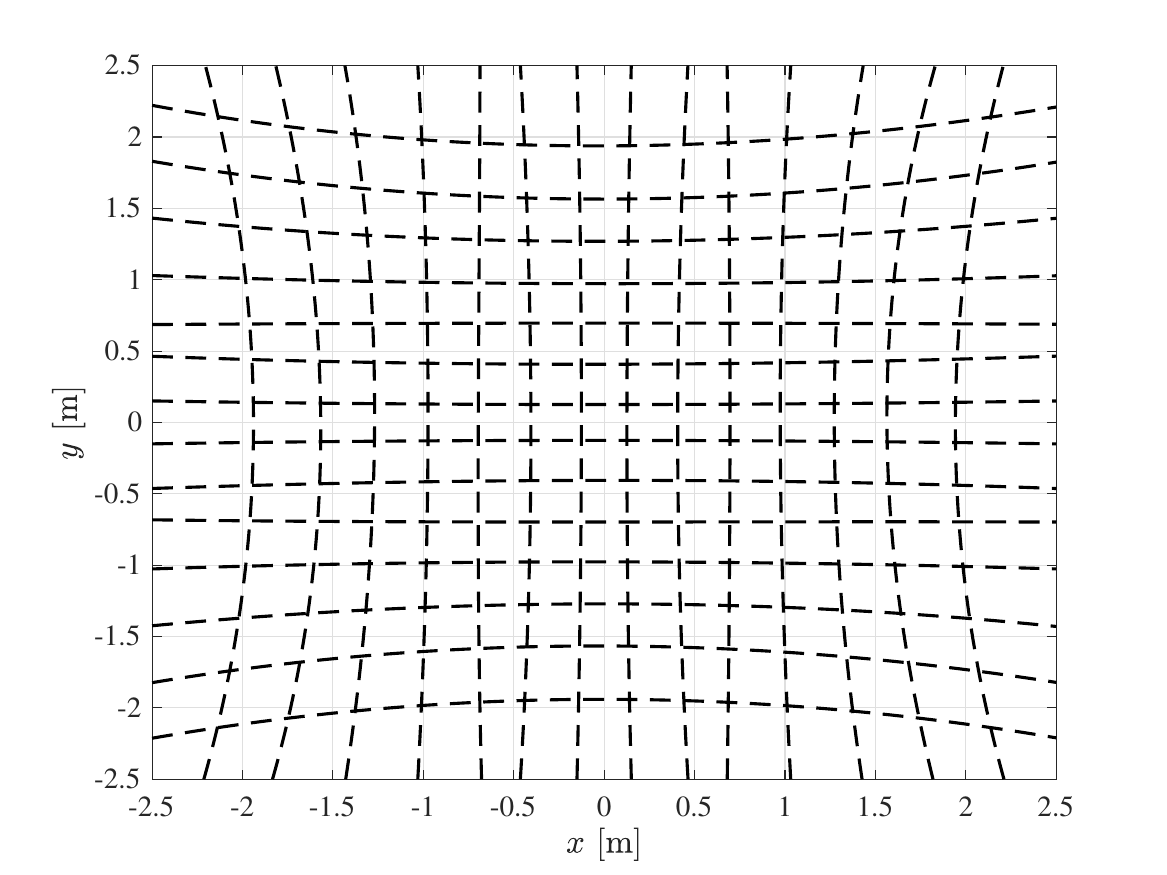}}
    \subfloat[SINR \label{Fig:6_NoClustering_b}]{\includegraphics[width=0.29\linewidth, keepaspectratio=true]{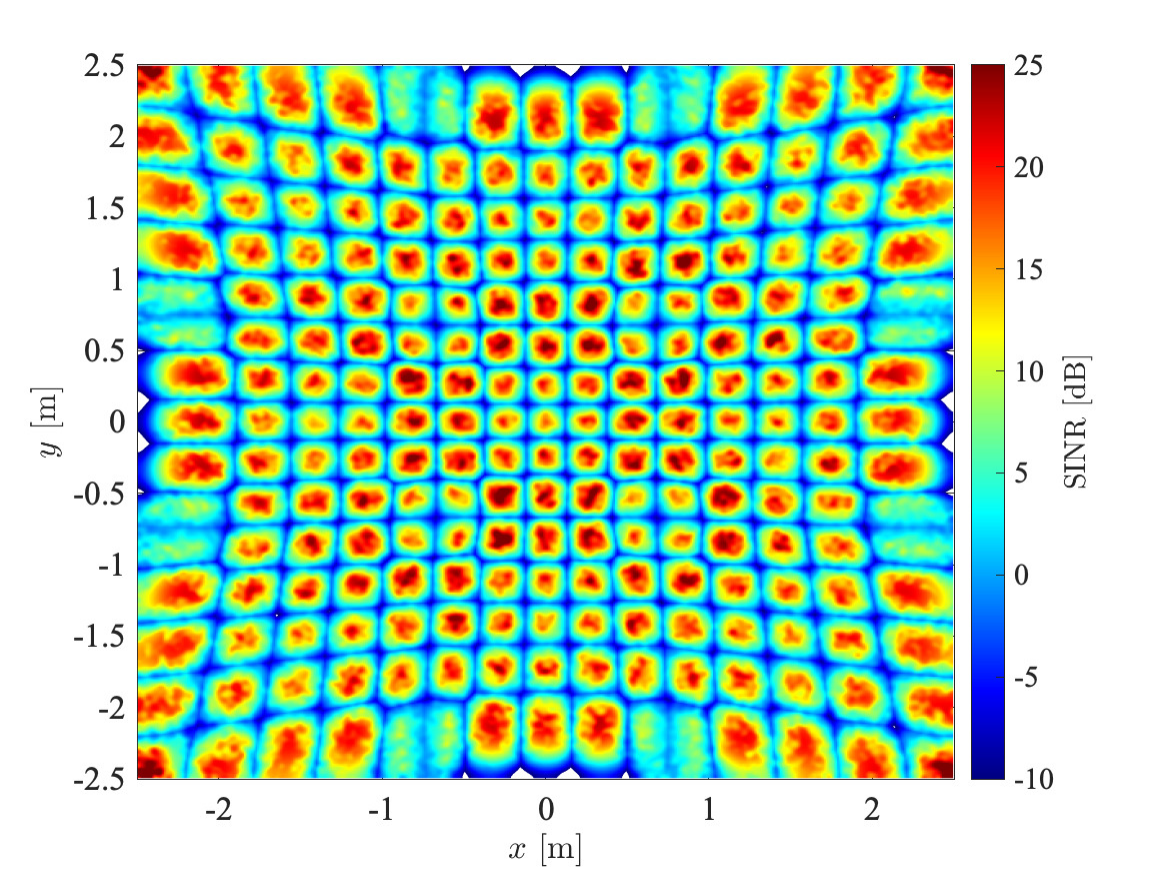}}
    \subfloat[Achievable rate \label{Fig:6_NoClustering_c}]{\includegraphics[width=0.29\linewidth, keepaspectratio=true]{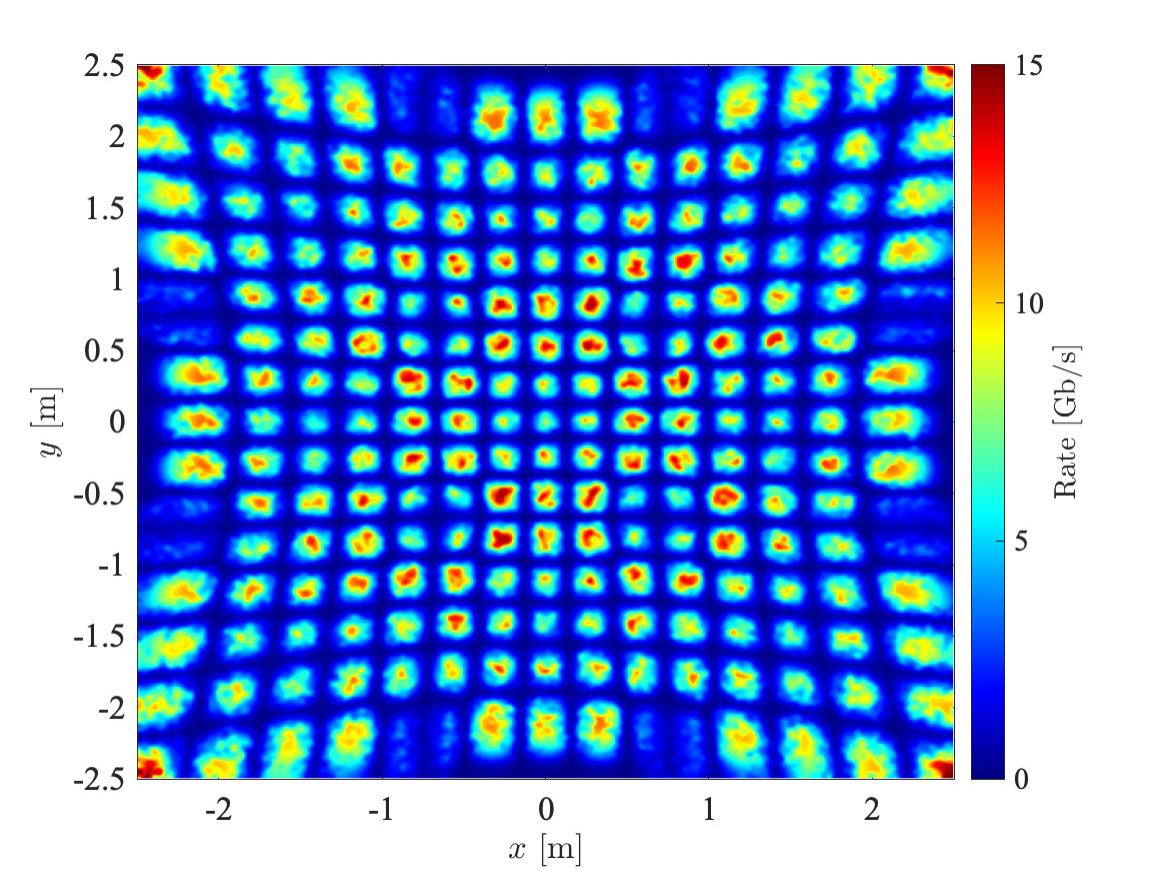}}
    \caption{Baseline SDMA clustering and the resulting spatial distributions of the received SINR and data rate on the receiver plane.}
    \label{Fig:6_NoClustering}
    \vspace{-15pt}
\end{figure*}

\section{Numerical Results and Discussions} \label{NumericalResults}
In this section, we evaluate the proposed double-tier \ac{AP} design for an indoor multi-user access network based on static clustering. The simulation setup consists of an indoor environment as depicted in Fig.~\ref{Fig:NetworkArchitecture} with dimensions of $5 \times 5 \times 3$~m$^3$. The remaining simulation parameters are listed in Table~\ref{Tab:5_1}, unless otherwise specified. Assuming that the receiver is oriented upwards, it requires a half-angle \ac{FOV} of at least $40^\circ$ to be able to receive signals in regions close to the walls. To this end, the half-angle \ac{FOV} of the \ac{ADR} is assumed to be $\mathrm{FOV} = 50^\circ$. The transmit optical power of $P_\mathrm{t}=10$~mW per \ac{VCSEL} is chosen on account of eye safety for each transmitter element based on \ac{IEC} 60825-1 and \ac{ANSI} Z136.1 standards \cite{IEC_Std608251,ANSI-Z136.4,Henderson2003}. The full treatment of eye safety is beyond the scope of this paper. In an earlier study \cite{HKazemi2023MultiBeam}, we have developed an algorithm for eye safety analysis of a \ac{VCSEL} array combined with a lens, which computes the maximum transmit power per \ac{VCSEL} for this purpose.

\vspace{-8pt}

\subsection{Clustering Scenarios: Spatial Distribution of Data Rate}
In the baseline clustering scenario, every single beam is considered to be carrying independent data for its coverage spot. In this case, clusters are of size one, creating a pure \ac{SDMA} network with the highest possible spatial reuse of the spectral resources. Since the bandwidth is fully reused across the network, the adjacent beams interfere with one another. Fig.~\subref*{Fig:6_NoClustering_a} illustrates the approximate borders of the beam spots on the receiver plane. Fig.~\subref*{Fig:6_NoClustering_b} demonstrates the spatial distribution of the received \ac{SINR} on the same plane. The maximum \ac{SINR} of $25$ dB is realized at the centers of the beam spots and the \ac{SINR} value is on the order of $7$ dB at the edge of the network, i.e., in proximity to the walls. To obtain the spatial distribution of the achievable rate, the \ac{SINR} values are mapped onto their respective data rates based on \eqref{Eq:RatePerUE} by assuming that each cluster serves a single\footnote{Note that when there is a single \ac{UE} per cluster, achievable rates for \ac{NOMA} and \ac{OFDMA} are the same according to \eqref{Eq:RatePerUE}.} \ac{UE}. This hypothetical assumption helps to understand the performance bounds in different locations over the coverage area of the \ac{AP}. According to Fig.~\subref*{Fig:6_NoClustering_c}, data rates are about $10$~Gb/s to $15$~Gb/s at the centers of the beam spots except in locations near the walls where still a data rate of $8$~Gb/s is provided. For the baseline \ac{SDMA}, the total number of clusters equals the total number of \acp{VCSEL} in the \ac{AP}, i.e., $N_\mathrm{c}=225$.

\renewcommand{\arraystretch}{1.0}%
\begin{table}[t!]
	\centering
	\caption{Simulation Parameters}
	\begin{tabular}{c|l|l}
		\textbf{Parameter}    & \textbf{Description}                & \textbf{Value} \\ \hline
		$h_\mathrm{DL}$       & Vertical separation                 & $3$ m          \\ 
		$w_0$                 & Beam waist radius                   & $5$ \textmu m  \\
		$\lambda$             & VCSEL wavelength                    & $950$ nm       \\ 
        $P_\mathrm{t}$        & Transmit optical power per VCSEL    & $10$ mW        \\
        $\mathrm{RIN}$        & RIN PSD 	                        & $-155$ dB/Hz   \\ 
 	$n_\mathrm{lens}$     & Lens refractive index               & $1.55$         \\
 	$n_\mathrm{CPC}$      & CPC refractive index                & $1.77$         \\
		$R_\mathrm{PD}$       & PD responsivity                     & $0.7$ A$/$W    \\
 	$N_\mathrm{PD}$       & Number of PDs per PD array          & $16$           \\
 	$\mathrm{FOV}$        & ADR half-angle FOV                  & $30^\circ$     \\
 	$F_\mathrm{n}$   	  & Preamplifier noise figure           & $5$ dB         \\
		$B$                   & System bandwidth                    & $2$ GHz        \\
        $N$                   & IFFT/FFT size                       & $1024$         \\
		$\mathrm{BER}$        & Pre-FEC BER                         & $10^{-3}$      \\ \hline
	\end{tabular}
	\label{Tab:5_1}
	\vspace{-10pt}
\end{table}

% \vspace{-5pt}

\begin{figure*}
\vspace{-20pt}
        \centering
        \subfloat[Scenario 1 ($N_\mathrm{c}=9$) \label{Fig:6_Scenarios_a}]{
            \begin{minipage}{0.99\linewidth}
            \centering
                \includegraphics[height=0.165\textheight, keepaspectratio=true]{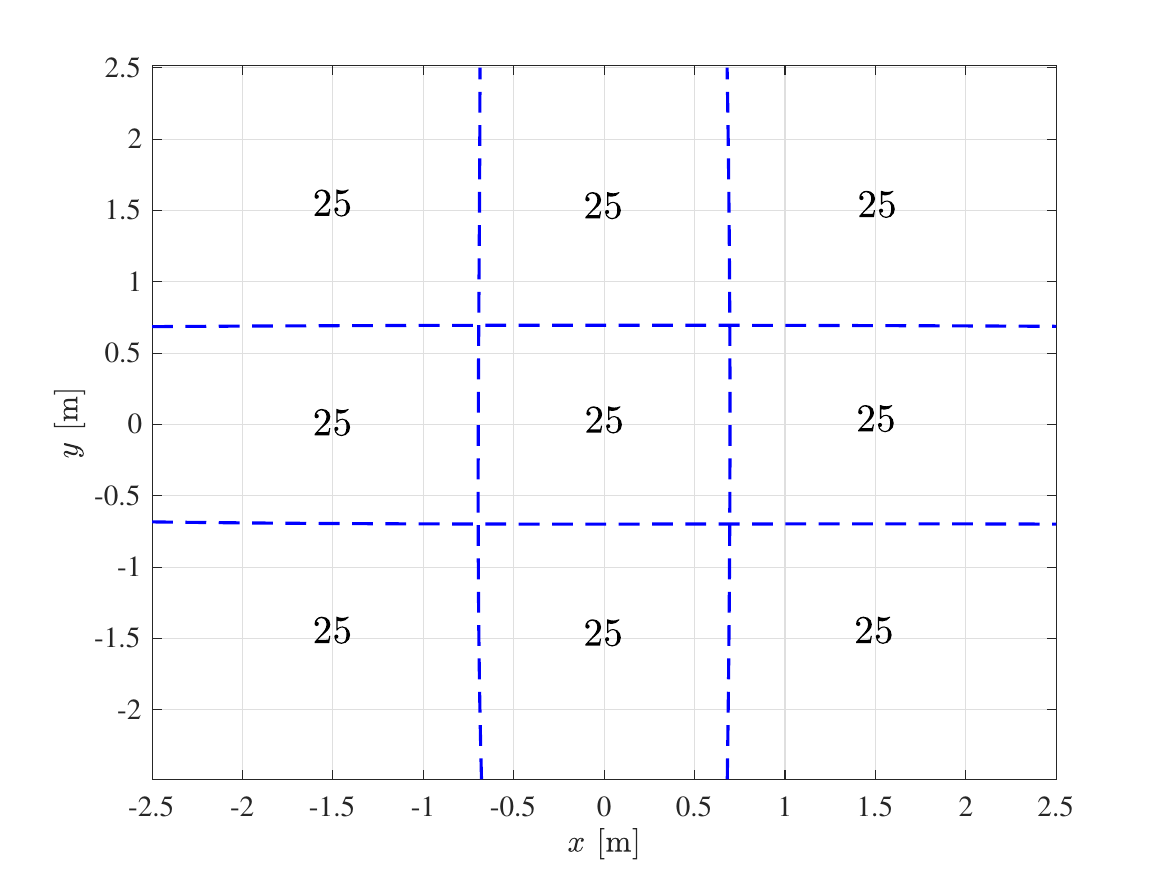}
                \includegraphics[height=0.165\textheight, keepaspectratio=true]{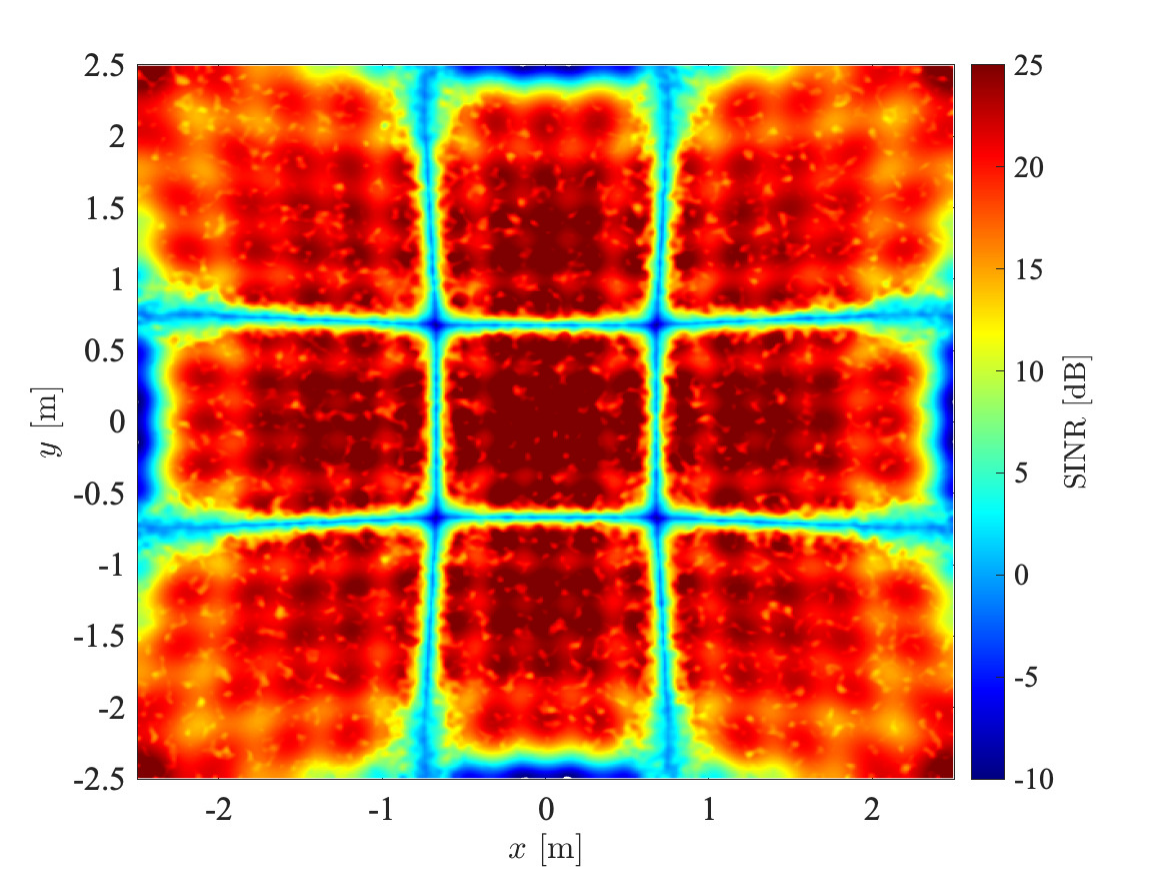}
                \includegraphics[height=0.165\textheight, keepaspectratio=true]{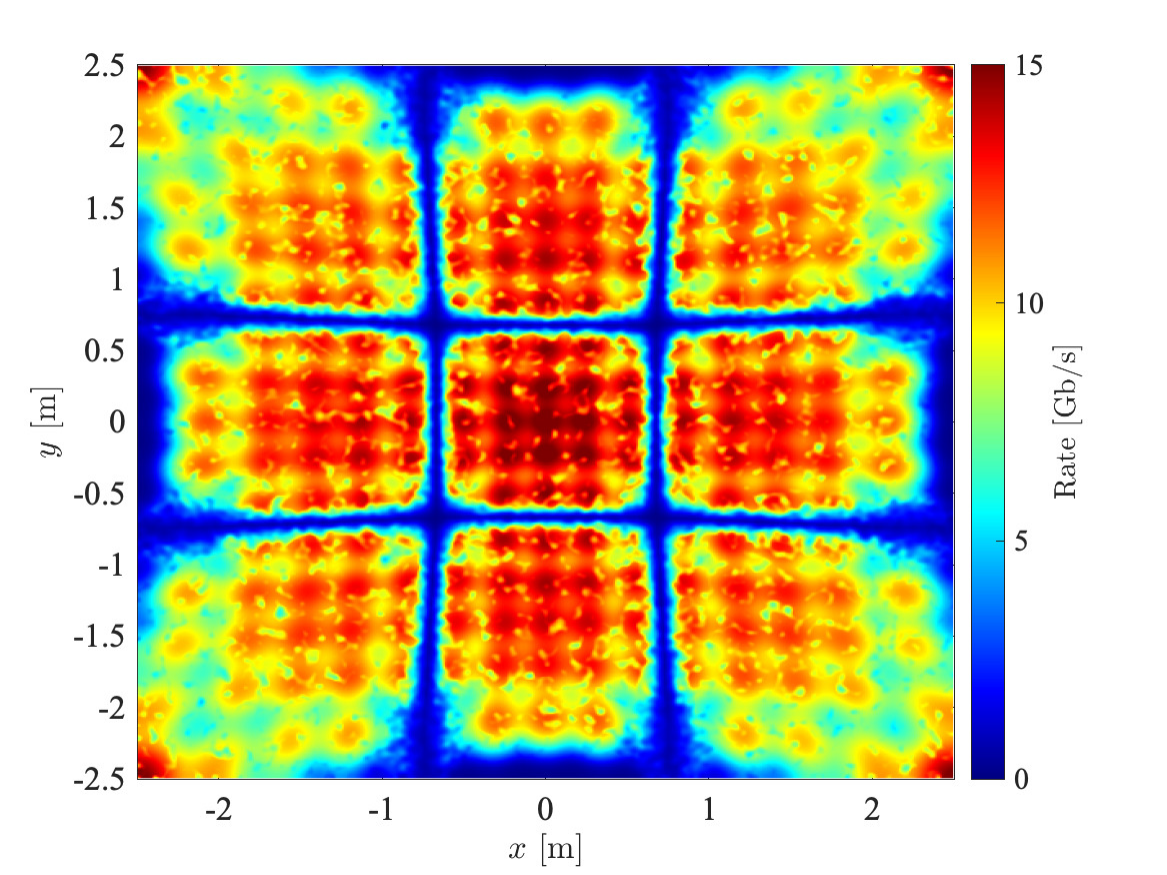}
            \end{minipage}} 
        \vspace{-1pt}    
        \subfloat[Scenario 2 ($N_\mathrm{c}=25$) \label{Fig:6_Scenarios_b}]{
            \begin{minipage}{0.99\linewidth}
            \centering
                \includegraphics[height=0.165\textheight, keepaspectratio=true]{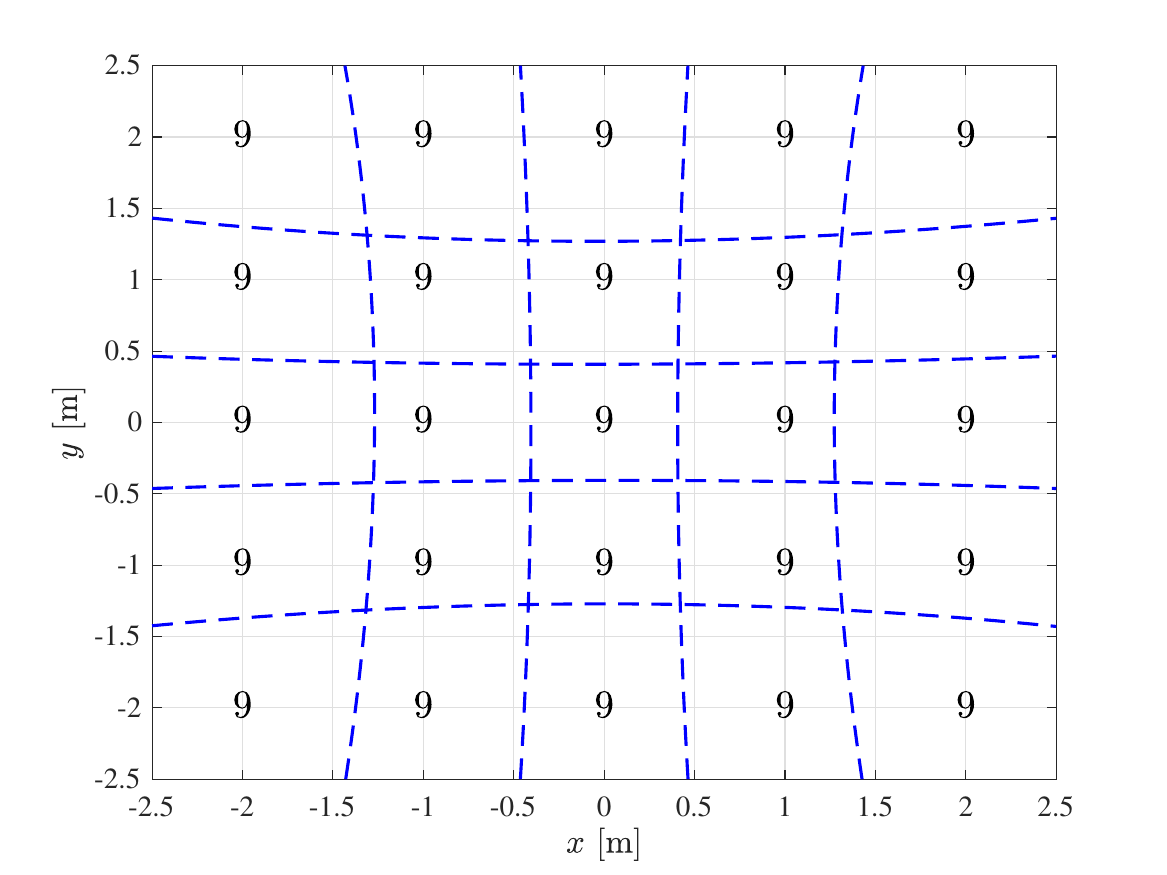}
                \includegraphics[height=0.165\textheight, keepaspectratio=true]{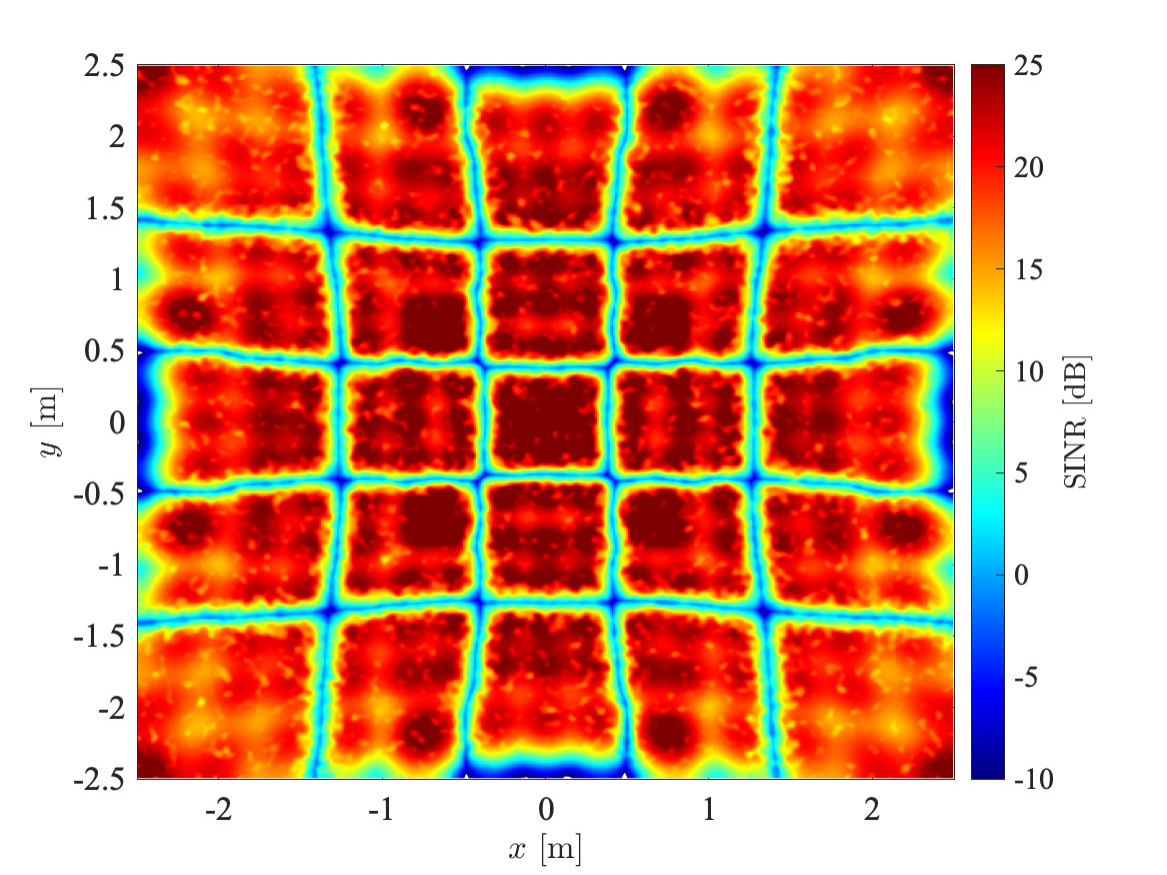}
                \includegraphics[height=0.165\textheight, keepaspectratio=true]{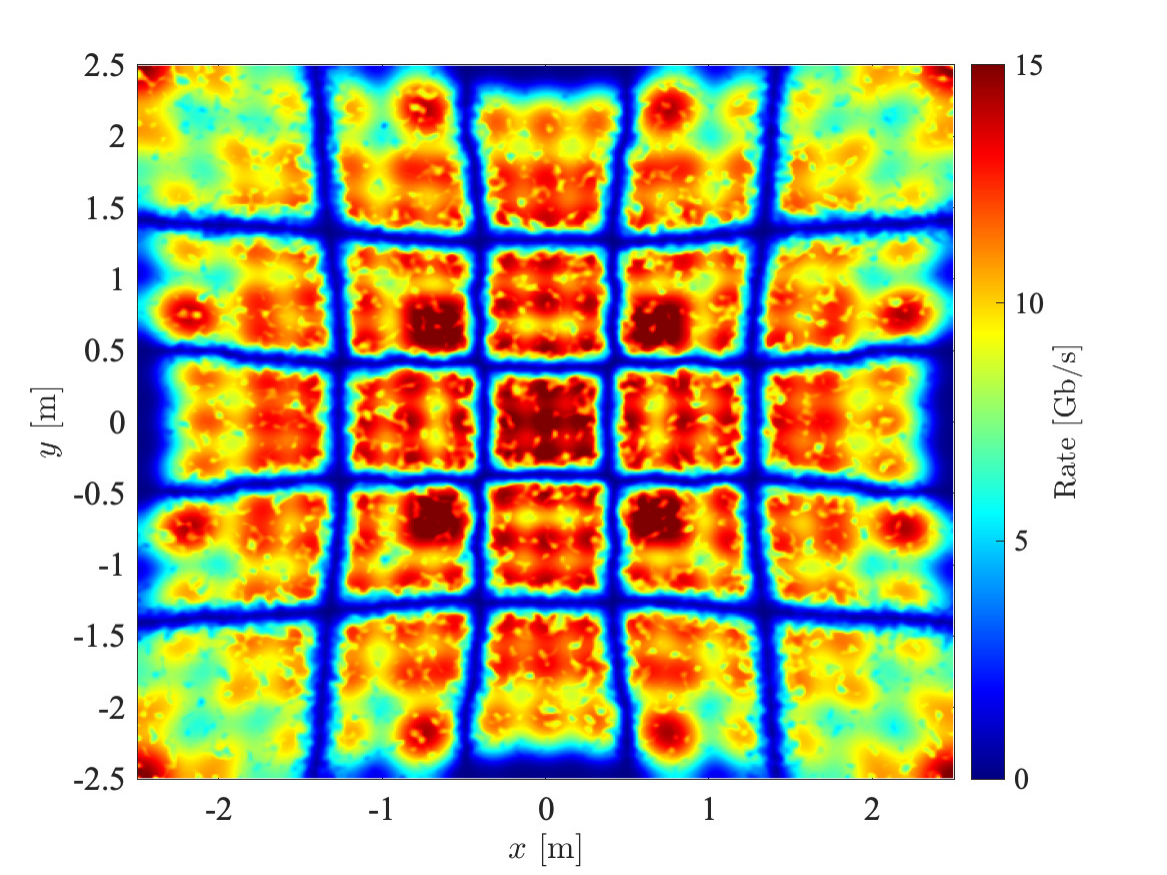}
            \end{minipage}} 
            \vspace{-1pt}  
        \subfloat[Scenario 3 ($N_\mathrm{c}=49$) \label{Fig:6_Scenarios_c}]{
            \begin{minipage}{0.99\linewidth}
            \centering
                \includegraphics[height=0.165\textheight, keepaspectratio=true]{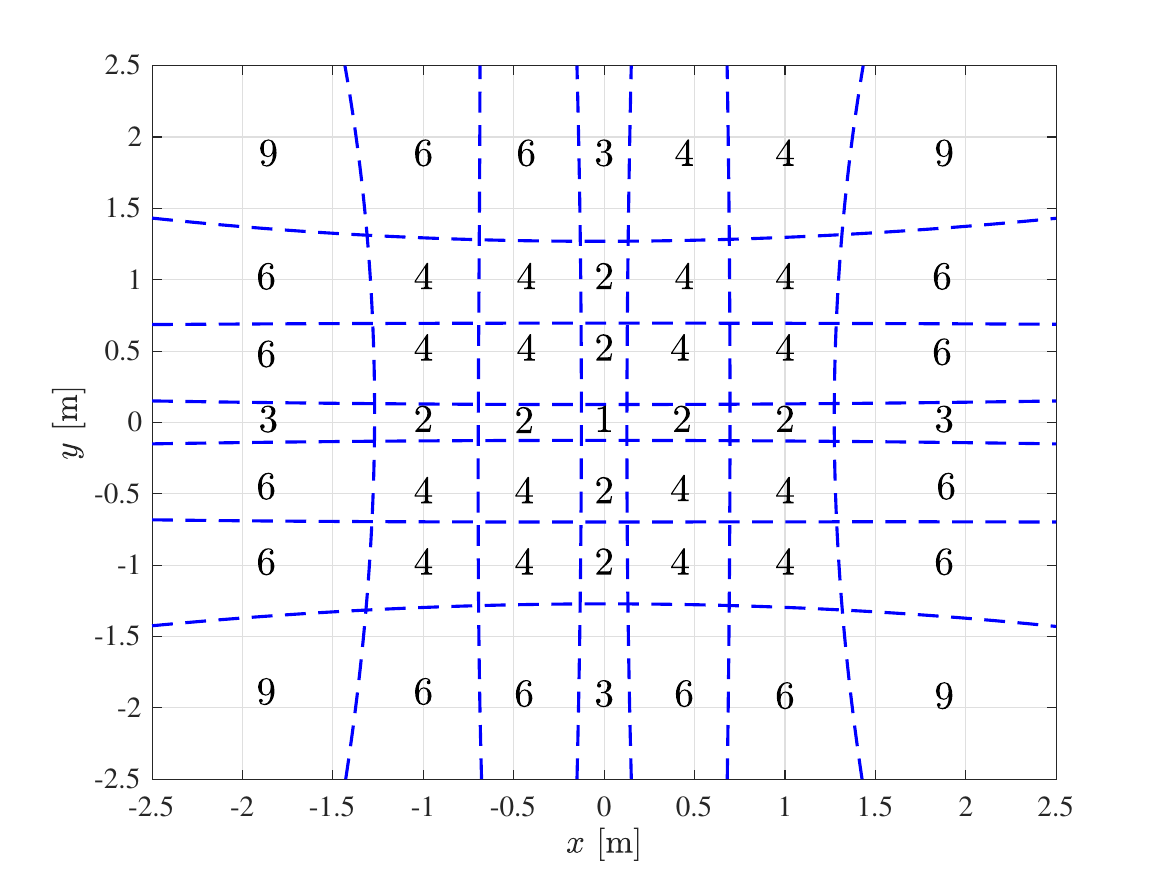}
                \includegraphics[height=0.165\textheight, keepaspectratio=true]{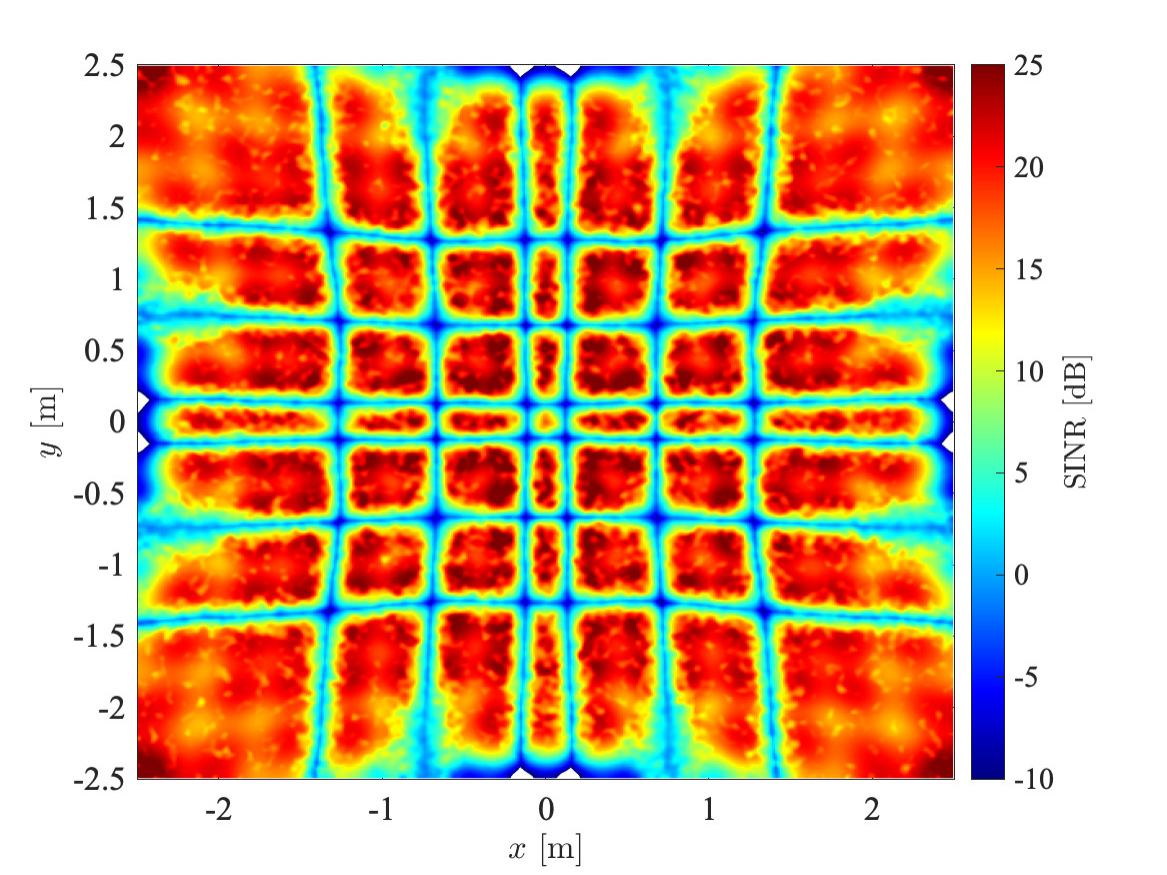}
                \includegraphics[height=0.165\textheight, keepaspectratio=true]{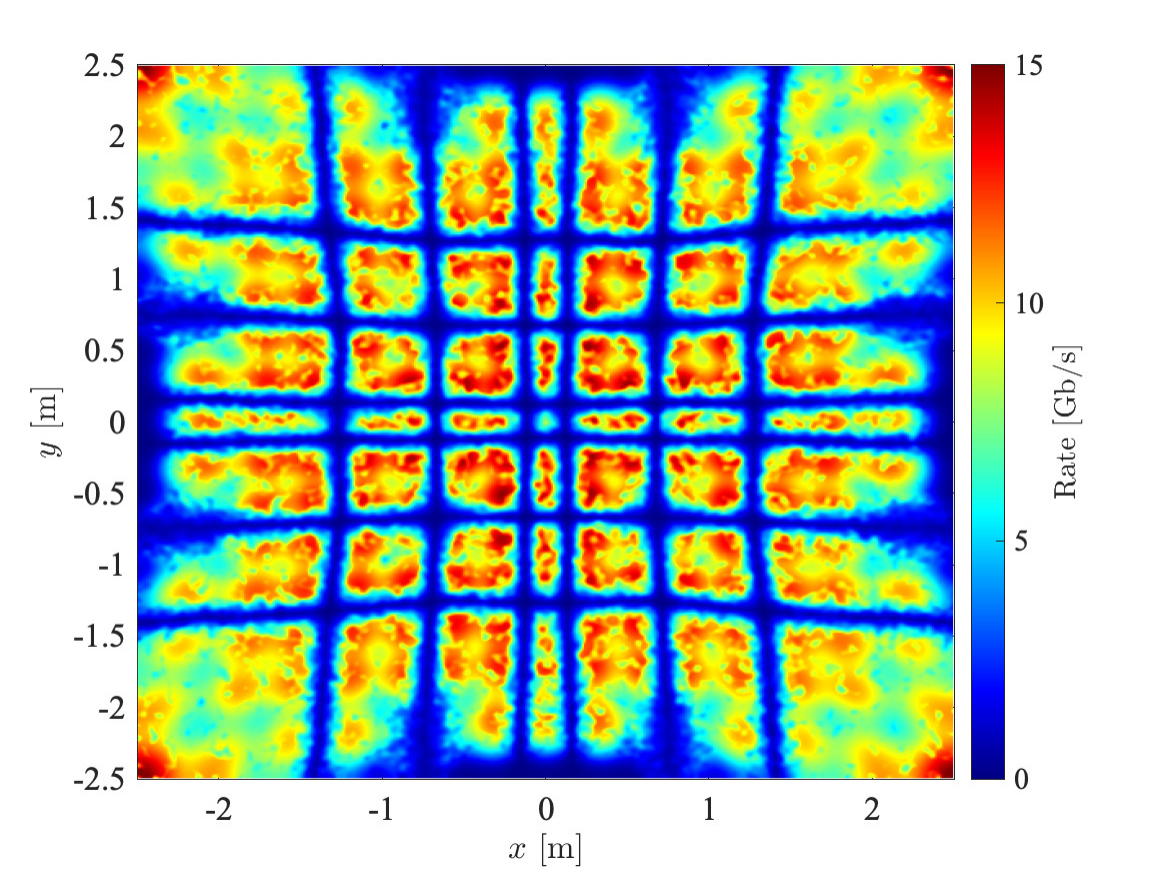}
            \end{minipage}}  
        \vspace{-1pt}    
        \subfloat[Scenario 4 ($N_\mathrm{c}=49$) \label{Fig:6_Scenarios_d}]{
            \begin{minipage}{0.99\linewidth}
            \centering
                \includegraphics[height=0.165\textheight, keepaspectratio=true]{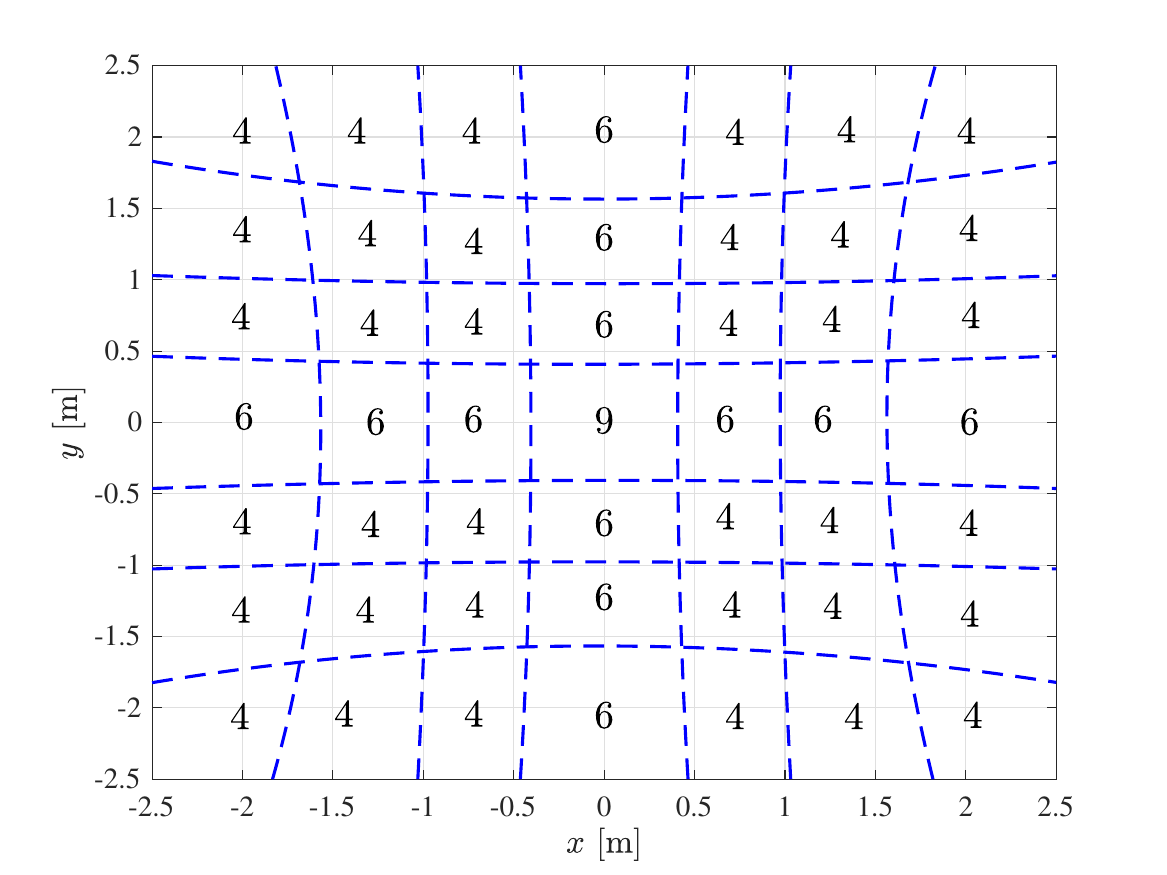}
                \includegraphics[height=0.165\textheight, keepaspectratio=true]{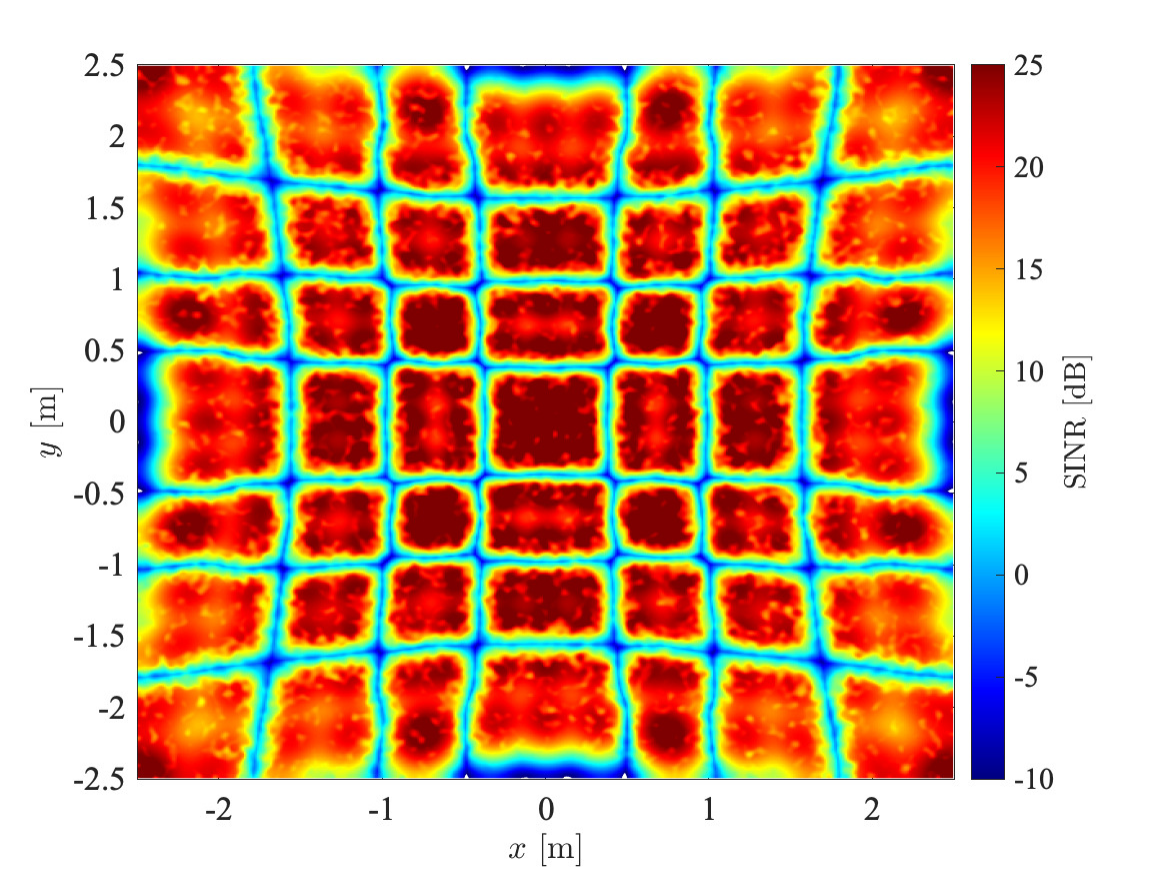}
                \includegraphics[height=0.165\textheight, keepaspectratio=true]{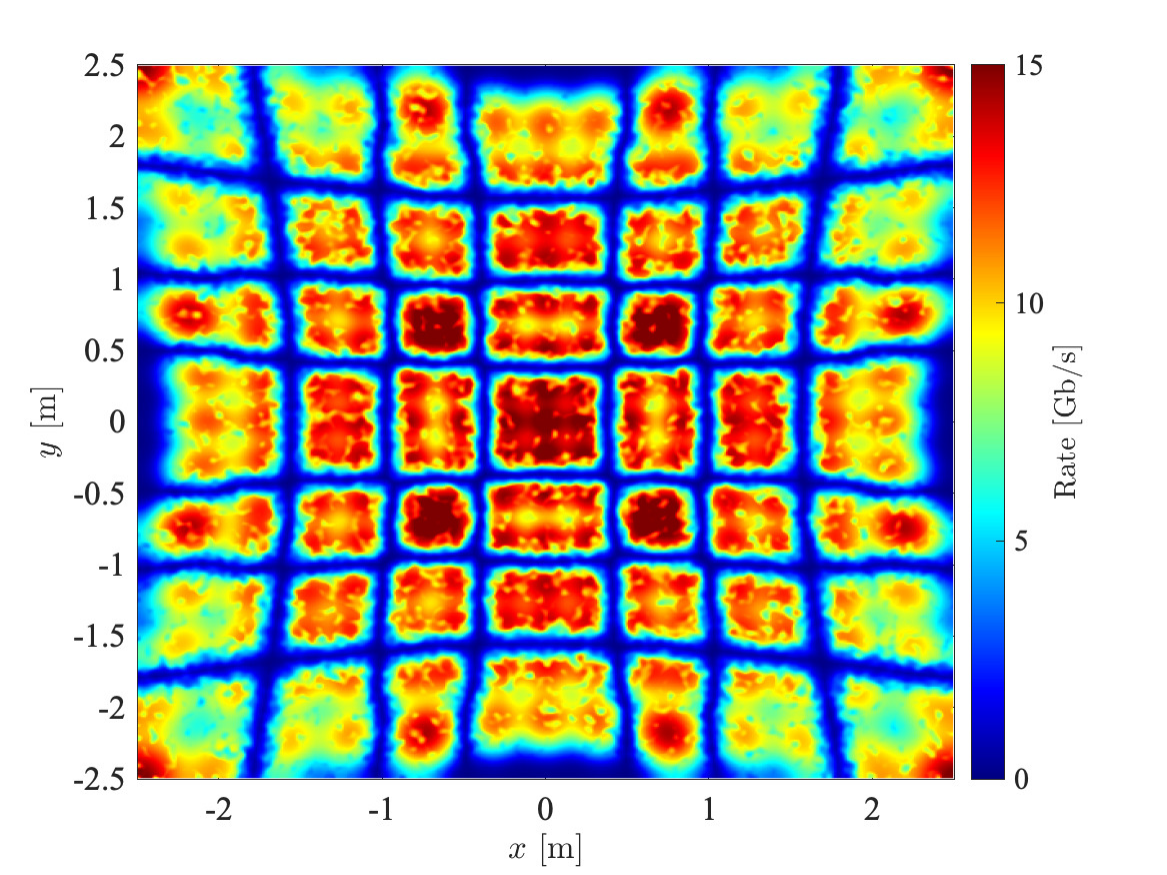}
            \end{minipage}} 
            \caption{Clustering layout (left), spatial distribution of \ac{SINR} (middle) and spatial distribution of data rate (right) on the receiver plane for different scenarios.}
            \label{Fig:6_Scenarios}
            \vspace{-10pt}
\end{figure*}

Besides the baseline \ac{SDMA}, we put forward four static clustering scenarios aiming to address interference management for the multi-beam optical wireless access network. The proposed clustering scenarios along with their corresponding \ac{SINR} and data rate distributions are presented in Fig.~\ref{Fig:6_Scenarios}. For each scenario, the left diagram depicts the clustering layout, and the displayed numbers indicate the cluster size. The total number of clusters considered for scenarios $1$ to $4$, respectively, is $N_\mathrm{c}=9,25,49,49$. In scenario~$1$, as shown in Fig.~\subref*{Fig:6_Scenarios_a}, each cluster consists of $25$ coordinated \acp{VCSEL}. In this case, each \ac{VCSEL} array at the \ac{AP} constitutes a separate cluster of size $25$. In scenario~$2$, as shown in Fig.~\subref*{Fig:6_Scenarios_b}, clusters are of a smaller size, with each one having $9$ coordinated \acp{VCSEL}. In scenarios $3$ and $4$, shown in Figs.~\subref*{Fig:6_Scenarios_c} and \subref*{Fig:6_Scenarios_d}, respectively, the size of clusters varies across the room. It can be observed that for all these scenarios, data rate values range from $8$~Gb/s to $15$~Gb/s, similar to the baseline \ac{SDMA}. The difference of the scenarios is in the distribution of their `hot' (i.e., high data rate) and `cold' (i.e., low data rate) regions as a result of the different clustering layouts used.

Moreover, in scenarios $1$ and $2$ in which clusters of an equal size are employed, the central cluster yields the highest data rate level, while the overall data rate values reduce for the neighboring clusters. Due to the symmetrical structure of the array of arrays of \acp{VCSEL} and the spatial tilting patterns of the beams for the $8$ arrays around the central array in the \ac{AP}, the clusters encompassing the $x$ and $y$ axes on the receiver plane produce higher data rates than those lying diagonally on the $xy$ plane. This can be explained by the fact that the clusters formed on the diagonal bisectors of the $xy$ plane are subject to a higher link distance from the \ac{AP}. Furthermore, compared to scenario $2$, scenario $1$ brings about a more even data rate distribution because of using a larger size for its clusters. In scenario $3$, different numbers of \acp{VCSEL} are assigned to clusters, as shown in Fig.~\subref*{Fig:6_Scenarios_c}. The rationale behind choosing these numbers is that we need to enlarge the size of clusters in proportion to their distance from the room center, in horizontal, vertical and diagonal directions on the $xy$ plane, so as to compensate for reduction in the received power by incorporating more \acp{VCSEL} in \ac{CoMP}-\ac{JT}. Note that the central region is mainly covered by the middle \ac{VCSEL} array of the \ac{AP} which has no tilt angle, while other regions are covered by the tilted \ac{VCSEL} arrays. It is observed that this scenario performs worse than the first two scenarios with uniformly sized clusters in terms of the data rate distribution. By contrast, in scenario $4$, shown in Fig.~\subref*{Fig:6_Scenarios_d}, when the size of the central cluster is changed to $9$ and the size of the clusters along the $x$ and $y$ axes is equally increased to $6$ while the same size of $4$ is kept for the rest of clusters, the resulting data rate distribution is better than that obtained by scenario $3$.

\vspace{-8pt}

\begin{figure*}[t!]
\begin{minipage}{0.3\linewidth}
    \vspace{-10pt}
    \centering
    \subfloat[NOMA \label{Fig:6_CDF1_10_SumRate_a}] {\includegraphics[width=\textwidth, keepaspectratio=true]{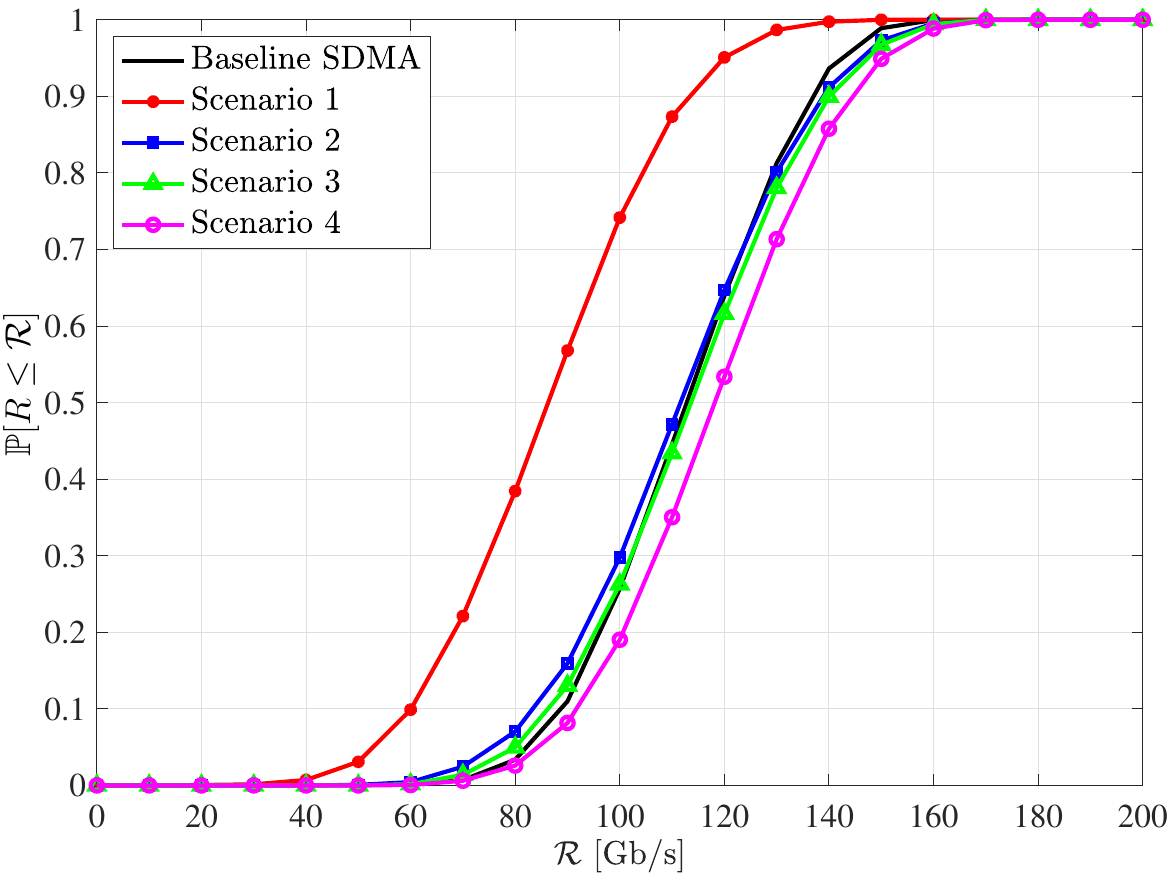}}\hfill
    \subfloat[OFDMA \label{Fig:6_CDF1_10_SumRate_b}] {\includegraphics[width=\textwidth, keepaspectratio=true]{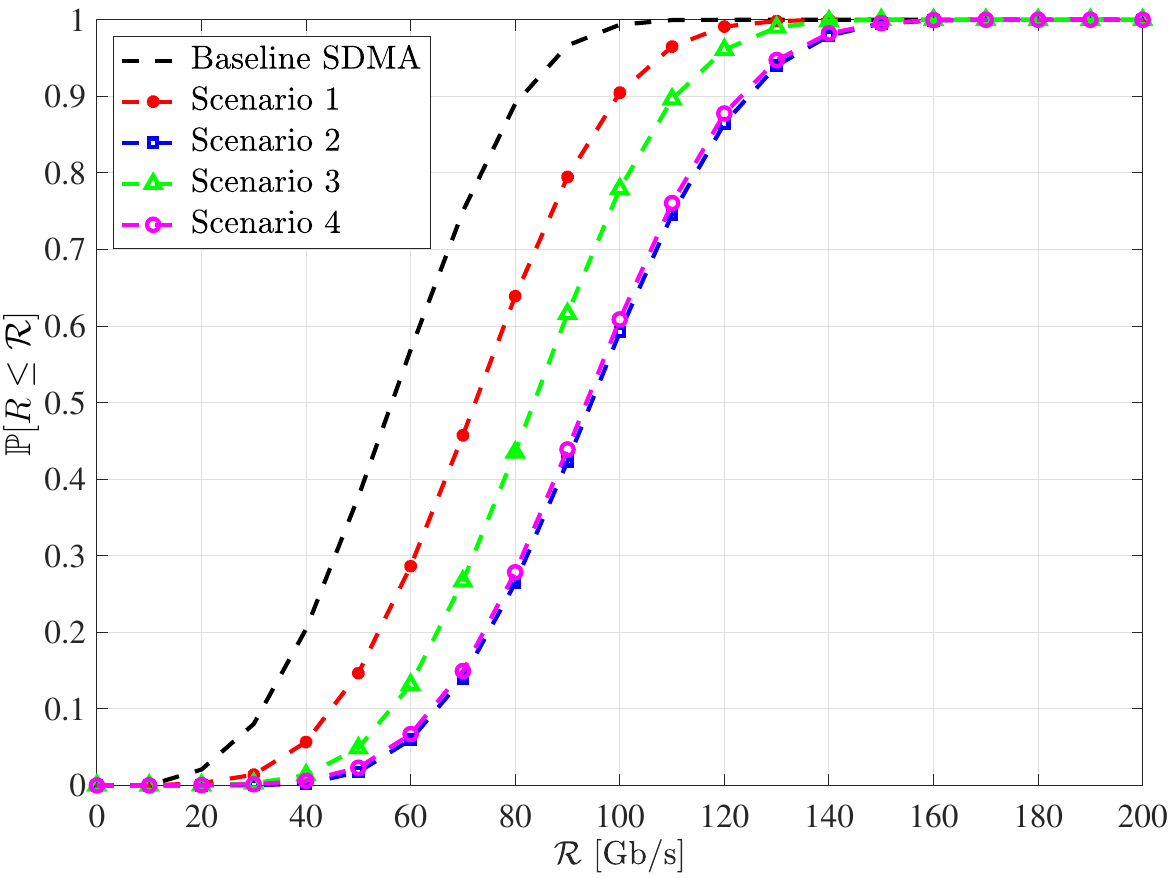}}
    \caption{CDF of the sum rate for different scenarios under NOMA and OFDMA for $K=10$.}
    \label{Fig:6_CDF1_10_SumRate}
\end{minipage}\hfill
\begin{minipage}{0.3\linewidth}
    \centering
    \subfloat[NOMA \label{Fig:6_CDF1_10_Fairness_a}] {\includegraphics[width=\textwidth, keepaspectratio=true]{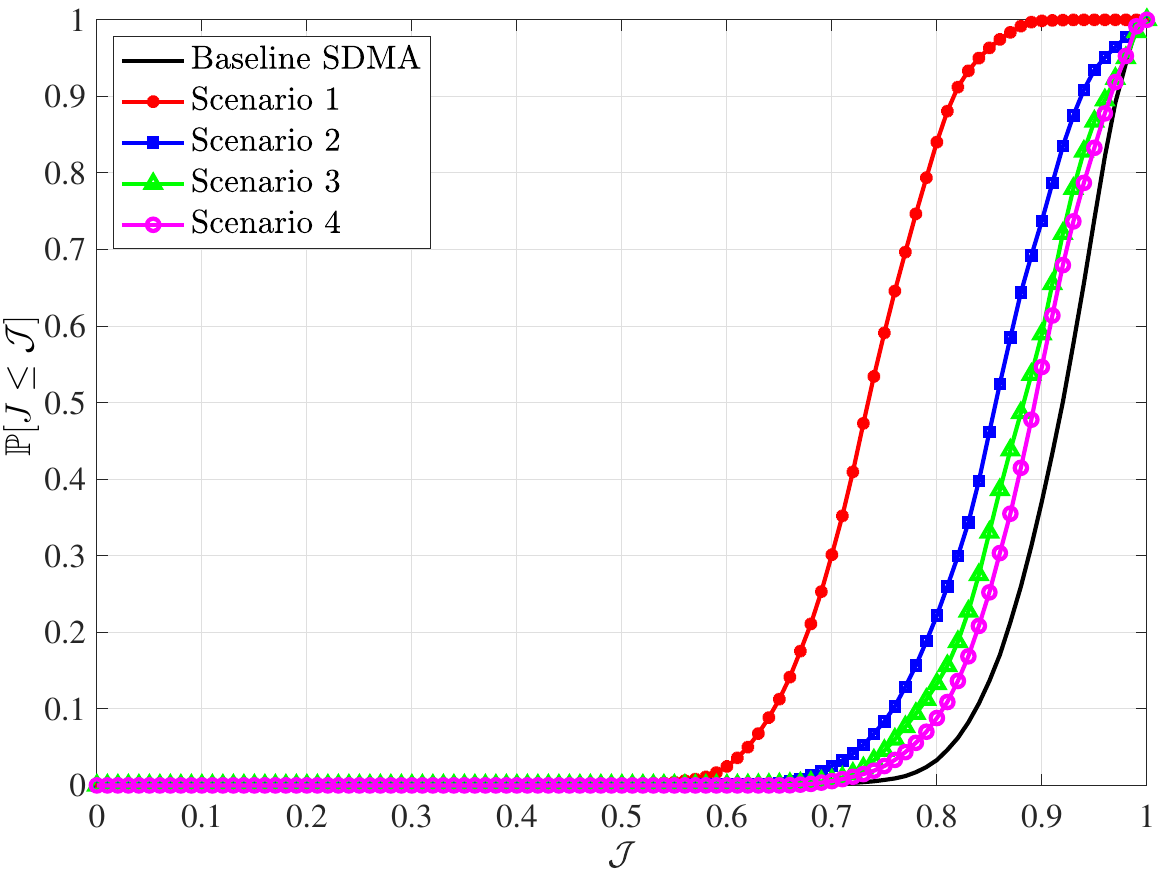}}\hfill
    \subfloat[OFDMA\label{Fig:6_CDF1_10_Fairness_b}] {\includegraphics[width=\textwidth, keepaspectratio=true]{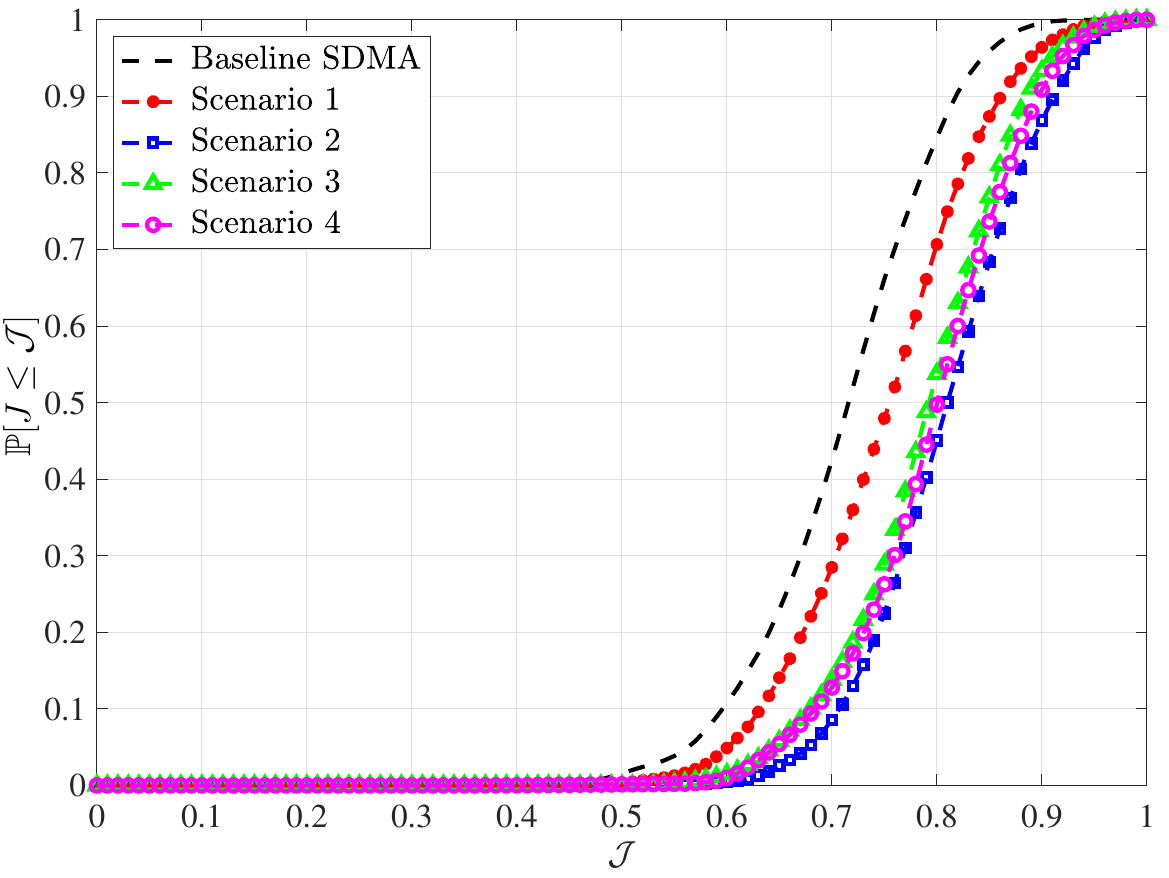}}
    \caption{CDF of the fairness index for different scenarios under NOMA and OFDMA for $K=10$.}
    \label{Fig:6_CDF1_10_Fairness}
\end{minipage}\hfill
\begin{minipage}{0.3\linewidth}
    \centering
    \subfloat[Average sum rate \label{Fig:6_AvgPerformance1_a}]{\includegraphics[width=\textwidth, keepaspectratio=true]{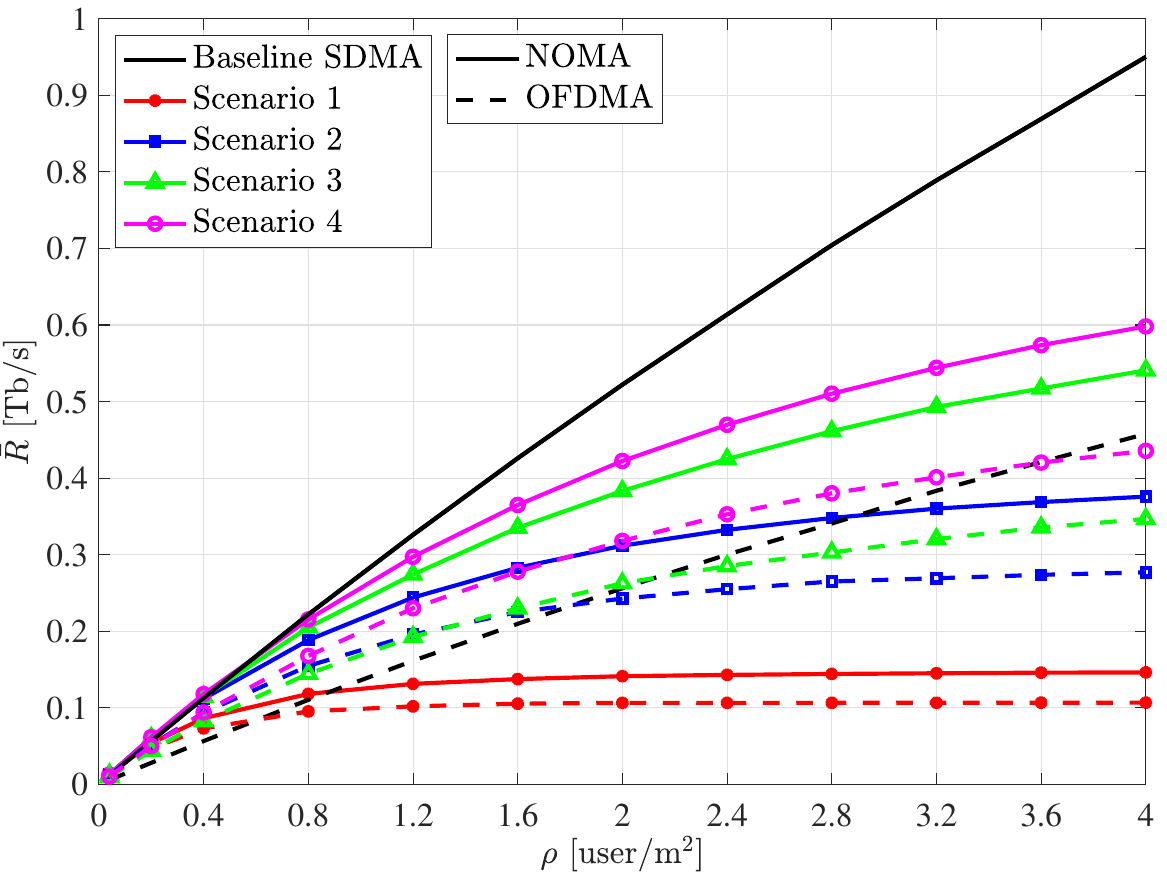}}\hfill
    \subfloat[Average fairness index \label{Fig:6_AvgPerformance1_b}]{\includegraphics[width=\textwidth, keepaspectratio=true]{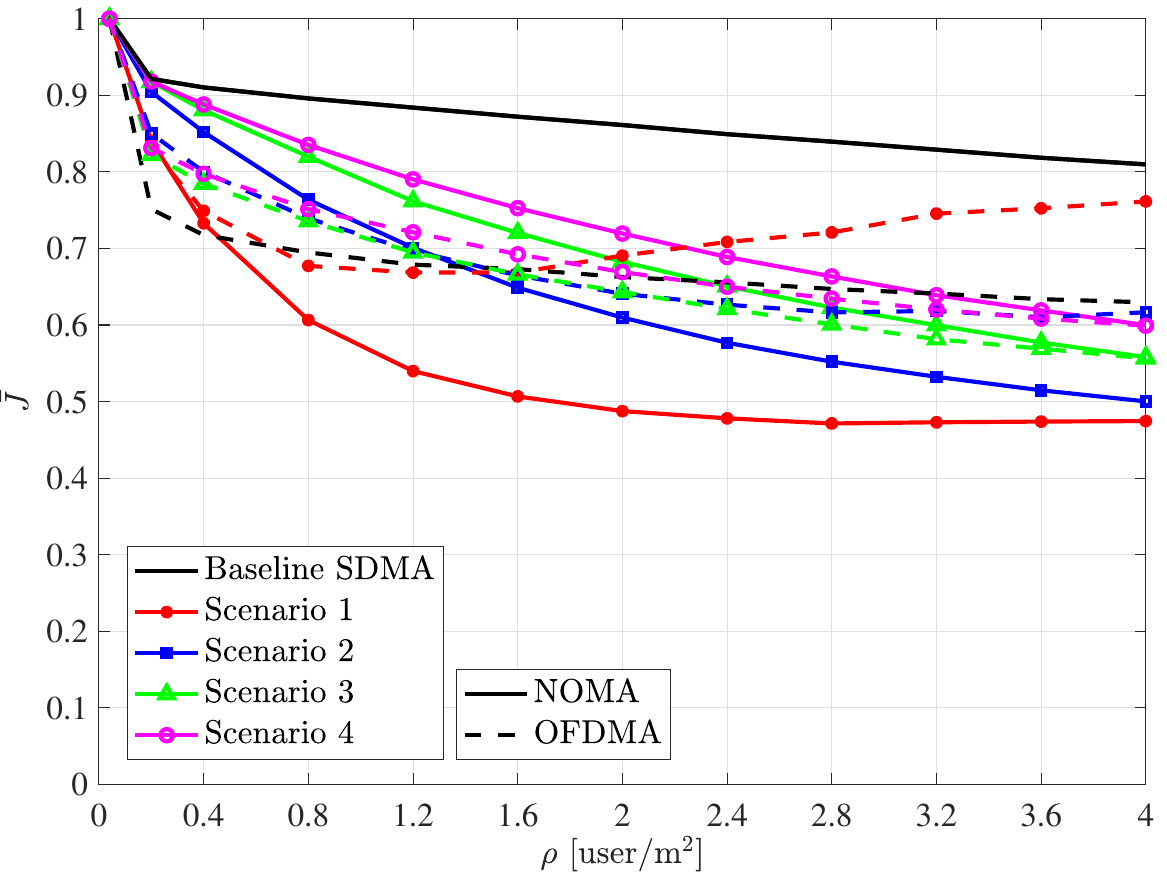}}
    \caption{Overall network performance versus the \ac{UE} density $\rho$. Solid curves correspond to NOMA and dashed curves represent OFDMA.}
    \label{Fig:6_AvgPerformance1}
\end{minipage}
\vspace{-15pt}
\end{figure*}

% \vspace{-7pt}

\subsection{Clustering Scenarios: Multi-User Performance Evaluation}
We study the multi-user performance of the proposed clustering scenarios based on Monte Carlo simulations. To this end, the $xy$ coordinates of $K$ \acp{UE} are randomly distributed in the room with a uniform distribution, and the statistics of the sum rate and the fairness index of the network are evaluated by using \eqref{Eq:SumRate} and \eqref{Eq:FairnessIndex} over $10^4$ random realizations. Note that the received \ac{SINR} in \eqref{Eq:SINR_NOMA} and \eqref{Eq:SINR_OFDMA} and hence the achievable rate in \eqref{Eq:RatePerUE} primarily depend on the $xy$ coordinates of \acp{UE} through the channel gain $H_{kj}^{\mathcal{C}_u}$, which is given by \eqref{Eq:Hk_Cu}, since $H_{kj}^{(i)}$ is related to the $xy$ coordinates of the $k$th \ac{UE} via \eqref{Eq:Hkj_i}.

First, let us consider the case where the number of \acp{UE} is fixed at $K=10$ as a representative example of low number of \acp{UE}. The \acp{CDF} of the sum rates offered by the clustering scenarios under \ac{NOMA} and \ac{OFDMA} are plotted in Fig.~\ref{Fig:6_CDF1_10_SumRate}, where $\mathbb{P}$ denotes the probability of an event. In the case of \ac{NOMA}, it can be observed from Fig.~\subref*{Fig:6_CDF1_10_SumRate_a} that scenarios $1$ and $2$ result in a worse sum rate performance relative to the baseline \ac{SDMA}, whereas scenarios $3$ and $4$ improve the performance. The sum rate for scenario $2$ closely follows that delivered by \ac{SDMA}. However, for scenario $1$, there is a large gap as compared to \ac{SDMA}. Unlike \ac{NOMA}, when \ac{OFDMA} is applied, as shown in Fig.~\subref*{Fig:6_CDF1_10_SumRate_b}, it is evident that the sum rate for all four scenarios is always improved with respect to \ac{SDMA}, as in \ac{OFDMA}, the \ac{SINR} improvement by \ac{CoMB}-\ac{JT} is a dominant factor leading to better data rates per \ac{UE} when the size of clusters $>1$.

The \acp{CDF} of the fairness indices are presented in Fig.~\ref{Fig:6_CDF1_10_Fairness}. For \ac{NOMA}, it can be observed from Fig.~\subref*{Fig:6_CDF1_10_Fairness_a} that the fairness performance for all the clustering scenarios is poorer than that achieved by \ac{SDMA}. By contrast, when \ac{OFDMA} is in use, the fairness index is effectively heightened by the clustering scenarios and a better performance is attained in comparison with \ac{SDMA}, likewise the sum rate results for \ac{OFDMA}.

To further shed light into the multi-user performance, let us turn our attention to the average performance metrics. The average sum rate $\Bar{R}$ is defined as the ergodic average of the sum rate, i.e., $\Bar{R}=\mathbb{E}[R]$ where $\mathbb{E}$ denotes the expected value of a random variable. The average fairness index $\Bar{J}$ is computed as $\Bar{J}=\mathbb{E}[J]$. The results are mainly presented as a function of \ac{UE} density $\rho$, which is defined as the normalized number of \acp{UE} per unit area of the room:
\begin{equation}
    \rho \triangleq \frac{K}{A_\mathrm{room}}.
\end{equation}

% \vspace{-8pt}

Fig.~\ref{Fig:6_AvgPerformance1} shows the average performance metrics for both \ac{NOMA} and \ac{OFDMA} for comparison over a range of \ac{UE} densities up to $\rho=4$ $\text{\ac{UE}}/\text{m}^2$, equivalent to $K=100$ \acp{UE}. From Fig.~\subref*{Fig:6_AvgPerformance1_a}, it can be observed that in both cases, the average sum rate is a monotonically increasing function of $\rho$. Also, \ac{NOMA} clearly outperforms \ac{OFDMA} in terms of the average sum rate for all the clustering scenarios. In the case of \ac{NOMA}, for low \ac{UE} densities of $\rho\leq0.4$ $\text{\ac{UE}}/\text{m}^2$, scenarios $3$ and $4$ slightly improve the performance over \ac{SDMA}. By increasing $\rho$, \ac{SDMA} takes the lead and its average sum rate approaches $1$ Tb/s for high \ac{UE} densities. Under \ac{OFDMA}, all the scenarios provide gain with respect to \ac{SDMA} over a range of $\rho\leq\rho_\mathrm{th}$, where $\rho_\mathrm{th}=0.6,1.8,2.1,3.6$ $\text{\ac{UE}}/\text{m}^2$ for scenarios $1$ to $4$, respectively. For both \ac{NOMA} and \ac{OFDMA}, as the \ac{UE} density increases, the average sum rate performance for all four scenarios converges to a constant value, which is scaled up with the number of clusters. Among the devised scenarios, scenario $4$ with $N_\mathrm{c}=49$ has the best performance and scenario $3$, though having the same number of clusters, holds the second best place. This highlights the point that the strategy used for adjusting the size of clusters in scenario $4$ is more efficient than that used in scenario $3$. We note that for both cases of \ac{NOMA} and \ac{OFDMA}, the sum rate is primarily limited in scenario $1$ in comparison with the other scenarios. The reason for this relates to the use of $9$ large clusters as shown in Fig.~\ref{Fig:6_Scenarios}, which leads to a significant underutilisation of the spectral resources over the network. Also, since the channel gain is almost uniformly distributed over the area of each cluster in scenario $1$, the gain of \ac{NOMA} over \ac{OFDMA} for high \ac{UE} densities is less than when scenarios with smaller clusters are deployed.

From the perspective of the average fairness index as shown in Fig.~\subref*{Fig:6_AvgPerformance1_b}, \ac{SDMA} with \ac{NOMA} retains the highest performance. For \ac{NOMA}, all the scenarios exhibit a monotonically decreasing behavior with $\rho$. Under \ac{OFDMA}, a similar trend is observed for all the scenarios excluding scenario 1, by which the fairness index raises for $\rho\geq1.2$ $\text{\ac{UE}}/\text{m}^2$ and is notably greater than that achieved by the baseline \ac{SDMA} at $\rho=4$ $\text{\ac{UE}}/\text{m}^2$. Furthermore, \ac{OFDMA} performs better than \ac{NOMA} in scenarios $1$ and $2$ and the performance gap of the two schemes becomes larger as $\rho$ increases. By comparison, \ac{NOMA} brings about higher fairness indices in scenarios $3$ and $4$, and the performance gap between \ac{NOMA} and \ac{OFDMA} is consistently reduced when $\rho$ increases.

\vspace{-10pt}

%%%%%%%%%%%%%%%%%%%%%%%%%%%%%%%%%%%%%%%%%%%%%%%%%%%%%%%%%%%%%%%%%%%%%%%%%%%%%%%%%%%%%%%%%%%%%%%%%%%%
%%%%%%%%%%%%%%%%%%%%%%%%%%%%%%%%%%%%%%%%%%%%%%%%%%%%%%%%%%%%%%%%%%%%%%%%%%%%%%%%%%%%%%%%%%%%%%%%%%%%
\section{Conclusions} \label{Conclusions}
We proposed the design of a novel double-tier \ac{AP} architecture built upon a $3\times3$ array of $5\times5$ \ac{VCSEL} arrays to establish an indoor grid-of-beam optical wireless access network. A full beam coverage with minimum overlaps between the beam spots were achieved by properly configuring the orientation angle of individual arrays. We introduced various beam clustering scenarios for downlink interference management and studied the multi-user performance of the network based on \ac{NOMA} and \ac{OFDMA} by considering a baseline \ac{SDMA} scheme whereby the system bandwidth is reused for every single beam. The results evince the superior performance of \ac{SDMA} with \ac{NOMA} in terms of the sum rate and the fairness index especially for higher \ac{UE} densities. In this case, the average sum rate of the network approaches $1$ Tb/s when the \ac{UE} density exceeds $4$ $\text{\ac{UE}}/\text{m}^2$. Instead, performance gains of the clustering scenarios are evident for \ac{OFDMA}. Particularly, scenario $4$ with $49$ clusters of balanced sizes yields $40\%$ improvement in the average sum rate performance compared to the baseline \ac{SDMA} for a \ac{UE} density of $1.2$ $\text{\ac{UE}}/\text{m}^2$. On the other hand, scenario $1$ in which every \ac{VCSEL} array is considered as a cluster of size $25$ improves the average fairness performance by $25\%$ relative to the baseline \ac{SDMA} for a \ac{UE} density of $4$ $\text{\ac{UE}}/\text{m}^2$. These results are indicative of the effectiveness of clustering-based interference management in the \ac{OFDMA} network. Future works include investigating the impact of an imperfect knowledge of the \acp{UE}' positions based on position estimation on the beam clustering performance.

% \vspace{-10pt}

%%%%%%%%%%%%%%%%%%%%%%%%%%%%%%%%%%%%%%%%%%%%%%%%%%%%%%%%%%%%%%%%%%%%%%%%%%%%%%%%%%%%%%%%%%%%%%%%%%%%
%%%%%%%%%%%%%%%%%%%%%%%%%%%%%%%%%%%%%%%%%%%%%%%%%%%%%%%%%%%%%%%%%%%%%%%%%%%%%%%%%%%%%%%%%%%%%%%%%%%%
\appendices
% \appendix
%------------------------------------------------
% Appendix A: Derivation of the Refraction Vector
%------------------------------------------------
\section{Derivation of the Refraction Vector in \eqref{Eq:Snell_v2_general}}\label{Appendix:A}
\begin{figure}[!t]
     \centering
     \subfloat[\label{Fig:VectorDecomposition} Vector decomposition in $\mathbb{R}^3$]{\includegraphics[width=0.5\linewidth]{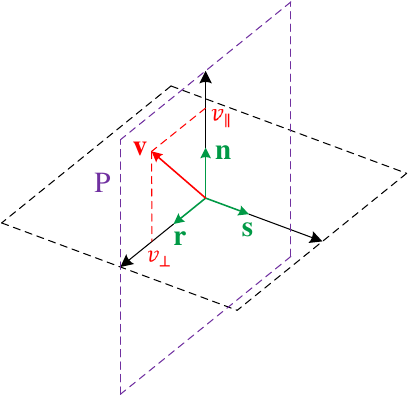}}
     \quad
     \subfloat[\label{Fig:Law_of_Refraction} The law of refraction]{\includegraphics[width=0.4\linewidth]{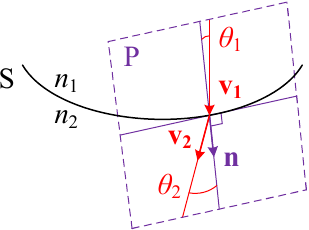}}
    \caption{Vector representations used for the analysis: (a) The decomposition of a vector $\mathbf{v}$ using a unit vector $\mathbf{n}$ in a given direction; (b) The refraction of a light ray at the boundary surface $S$ of two homogeneous media of refractive indices $n_1$ and $n_2$ according to the law of refraction.}
	\label{Fig:VectorRepresentations}
    \vspace{-10pt}
\end{figure}

In order to derive \eqref{Eq:Snell_v2_general}, first, we establish a key identity that allows for the decomposition of any arbitrary vector $\mathbf{v}\in\mathbb{R}^3$ using a unit vector $\mathbf{n}\in\mathbb{R}^3$ in a given direction. Fig.~\subref*{Fig:VectorDecomposition} depicts a right-handed Cartesian coordinate system with three unit and mutually orthogonal vectors of $\mathbf{r}$, $\mathbf{s}$ and $\mathbf{n}$, constituting an orthonormal basis for $\mathbb{R}^3$. Based on the right-hand rule, $\mathbf{s}=\mathbf{n}\times\mathbf{r}$ and $\mathbf{r}=\mathbf{s}\times\mathbf{n}$, and as a result:
\begin{equation}
    \mathbf{r} = \left(\mathbf{n}\times\mathbf{r}\right)\times\mathbf{n}.
    \label{Eq:RightHandRule}
\end{equation}
Note that the vector $\mathbf{v}$ lies in the plane spanned by $\mathbf{n}$ and $\mathbf{r}$, as represented by $P$ in Fig.~\subref*{Fig:VectorDecomposition}. Therefore, it can be expressed in the form:
\begin{equation}
    \mathbf{v} = v_\parallel\mathbf{n}+v_\perp\mathbf{r},
    \label{Eq:VectorComponents}
\end{equation}
where $v_\parallel$ and $v_\perp$ are the components of $\mathbf{v}$ along the directions of $\mathbf{n}$ and $\mathbf{r}$, respectively. The projection of $\mathbf{v}$ onto the direction of $\mathbf{n}$ is given by:
\begin{equation}
\mathbf{n}\cdot\mathbf{v} = v_\parallel\left(\mathbf{n}\cdot\mathbf{n}\right)+v_\perp\left(\mathbf{n}\cdot\mathbf{r}\right) = v_\parallel,
\label{Eq:n_dot_v}
\end{equation}
since $\mathbf{n}\cdot\mathbf{n}=1$ and $\mathbf{n}\cdot\mathbf{r}=0$. Also, the cross product of $\mathbf{n}$ and $\mathbf{v}$ is obtained as:
\begin{equation}
\mathbf{n}\times\mathbf{v} = v_\parallel\left(\mathbf{n}\times\mathbf{n}\right)+v_\perp\left(\mathbf{n}\times\mathbf{r}\right) = v_\perp\left(\mathbf{n}\times\mathbf{r}\right),
\label{Eq:n_cross_v}
\end{equation}
through the use of $\mathbf{n}\times\mathbf{n}=0$. Based on \eqref{Eq:RightHandRule}, it follows that:
\begin{equation}
\left(\mathbf{n}\times\mathbf{v}\right)\times\mathbf{n} = v_\perp\left(\mathbf{n}\times\mathbf{r}\right)\times\mathbf{n} = v_\perp\mathbf{r}.
\label{Eq:n_cross_v_cross_n}
\end{equation}
Combining \eqref{Eq:n_cross_v} and \eqref{Eq:n_cross_v_cross_n} with \eqref{Eq:VectorComponents}, we arrive at the following vector identity:
\begin{equation}
    \mathbf{v} = \left(\mathbf{n}\cdot\mathbf{v}\right)\mathbf{n}+\left(\mathbf{n}\times\mathbf{v}\right)\times\mathbf{n}.
    \label{Eq:VectorDecomposition}
\end{equation}

% \vspace{-12pt}

Now we can proceed with deriving the refraction vector $\mathbf{v}_2$ in terms of the incidence vector $\mathbf{v}_1$ and the normal vector $\mathbf{n}$ to the boundary surface $S$ between medium~$1$ and medium~$2$ with refractive indices $n_1$ and $n_2$, as shown in Fig.~\subref*{Fig:Law_of_Refraction}. Note that $\mathbf{v}_1$ and $\mathbf{v}_2$ are unit vectors and they both lie on the plane of incidence $P$, forming the angle of incidence $\theta_1$ and the angle of refraction $\theta_2$ with respect to $\mathbf{n}$. By applying the vector decomposition in \eqref{Eq:VectorDecomposition} and the law of refraction in \eqref{Eq:Law_of_Refraction}, $\mathbf{v}_2$ can be expressed as:
\begin{equation}
\begin{aligned}
    \mathbf{v}_2 &= \left(\mathbf{n}\cdot\mathbf{v}_2\right)\mathbf{n}+\left(\mathbf{n}\times\mathbf{v}_2\right)\times\mathbf{n}, \\
                 &= \left(\mathbf{n}\cdot\mathbf{v}_2\right)\mathbf{n}+\mu\left(\mathbf{n}\times\mathbf{v}_1\right)\times\mathbf{n}.
\label{Eq:v2}
\end{aligned}
\end{equation}
By using Snell's law in \eqref{Eq:Snell_Scalar} and considering that $0\leq\theta_2\leq\dfrac{\pi}{2}$ as observed from Fig.~\subref*{Fig:Law_of_Refraction},  and hence $\cos\theta_2\geq0$, we obtain:
\begin{equation}
\begin{aligned}
    \mathbf{n}\cdot\mathbf{v}_2 &= \cos\theta_2, \\
                                &= \sqrt{1-\sin^2\theta_2}, \\
                                &= \sqrt{1-\mu^2\sin^2\theta_1}, \\
                                &= \sqrt{1-\mu^2\left(1-\cos^2\theta_1\right)}, \\
                                &= \sqrt{1-\mu^2\left(1-\left(\mathbf{n}\cdot\mathbf{v}_1\right)^2\right)}.
\end{aligned}
\label{Eq:n_dot_v2}
\end{equation}
Furthermore, based on \eqref{Eq:VectorDecomposition}, we have:
\begin{equation}
    \left(\mathbf{n}\times\mathbf{v}_1\right)\times\mathbf{n} = \mathbf{v}_1-\mathbf{n}\cdot\mathbf{v}_1.
    \label{Eq:v1-n_dot_v1}
\end{equation}
Substituting \eqref{Eq:n_dot_v2} and \eqref{Eq:v1-n_dot_v1} into \eqref{Eq:v2}, $\mathbf{v}_2$ is derived in terms of $\mathbf{n}$ and $\mathbf{v}_1$, resulting in the desired expression as given by \eqref{Eq:Snell_v2_general}:
\begin{equation}
   \mathbf{v}_2 = \sqrt{1-\mu^2\left(1-\left(\mathbf{n}{\cdot}\mathbf{v}_1 \right)^2\right)}\mathbf{n}+\mu\left(\mathbf{v}_1-\left(\mathbf{n}{\cdot}\mathbf{v}_1\right)\mathbf{n}\right).
\end{equation}
%
%------------------------------------------------
% End of Appendix A
%------------------------------------------------

%%%%%%%%%%%%%%%%%%%%%%%%%%%%%%%%%%%%%%%%%%%%%%%%%%%%%%%%%%%%%%%%%%%%%%%%%%%%%%%%%%%%%%%%%%%%%%%%%%%%
%%%%%%%%%%%%%%%%%%%%%%%%%%%%%%%%%%%%%%%%%%%%%%%%%%%%%%%%%%%%%%%%%%%%%%%%%%%%%%%%%%%%%%%%%%%%%%%%%%%%
\section*{Acknowledgement}
The authors acknowledge financial support by the Engineering and Physical Sciences Research Council (EPSRC) under grant EP/S016570/1 `Terabit Bidirectional Multi-User Optical Wireless System (TOWS) for 6G LiFi'. The authors also acknowledge discussions with Dr. Mohammad Dehghani Soltani during early stages of this work in 2021.

\bibliographystyle{IEEEtran}
\bibliography{IEEEabrv,references}


% \end{IEEEbiography}


%\newpage



%\vfill

% \enlargethispage{-5in}

\begin{IEEEbiography}[{\includegraphics[width=1in,height=1.25in,clip,keepaspectratio]{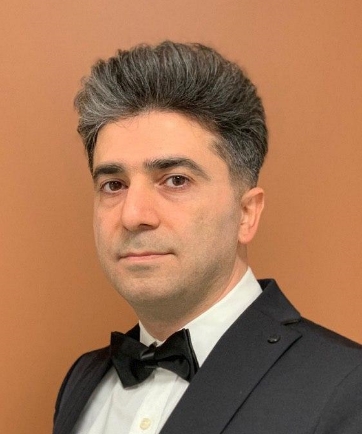}}]{Hossein Kazemi}
(Member, IEEE) received the Ph.D. degree in Electrical Engineering from The University of Edinburgh, U.K., in 2019. He also received the M.Sc. degree in Electrical Engineering (Microelectronic Circuits) from Sharif University of Technology, Tehran, Iran, in 2011, and the M.Sc. degree (Hons.) in Electrical Engineering (Wireless Communications) from Ozyegin University, Istanbul, Turkey, in 2014. He is a Postdoctoral Research Associate at the LiFi Research and Development Center, University of Cambridge, U.K. Dr Kazemi was the recipient of the Best Paper Award for the 2022 IEEE Global Communications Conference (GLOBECOM). His current research interests include the design, analysis and optimization of ultra-high-speed optical wireless communication systems for 6G and beyond networks. 
\end{IEEEbiography}

% \vfill

\begin{IEEEbiography}[{\includegraphics[width=1in,height=1.25in,clip,keepaspectratio]{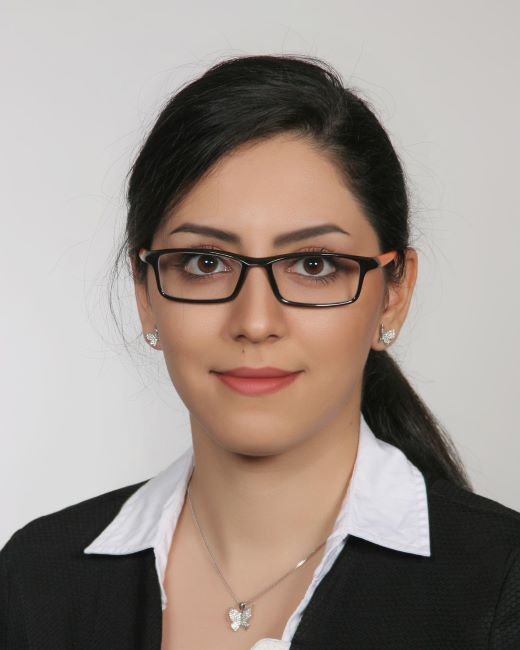}}]{Elham Sarbazi} received the Ph.D. degree in Electrical Engineering from the University of Edinburgh, U.K., in 2019. She also received the B.Sc. degree in Electrical and Computer Engineering from the University of Tehran, Iran, in 2011, and the M.Sc. degree (Hons.) in Electrical Engineering from Ozyegin University, Turkey, in 2014. Since 2019, she was with the LiFi Research and Development Centre as a Postdoctoral Research Associate, at the University of Edinburgh, Edinburgh, U.K. and the University of Strathclyde, Glasgow, U.K. Dr Sarbazi was the recipient of the Best Paper Award for the 2022 IEEE Global Communications Conference (GLOBECOM). Her current research interests focus on the modelling, design and optimization of ultra-high speed optical wireless transmitters and receivers for 6G and beyond networks.
\end{IEEEbiography}

% \vfill

\begin{IEEEbiography}[{\includegraphics[width=1in,height=1.25in,clip,keepaspectratio]{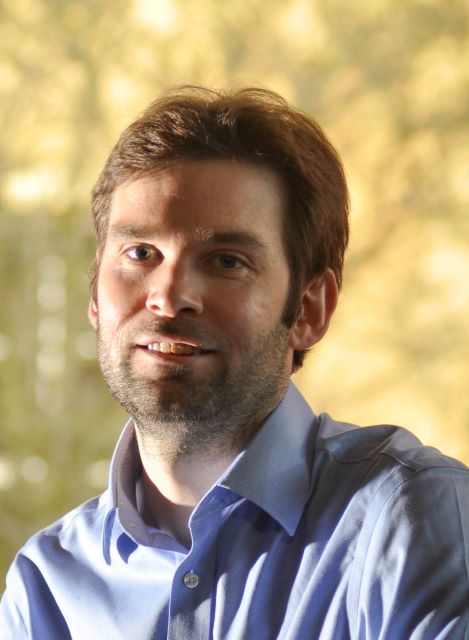}}]{Michael Crisp}
(Member, IEEE) received his M.Eng. and Ph.D. degrees in from the University of Cambridge, Cambridge, U.K., in 2005 and 2010, respectively. He is currently an Associate Professor of RF systems with the Engineering Department at University of Cambridge. His current research interests include radio over fibre systems, microwave photonics, optical wireless communications and RF systems for next generation battery-less sensors. He has authored and co-authored over 80 conference and journal publications and co-founded a spin out company PervasID Ltd. He was a recipient of the Royal Academy of Engineering Young Entrepreneurs Award in 2011.
\end{IEEEbiography}

% \vfill

\begin{IEEEbiography}[{\includegraphics[width=1in,height=1.25in,clip,keepaspectratio]{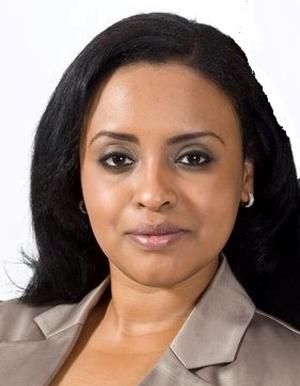}}]{Taisir Elgorashi} 
received the B.S. degree (Hons.) in electrical and electronic engineering from the University of Khartoum, Khartoum, Sudan, in 2004, the M.Sc. degree (Hons.) in photonic and communication systems from the University of Wales, Swansea, UK, in 2005, and the Ph.D. degree in optical networking from the University of Leeds, Leeds, UK, in 2010. She is currently a Senior Lecturer (Associate Professor) with the Department of Engineering, King’s College London, London, UK. Before joining King’s College London, she was a lecturer in optical networks with the School of Electronic and Electrical Engineering, University of Leeds from 2015 to 2022 and held a postdoctoral research position at the same school from 2010 to 2014, where she focused on the energy efficiency of optical networks investigating the use of renewable energy in core networks, green data centres, distributed cloud computing and network virtualization. The energy efficiency techniques developed by her contributed three out of eight carefully chosen core network energy efficiency improvement measures recommended by the GreenTouch consortium, a consortium of 55 leading ICT industry and academic research organisations, for every operator network worldwide. She was a BT Research Fellow in in 2012 studying hybrid wireless optical broadband access networks. Her work led to several invited talks at GreenTouch, Bell Labs, the Optical Network Design and Modelling Conference, the Optical Fiber Communications Conference, the International Conference on Computer Communications, and the EU Future Internet Assembly in collaboration with Alcatel-Lucent and Huawei. She was awarded the IET 2016 Premium Award for best paper in IET Optoelectronics. She co-chaired multiple times the Green Communication Systems and Networks (GCSN) symposium in the IEEE Communication Society (ComSoc) flagship conferences, ICC (2020, 2023) and GLOBECOM (2016, 2018).  She is currently an associate Editor in IET Optoelectronics and IEEE Transactions on Green Communications and Networking.
\end{IEEEbiography}

% \vfill

\begin{IEEEbiography}[{\includegraphics[width=1in,height=1.25in,clip,keepaspectratio]{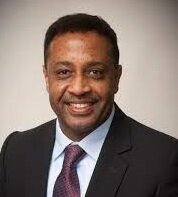}}]{Jaafar Elmirghani}
(Fellow, IEEE) is Fellow of IEEE, Fellow of IET and Fellow of Institute of Physics. He is Director of Technology, Research and Innovation in Neom; Professor of Communication Networks and Systems at Kings College London. He was the Director of the Institute of Communication and Power Networks and Professor of Communication Networks and Systems within the School of Electronic and Electrical Engineering, University of Leeds 2007-2022. He joined Leeds in 2007 having been full Professor and chair in optical communications at the University of Wales Swansea 2000-2007. He has given over 120 invited and keynote talks. He has provided outstanding leadership in a number of large research projects, secured over £50m in grants and was PI of the £6m EPSRC Intelligent Energy Aware Networks (INTERNET) Programme Grant, 2010-2016. He is Co-Chair of the IEEE Sustainable ICT initiative, a pan IEEE Societies initiative responsible for Green ICT activities across IEEE, 2012-present. He received the IEEE Communications Society 2005 Hal Sobol award for exemplary service to meetings and conferences, the IEEE Communications Society 2005 Chapter Achievement award, the University of Wales Swansea inaugural ‘Outstanding Research Achievement Award’, 2006, the IEEE Communications Society Signal Processing and Communication Electronics outstanding service award, 2009, the IEEE Comsoc Transmission Access and Optical Systems outstanding Service award 2015 in recognition of “Leadership and Contributions to the Area of Green Communications”, the GreenTouch 1000x award in 2015 for “pioneering research contributions to the field of energy efficiency in telecommunications", the IET 2016 Premium Award for best paper in IET Optoelectronics, shared the 2016 Edison Award in the collective disruption category with a team of 6 from GreenTouch for their joint work on the GreenMeter, the IEEE Communications Society Transmission, Access and Optical Systems technical committee 2020 Outstanding Technical Achievement Award for outstanding contributions to the “energy efficiency of optical communication systems and networks”. He was elected Fellow of IEEE for “Contributions to Energy-Efficient Communications,” 2021. His work led to 5 IEEE standards with a focus on cloud and fog computing and energy efficiency, where he currently heads the work group responsible for IEEE P1925.1, IEEE P1926.1, IEEE P1927.1, IEEE P1928.1 and IEEE P1929.1 standards; this resulting in significant impact through industrial and academic uptake. He was named among the top 2\% of scientists in the world by citations, in 2019, in Elsevier Scopus, Stanford University database, named ever since. He is/has been on the Technical Programme Committee of 43 IEEE ICC/GLOBECOM Conferences, between 1995 and 2022, including 21 times as the Symposium Chair. He was Area Editor of IEEE JSAC series on Machine Learning for Communications and is PI of the EPSRC £6.6m Terabit Bidirectional Multi-user Optical Wireless System (TOWS) for 6G LiFi, 2019-2025. He has published over 600 technical papers, supervised over 75 PhDs and has research interests in energy efficiency, optical wireless systems, optimisation, machine learning, cloud and fog computing.
\end{IEEEbiography}

% \vfill

\begin{IEEEbiography}[{\includegraphics[width=1in,height=1.25in,clip,keepaspectratio]{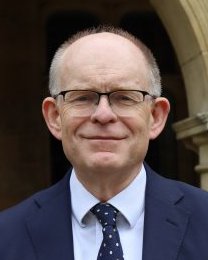}}]{Richard V. Penty}
(Senior Member, IEEE) received the Ph.D. degree in engineering from the University of Cambridge, U.K., in 1989. He was a Science and Engineering Research Council IT Fellow researching all optical nonlinearities in waveguide devices in Cambridge. In 2019, he was appointed as the Deputy Head of the School of Technology, University of Cambridge, and the Deputy Vice Chancellor in 2020, becoming the Head of the School of Technology in 2023. He is currently a Professor of photonics with the University of Cambridge, having previously held academic posts with the University of Bath and the University of Bristol. His research interests include high-speed optical communication systems, photonic integration, optical switching, and sensing systems and quantum communications. He is a fellow of the Royal Academy of Engineering and the IET. He was the Editor-in-Chief of the IET Optoelectronics journal from 2006 to 2019.
\end{IEEEbiography}

% \vfill

\begin{IEEEbiography}[{\includegraphics[width=1in,height=1.25in,clip,keepaspectratio]{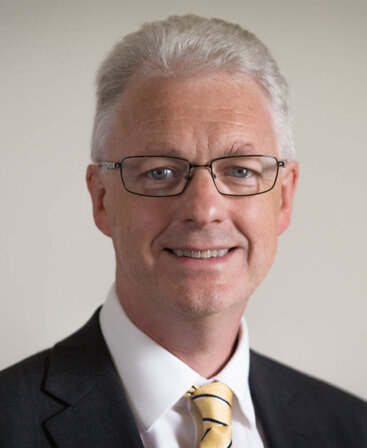}}]{Ian H. White} 
(Fellow, IEEE) is a Professorial Fellow at the University of Bath, having previously served as Vice-Chancellor and President of the University since 2019. Prior to that he was a Pro-Vice-Chancellor of the University of Cambridge and then Master of Jesus College, Cambridge. Professor White’s research interests are in photonics, communications and RFID tracking, his work having led to him receive the IEEE Aron Kressel Award in 2011. He is a co-founder of Zinwave Ltd, a company providing wireless coverage solutions, and PervasID Ltd which develops RFID readers, and is currently an editor-in-chief of Nature Microsystems and Nanoengineering. At Cambridge, Ian led the Photonics activity in the Electrical Division of the Department of Engineering where he led a series of major projects funded by EPSRC. His research interests in photonics include optical data communications, quantum communications and laser diode-based devices. Arising from this research, he has studied advanced RF Access Systems and RfID detection. In respect of technical standards, Ian led the 10GbE IEEE 802.3aq channel ad hoc study group following the invention of offset launch for Gigabit Ethernet, this standard now having several billion links worldwide, with devices arising from the work in this standard now finding wide use in electronic consumer products. Ian has approximately 1000 publications and is an editor-in-chief of Nature Microsystems and Nanoengineering. He is a Fellow of the Institute of Electrical and Electronic Engineers (IEEE), the UK Royal Academy of Engineering, and the Institution of Engineering and Technology. He was appointed CBE in the King’s Birthday Honours 2024.
\end{IEEEbiography}

% \vfill

\begin{IEEEbiography}[{\includegraphics[width=1in,height=1.25in,clip,keepaspectratio]{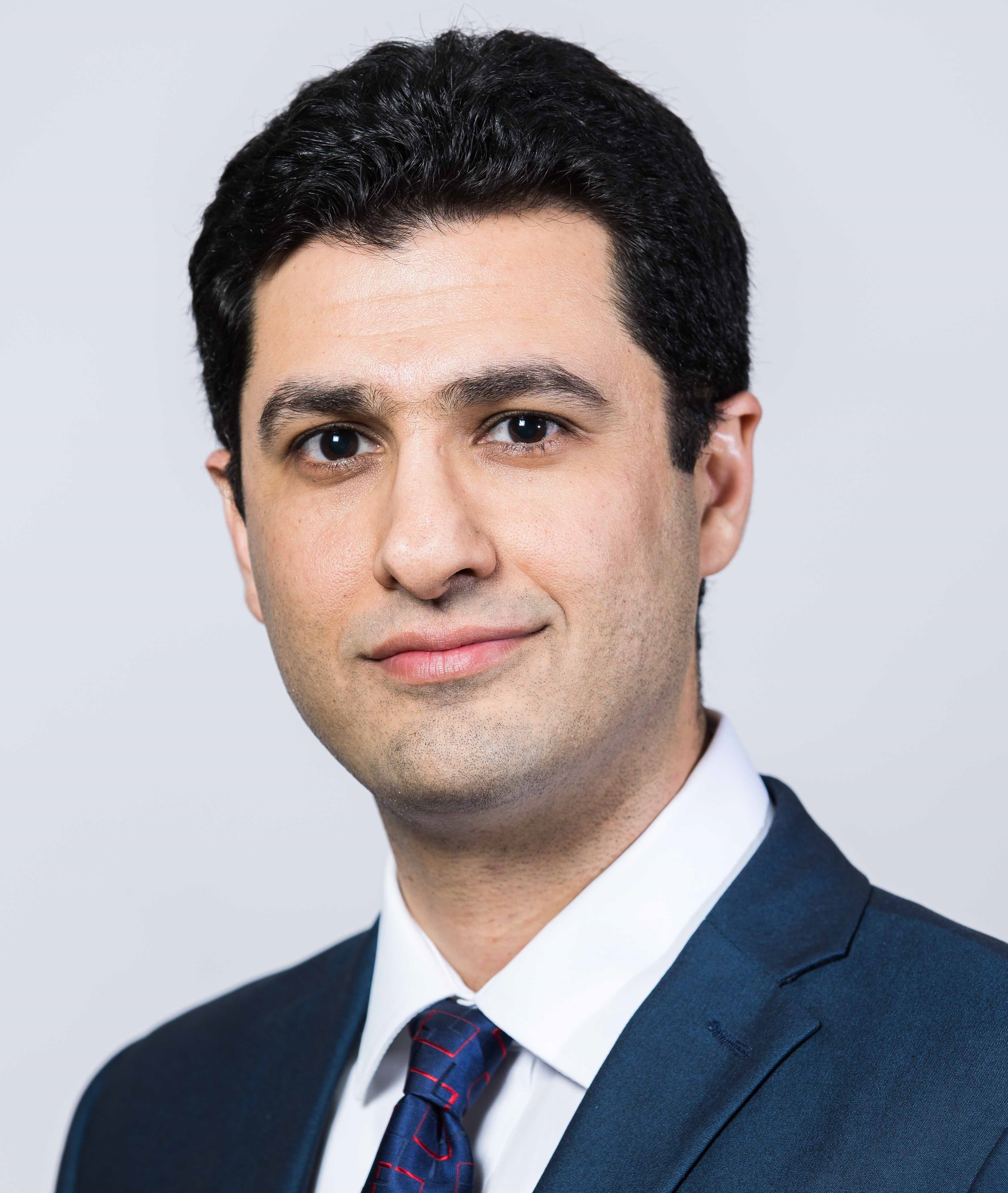}}]{Majid Safari} 
(Senior Member, IEEE) received Ph.D. in Electrical and Computer Engineering from the University of Waterloo, Canada in 2011. He is currently a Professor of Optical and Wireless Communications and the Deputy Head of Institute for Imaging, Data, and Communications (IDCOM) at the University of Edinburgh. He is a recipient of Mitacs Fellowship, Canada and prestigious grants from Leverhulme Trust and EPSRC, UK. He has published more than 150 papers and received best paper awards from IEEE GLOBECOM 2022 and IEEE ICC 2023. Prof Safari's main research interests include the application of optics, information theory, and signal processing in optical, wireless, and quantum communications. Prof Safari is a senior member of IEEE and has served as an Associate Editor of IEEE Transactions on Communications and IEEE Communication Letters. He has also served as the TPC co-chair of the OWC workshop at IEEE WCNC 2023 and IEEE GLOBECOM 2024.
\end{IEEEbiography}

% \vfill

\begin{IEEEbiography}[{\includegraphics[width=1in,height=1.25in,clip,keepaspectratio]{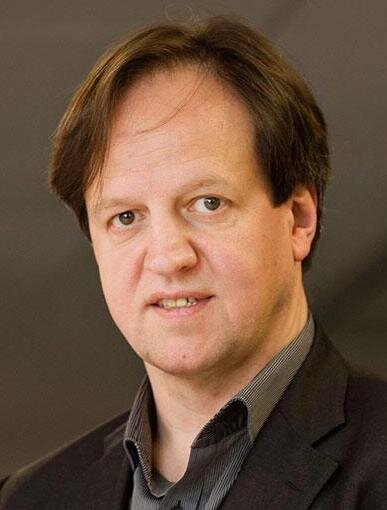}}]{Harald Haas} 
(Fellow, IEEE) received his Ph.D. from The University of Edinburgh, U.K., in 2001. He is the Van Eck Chair of Engineering at the University of Cambridge and the founder of pureLiFi Ltd., where he also serves as the Chief Scientific Officer (CSO). His recent research interests focus on photonics, communication theory and signal processing for optical wireless communication systems. Since 2017, he has been recognised as a highly cited researcher by Clarivate/Web of Science. He has delivered two TED talks and one TEDx talk. In 2016, he received the Outstanding Achievement Award from the International Solid State Lighting Alliance. He was awarded the Royal Society Wolfson Research Merit Award in 2017, the IEEE Vehicular Technology Society James Evans Avant Garde Award in 2019, and the Enginuity: The Connect Places Innovation Award in 2021. In 2022, he received the Humboldt Research Award for his research contributions. He is a Fellow of the Royal Academy of Engineering (RAEng), the Royal Society of Edinburgh (RSE), and the Institution of Engineering and Technology (IET). In 2023, he was shortlisted for the European Inventor Award.
\end{IEEEbiography}

\end{document}

%% file: acronyms.tex
\acrodef{1D}{one-dimensional}
\acrodef{2D}{two-dimensional}
\acrodef{3D}{three-dimensional}
\acrodef{4G}{4th generation}
\acrodef{5G}{5th generation}
\acrodef{6G}{sixth generation}
\acrodef{ADT}{angle diversity transmitter}
\acrodef{ADR}{angle diversity receiver}
\acrodef{AI}{artificial intelligence}
\acrodef{ANSI}{American national standards institute }
\acrodef{AP}{access point}
\acrodef{AR}{augmented reality}
\acrodef{AWGN}{additive white Gaussian noise}
\acrodef{AWGR}{arrayed waveguide grating router}
\acrodef{BDMA}{beam division multiple access}
\acrodef{BER}{bit error ratio}
\acrodef{BS}{base station}
\acrodef{CDF}{cumulative distribution function}
\acrodef{CSI}{channel state information}
\acrodef{CPC}{compound parabolic concentrator}
\acrodef{CoMB}{coordinated multi-beam}
\acrodef{CoMB{-}JT}{coordinated multi-beam joint transmission}
\acrodef{CoMP}{coordinated multi-point}
\acrodef{DC}{direct current}
\acrodef{DCO{-}OFDM}{DC-biased optical orthogonal frequency division multiplexing}
\acrodef{DFB}{distributed feedback laser}
\acrodef{DMT}{discrete multitone}
\acrodef{EGC}{equal gain combining}
\acrodef{FEC}{forward error correction}
\acrodef{FFT}{fast Fourier transform}
\acrodef{FF}{fill factor}
\acrodef{FOV}{field of view}
\acrodef{FSO}{free space optics}
\acrodef{IBI}{inter-beam interference}
\acrodef{IEC}{International electrotechnical commission}
\acrodef{IFFT}{inverse fast Fourier transform}
\acrodef{IM{/}DD}{intensity modulation and direct detection}
\acrodef{IM{-}DD}{intensity modulation and direct detection}
\acrodef{IoT}{Internet of Things}
\acrodef{IR}{infrared}
\acrodef{ISI}{inter-spot interference}
\acrodef{ICI}{inter-cluster interference}
\acrodef{JT}{joint transmission}
\acrodef{LD}{laser diode}
\acrodef{LED}{light-emitting diode}
\acrodef{LiFi}{light fidelity}
\acrodef{MIMO}{multiple input multiple output}
\acrodef{MPE}{maximum permissible exposure}
\acrodef{MPTP}{maximum permissible transmit power}
\acrodef{MHP}{most hazardous position}
\acrodef{MRC}{maximum ratio combining}
\acrodef{MUI}{multi-user interference}
\acrodef{NIR}{near infrared}
\acrodef{NOMA}{non-orthogonal multiple access}
\acrodef{OOK}{on-off keying}
\acrodef{PAM}{pulse amplitude modulation}
\acrodef{PSD}{power spectral density}
\acrodef{PD}{photodiode}
\acrodef{OFDM}{orthogonal frequency division multiplexing}
\acrodef{OFDMA}{orthogonal frequency division multiple access}
\acrodef{OCC}{optical camera communication}
\acrodef{OW}{optical wireless}
\acrodef{OWC}{optical wireless communication}
\acrodef{QAM}{quadrature amplitude modulation}
\acrodef{RF}{radio frequency}
\acrodef{RIN}{relative intensity noise}
\acrodef{RSS}{received signal strength}
\acrodef{SDMA}{space division multiple access}
\acrodef{SIC}{successive interference cancellation}
\acrodef{SiPM}{silicon photomultiplier}
\acrodef{SMF}{single mode fiber}
\acrodef{SINR}{signal-to-interference-plus-noise ratio}
\acrodef{SNR}{signal-to-noise ratio}
\acrodef{Tb{/}s}{Terabit/s}
\acrodef{TIA}{transimpedance amplifier }
\acrodef{TEM}{transverse electromagnetic}
\acrodef{UE}{user equipment}
\acrodef{UD}{user device}
\acrodef{UHD}{ultra-high-definition}
\acrodef{VCSEL}{vertical cavity surface emitting laser}
\acrodef{VLC}{visible light communication}
\acrodef{VR}{virtual reality}
\acrodef{WiFi}{wireless fidelity}

\acrodef{LOS}{line-of-sight}
\acrodef{NLOS}{non-line-of-sight}
\acrodef{GoB}{grid-of-beam}